\documentclass[11pt,a4paper]{article}
\usepackage[margin=25mm]{geometry}
\usepackage[T1]{fontenc}
\usepackage{lmodern,microtype,amsmath,amssymb,amsthm}
\usepackage[titles]{tocloft}
\usepackage{array,booktabs,tabularx,hyperref}
\cftsetindents{section}{0em}{2.8em}
\cftsetindents{subsection}{2.8em}{3.6em}
\cftsetindents{subsubsection}{6.4em}{4em}
\newcolumntype{L}[1]{>{\raggedright\arraybackslash}p{#1}}
\newcolumntype{Y}{>{\raggedright\arraybackslash}X}
\newtheorem{theorem}{Theorem}
\newtheorem{proposition}[theorem]{Proposition}
\newtheorem{lemma}[theorem]{Lemma}
\newtheorem{corollary}[theorem]{Corollary}
\theoremstyle{remark}
\newtheorem{remark}[theorem]{Remark}
\newcommand{\F}{\mathbb F}\newcommand{\Z}{\mathbb Z}
\newcommand{\PAut}{\operatorname{PAut}}\newcommand{\Aut}{\operatorname{Aut}}
\newcommand{\Sym}{\operatorname{Sym}}\newcommand{\Alt}{\operatorname{Alt}}
\newcommand{\core}{\operatorname{core}}\newcommand{\Rad}{\operatorname{Rad}}
\newcommand{\rad}{\operatorname{rad}}\newcommand{\ord}{\operatorname{ord}}
\newcommand{\PSL}{\operatorname{PSL}}\newcommand{\PGL}{\operatorname{PGL}}
\newcommand{\End}{\operatorname{End}}
\newcommand{\Av}{\operatorname{Av}}\newcommand{\Fix}{\operatorname{Fix}}
\newcommand{\N}{\mathbb N}
\title{Typical Permutation Automorphism Groups of Simple-Root Cyclic Codes}
\author{Guanghui Zhang}
\date{(School of Mathematics and Physics, Suqian University, Suqian, Jiangsu 223800, China)}
\begin{document}
\maketitle
\begin{abstract}
The automorphism group of a cyclic code measures the coordinate symmetries
that remain after its cyclic constituents have been selected.
We study simple-root cyclic codes sampled uniformly among all labelled
cyclic subspaces, allowing arbitrary sums of irreducible constituents.
For each fixed prime characteristic, we prove that the probability of
having automorphisms beyond the shift and Frobenius multiplier tends to
zero in length average, uniformly over all extension alphabets, with the
field supremum taken before the average. A separate arithmetic criterion
gives pointwise convergence when the length retains a sufficiently large
prime divisor on which the order of the characteristic equals its order
modulo the length. At lengths $q^m-1$, Stewart's theorem gives an
exceptional probability $o(m^{-A})$ for every fixed alphabet $q$ and
every fixed $A>0$, as $m$ tends to infinity through the positive integers.
These results also hold after conditioning on full shift period, and the
pointwise result persists in every fixed positive-width rate window about
one half. The proof combines exact cyclic-core descent, counts of actual
coordinate actions, local group-algebra operators, and composition-length
bounds. An exact distribution at prime-power lengths shows the necessary
limit of such statements: for binary length $3^j$, the largest probability
of any prescribed subgroup of the labelled symmetric group is $2^{1-j}$. Thus no deterministic
group can be a pointwise target along all simple-root lengths.
\end{abstract}


\clearpage
\tableofcontents
\clearpage

\section{Introduction}\label{sec:introduction}

Cyclic codes have coordinate symmetries that are prescribed by their
length and alphabet, as well as possible additional symmetries that depend
on the selected code. Determining when the latter occur connects the
structure of cyclic ideals with finite permutation groups. The
Berger--Charpin expectation, discussed by Charpin
\cite[Section 3.5]{Charpin} and recalled by Wu and Xiang \cite{WX},
is that a cyclic code generally has no permutation automorphisms beyond
the cyclic shift and an alphabet-dependent Frobenius multiplier.
A probabilistic formulation must specify what is sampled and how lengths
are weighted. We use the uniform distribution on all labelled cyclic codes
at each simple-root length, and distinguish length averages from assertions
along a prescribed sequence of lengths.

\subsection{Model and principal results}

Fix a prime power $q=p^u$. For $p\nmid n$, let $\mathcal C_q(n)$
be the set of cyclic subspaces of $\F_q^n$, with coordinates labelled by
$\Z_n$. Equivalently, these are the ideals of the semisimple algebra
$\F_q[x]/(x^n-1)$. A code chosen uniformly from $\mathcal C_q(n)$
selects each simple constituent independently with probability $1/2$.
The model includes reducible codes and the zero and full codes.
It does not assign equal probability to dimensions or to
permutation-equivalence classes.

Let
\[
 S_n=\langle\sigma_n\rangle,\qquad \sigma_n(x)=x+1,
 \qquad A(n)=N_{\Sym(n)}(S_n),
 \qquad A_q(n)=\langle\sigma_n,x\mapsto qx\rangle.
\]
The group $A(n)$ consists of all affine permutations of $\Z_n$,
and $A_q(n)$ preserves every code in $\mathcal C_q(n)$.
Write
\[
 \epsilon_q(n)=\Pr_{C\in\mathcal C_q(n)}
       \bigl(\PAut(C)\ne A_q(n)\bigr),
 \qquad N_q(X)=\#\{n\le X:(n,q)=1\}.
\]
For a function $v$ on the eligible lengths, put
\[
 \Av_{q,X}v=\frac{1}{N_q(X)}
      \sum_{\substack{n\le X\\(n,q)=1}}v(n).
\]
Thus the length average first chooses an eligible length uniformly up to
$X$, then chooses a code uniformly at that length. Throughout this paper,
\emph{nondegenerate} means that the code has full shift period:
\[
 \min\{d\ge1:\sigma_n^dc=c\text{ for every }c\in C\}=n.
\]
This is the absence of a shorter periodic representation of the code.

\begin{theorem}\label{thm:main}
For every fixed prime $p$,
\begin{equation}\label{eq:uniformtarget}
 \frac{1}{N_p(X)}\sum_{\substack{n\le X\\p\nmid n}}
          \sup_{u\ge1}\epsilon_{p^u}(n)\longrightarrow0
 \qquad(X\longrightarrow\infty).
\end{equation}
The same assertion holds when each code distribution is conditioned
on nondegeneracy.
\end{theorem}

The field supremum in~\eqref{eq:uniformtarget} precedes the length
average. In particular, the extension degree may vary arbitrarily with
the length, provided that the characteristic remains fixed. For each
fixed alphabet $q$, the theorem gives
\begin{equation}\label{eq:target}
 \Av_{q,X}\epsilon_q\longrightarrow0.
\end{equation}
This conclusion concerns the proportion of exceptional codes at each
length. It does not assert that all codes at most lengths have the same
automorphism group.

A length average gives no information on an arbitrary sparse sequence.
The next result supplies a pointwise criterion in terms of one prime
divisor of the length.

\begin{theorem}\label{gl:criterion}
Fix a prime $p$, and let $n$ tend to infinity through integers coprime
to $p$. Put $m=\ord_n(p)$. Suppose that, at each length, there is a
prime divisor $z$ of $n$ such that
\begin{equation}\label{gl:condition}
 \ord_z(p)=m,
 \qquad \frac{z}{m(\log m)^2}\longrightarrow\infty.
\end{equation}
Then
\[
 \sup_{u\ge1}\epsilon_{p^u}(n)\longrightarrow0.
\]
The same conclusion holds when the original code laws are conditioned
on nondegeneracy.
\end{theorem}

The equality $\ord_z(p)=m$ says that $z$ has the full multiplicative
order of the length. Stewart's greatest-prime-factor theorem
\cite[Theorem 1.1]{Stewart} supplies such primes for all sufficiently
large exponents, yielding the following uniform statement.

\begin{corollary}\label{gl:primitive-divisors}\label{allm:completion}
Fix a prime $p$. For every sufficiently large integer $m$, let
$z_m=P(\Phi_m(p))$, where $P(N)$ denotes the greatest prime divisor
of $N$ and $\Phi_m$ is the $m$th cyclotomic polynomial. Then
\[
 \sup_{\substack{n\mid p^m-1\\z_m\mid n}}\;
 \sup_{u\ge1}\epsilon_{p^u}(n)\longrightarrow0
 \qquad(m\longrightarrow\infty).
\]
The same uniform assertion holds under nondegenerate conditioning.
In particular, this includes every choice of lengths
$\rad(p^m-1)\mid n_m\mid p^m-1$. For every fixed prime power $q$,
it also gives
\[
 \Pr_{C\in\mathcal C_q(q^m-1)}
       \bigl(\PAut(C)=A_q(q^m-1)\bigr)\longrightarrow1,
\]
with the same limit after conditioning on nondegeneracy.
\end{corollary}

There is no primality restriction on the exponent $m$ in this corollary.
The quantitative estimate in Theorem~\ref{gl:quantitative} further gives,
for every fixed $A>0$,
\[
 \sup_{\substack{n\mid p^m-1\\z_m\mid n}}\;
 \sup_{u\ge1}\epsilon_{p^u}(n)=o(m^{-A}),
\]
for both the unconditioned and the nondegenerate laws; see
Corollary~\ref{gl:superpolynomial}. The rate here is expressed in the
exponent $m$, not in the code length. Proposition~\ref{prop:rate-conditioning}
also shows that pointwise convergence under~\eqref{gl:condition} persists
when codes are conditioned to have rate in any fixed positive-width
window about $1/2$. Conditioning on one exact dimension is a different
problem.

\subsection{Relation to previous results}

Feng, Hollmann, Li, and Xiang classified the nonstandard nondegenerate
irreducible cyclic codes \cite[Theorem 1.1]{FHLXClassification}.
Their subsequent density theorem \cite[Theorem 1.1]{FHLXDensity}
states that, outside a density-zero set of lengths, every nondegenerate
irreducible cyclic code over every extension alphabet of a fixed
characteristic is standard. This universal assertion for irreducible codes
does not directly control a random sum of constituents. A permutation
preserving their sum need not preserve the selected constituents
individually. The estimates below therefore apply to the whole code
rather than to the automorphism groups of its irreducible summands.

Wu and Xiang \cite[Theorems A--C]{WX} study the universal group
\[
 B_q(n)=\bigcap_{C\in\mathcal C_q(n)}\PAut(C)
       =\Aut\bigl(\operatorname{Cyc}(\langle q\rangle,\Z_n)\bigr).
\]
They characterize the lengths where $B_q(n)>A_q(n)$ and show that
these lengths have density zero for each fixed alphabet. Their
Proposition 5.10 gives $B_q(q^m-1)=A_q(q^m-1)$ at every primitive
length. An intersection identity specifies the automorphisms shared by
all codes; it does not give the probability that a single code attains
that group. Table~\ref{tab:comparison} records the different quantifiers.

\begin{table}[htbp]
\centering\small
\caption{Objects and quantifiers in the relevant results.}
\label{tab:comparison}
\renewcommand{\arraystretch}{1.15}
\begin{tabularx}{\linewidth}{@{}L{0.23\linewidth}L{0.27\linewidth}Y@{}}
\toprule
Work & Object & Conclusion \\
\midrule
Feng--Hollmann--Li--Xiang \cite{FHLXClassification}
& Nondegenerate irreducible cyclic codes
& Classification of nonstandard codes. \\
Feng--Hollmann--Li--Xiang \cite{FHLXDensity}
& All such irreducible codes, over all alphabets of fixed characteristic
& The lengths admitting a nonstandard code have density zero. \\
Wu--Xiang \cite{WX}
& Intersection of all cyclic-code automorphism groups at a fixed length
& Exceptional lengths have density zero for fixed $q$; the intersection
is $A_q(n)$ at every length $q^m-1$. \\
Present paper
& One uniform labelled cyclic code, possibly reducible
& The exceptional proportion has characteristic-uniform length average
zero, with separate pointwise estimates under~\eqref{gl:condition}. \\
\bottomrule
\end{tabularx}
\end{table}

There is an obstruction to replacing these conclusions by pointwise
concentration at every eligible length. Wu and Xiang's arithmetic
criterion gives $B_2(3^j)>A_2(3^j)$ for $j\ge2$
\cite[Theorem~B, condition (E2), and Example~5.7]{WX}.
Corollary~\ref{tree:universal-probability} gives the stronger identity
\[
 \sup_{H\le\Sym(3^j)}
 \Pr_{C\in\mathcal C_2(3^j)}\bigl(\PAut(C)=H\bigr)
 =2^{1-j}\longrightarrow0,
\]
also under full shift-period conditioning. More generally, an exact
calculation holds at length $\ell^j$ whenever $\ell$ is odd and
$\ord_{\ell^2}(q)=\ell(\ell-1)$. It follows from the periodic-code
automorphism formula \cite[Lemma 2.4]{FHLXClassification} and the
cyclotomic decomposition. In the binary case the same wreath groups
occur in the constructions of Bienert and Klopsch
\cite[Theorem 5.1]{BienertKlopsch}. We use this calculation to determine
the limits of a pointwise probability formulation, rather than to claim
a new classification of those groups.

\subsection{Proof structure}

The proof separates two tasks: finding a structured action forced by an
additional automorphism, and bounding the probability that a code is
invariant under that action. Canonical translation and alternating
witnesses admit direct counts in terms of Frobenius orbits. Exact
reflection through cyclic cores then reduces the remaining events to
quotient actions. The quotient maps are fixed coordinate maps, and each
has a uniform image code before conditioning. A union bound over these
maps avoids any assumption that a quotient selected by an observed
witness has a uniform conditional law.

The group-theoretic reduction uses the primitive full-cycle classification
\cite[Theorem 1.2]{JZ}, which depends on the classification of finite
simple groups. Diagonal factors and their coordinate covers must be
retained in their actual permutation representations. Counting abstract
group isomorphism types would omit the multiplicities and embeddings
that determine the probability of the union. For natural cross-characteristic projective actions, the local
group algebra produces an alternating action on the blocks. For
nontrivial scalar covers in the same characteristic range, it
produces a nonconstant translation on prime-size blocks. The
remaining cases consist of fixed simple types and
defining-characteristic scalar covers.

For the length average, fixed-type amplification combines with
Wiertelak's unconditional multiplicative-order theorem, in the form
stated by Moree \cite[Theorem~1 and the paragraph following it]{Moree},
and summable degree tails.
For pointwise bounds, the prime $z$ in~\eqref{gl:condition} survives
every stripped quotient and yields many full-length orbits under
multiplication by $p$ on its Fourier labels. For $q=p^u$ a length-$m$
orbit can split into $\gcd(m,u)$ shorter orbits. The uniform estimates
use the monotonicity of the orbit defects and amplification counts
under this splitting. A logarithmic cutoff separates defining-characteristic
parameters controlled by amplification from those controlled by small
composition length. Bardoe--Sin's descriptions
\cite[Theorems A and C]{BardoeSin} give the latter bounds, including
a separate binary rank-two estimate. The resulting estimates are uniform
in the alphabet extension degree and in the retained-prime divisor
families.

Table~\ref{tab:dependencies} summarizes the proof dependencies. The
structural reduction is shared by the length-average and pointwise
arguments; the exact nonconcentration calculation is independent of
both positive limit theorems.

\begin{table}[htbp]
\centering\small
\caption{Organization and dependencies of the proof.}
\label{tab:dependencies}
\renewcommand{\arraystretch}{1.12}
\begin{tabularx}{\linewidth}{@{}L{0.08\linewidth}L{0.43\linewidth}Y@{}}
\toprule
Section & Main task & Role in the argument \\
\midrule
I & Model, results, and comparison & Specifies the probability laws. \\
II & Cyclic ideals and probability estimates & Supplies constituent selection and fixed quotient laws. \\
III & Canonical translation and alternating actions & Counts the first two obstruction events. \\
IV & Cyclic cores and coordinate covers & Establishes exact descent and localization. \\
V & Exhaustive structural reduction & Reduces every remaining nonaffine event to simple-group actions. \\
VI & Characteristic-uniform length averages & Proves Theorem~\ref{thm:main} using Sections II--V. \\
VII & Pointwise estimates and retained primes & Proves Theorem~\ref{gl:criterion} and its quantitative consequences. \\
VIII & Exact prime-power distribution & Shows that no deterministic target works at all lengths. \\
IX & Composite order carriers & Gives a pointwise family in which no prime divisor has the full order. \\
X & Conclusion & States the scope and remaining questions. \\
\bottomrule
\end{tabularx}
\end{table}

\section{Cyclic modules and probability tools}\label{sec:preliminaries}

Throughout the paper, $q=p^u$ is a prime power and $p\nmid n$.
All logarithms are natural. The functions $\varphi$, $\tau$, and
$\rad$ denote Euler's totient, the number of positive divisors, and
the product of the distinct prime divisors, respectively; in particular,
$\rad(1)=1$. The functions $\omega$ and $\Omega$ count prime factors
without and with multiplicity. When $\tau$ denotes a permutation, this
will be stated explicitly. We use the convention $\ord_1(q)=1$.
Unless otherwise specified, probabilities at length $n$ refer to the
uniform law on $\mathcal C_q(n)$ introduced in Section~\ref{sec:introduction}.

\subsection{Independent cyclic constituents and canonical events}

The simple-root hypothesis gives the semisimple algebra
\[
 \F_q[C_n]\cong\F_q[x]/(x^n-1)
       \cong\prod_{i=1}^{m_q(n)}\F_{q^{d_i}},
 \qquad
 m_q(n)=\sum_{d\mid n}\frac{\varphi(d)}{\ord_d(q)}.
\]
Here $d_1,\ldots,d_{m_q(n)}$ are the degrees of the distinct
irreducible factors of $x^n-1$. Each field factor has precisely two
ideals. Consequently, a uniform cyclic code is obtained by selecting
each simple constituent independently with probability $1/2$, and
\begin{equation}\label{eq:constituent-law}
 |\mathcal C_q(n)|=2^{m_q(n)},\qquad
 \sum_{i=1}^{m_q(n)}d_i=n,\qquad d_i\mid\ord_n(q).
\end{equation}
Over a splitting field the shift has $n$ distinct eigenlines, indexed by
$\Z_n$. The scalar extension of each simple $\F_q$-constituent is the
sum of the eigenlines in one multiplication-by-$q$ orbit.
The coordinate multiplier $x\mapsto qx$ permutes the Fourier
eigenlines within these orbits. Thus it preserves every cyclic code,
which proves $A_q(n)\le\PAut(C)$ directly from the constituent model.

Write $\sigma=\sigma_n$ when the length is fixed. For a prime
$\ell\mid n$, define the independent translations on the canonical
blocks of size $\ell$ by
\[
 V_\ell(n)=
 \left\{x\longmapsto x+(n/\ell)v_{x\bmod(n/\ell)}:
                  v\in\F_\ell^{\,n/\ell}\right\}.
\]
The condition of original-code translation rigidity is
\begin{equation}\label{eq:Q}
 Q(C):\quad \PAut(C)\cap V_\ell(n)
     =\langle\sigma_n^{n/\ell}\rangle
                 \quad\text{for every prime }\ell\mid n.
\end{equation}
We also write $\mathcal A(C)$ for the event that some group
$S_n\le G\le\PAut(C)$ preserves a coordinate-block system of block
size at least five and that the induced action of a block stabilizer
contains the natural alternating group. The whole coordinate set is
allowed as one block, and the blocks need not be minimal.

Two probabilities used throughout the structural reduction are
\begin{equation}\label{eq:nonaffine-rigidity-probabilities}
 f_q(n)=\Pr\bigl(\PAut(C)\not\le A(n)\bigr),\qquad
 t_q(n)=\Pr\bigl(\neg Q(C)\bigr).
\end{equation}
At length one, $Q$ holds vacuously, $\mathcal A$ is false, and
$f_q(1)=t_q(1)=\epsilon_q(1)=0$. This convention also agrees with
the trivial coordinate action.

\subsection{Fixed quotient codes}

The structural argument will produce a shorter code by identifying
coordinates in the orbits of a cyclic subgroup. It is important to
define every possible quotient before selecting a witnessing group.
For a fixed divisor $v\mid n$, put $c=n/v$ and
\begin{equation}\label{eq:fixed-quotient-definition}
 M_c=\langle\sigma_n^v\rangle,\qquad
 \iota_v:\F_q^v\longrightarrow\F_q^n,
 \quad (\iota_v(w))_x=w_{x\bmod v},\qquad
 D_v(C)=\iota_v^{-1}\bigl(C\cap\Fix(M_c)\bigr).
\end{equation}
Thus $\iota_v$ identifies $\F_q^v$ with the vectors constant on
the $M_c$-orbits. In particular, $D_n(C)=C$.

\begin{lemma}\label{lem:quotient-law}
For each fixed divisor $v\mid n$, the random code $D_v(C)$ is
uniform on $\mathcal C_q(v)$. If
$S_n\le G\le\PAut(C)$ and $M_c\lhd G$, the group induced by
$G$ on the $M_c$-orbits preserves $D_v(C)$.
\end{lemma}
\begin{proof}
The embedding $\iota_v$ intertwines the length-$v$ shift with
$\sigma_n$, so $D_v(C)$ is cyclic. To identify its distribution,
choose a primitive $n$th root of unity over a splitting field.
A Fourier label $j\in\Z_n$ belongs to $\Fix(M_c)$ if and only if
\[
 n\mid jv\quad\Longleftrightarrow\quad c\mid j.
\]
The map $\Z_v\to\Z_n$, $j\mapsto cj$, is therefore a
multiplication-by-$q$-equivariant bijection from the length-$v$
Fourier labels onto the labels of this fixed space. It identifies
the simple constituents at length $v$ with a specified subset of
those at length $n$. Restricting the independent selection bits in
\eqref{eq:constituent-law} to this subset gives independent
unbiased bits, proving the uniform law.

If $M_c\lhd G$, every element of $G$ permutes its orbits and hence
preserves $\Fix(M_c)$. It also preserves $C$, so it preserves their
intersection. Transport through $\iota_v$ gives the asserted
action on $D_v(C)$.
\end{proof}

The distribution in Lemma~\ref{lem:quotient-law} is unconditional
for each fixed $v$. Selecting a quotient after observing a code
does not assert a conditional uniform law. Later arguments instead
take a union over the fixed maps in
\eqref{eq:fixed-quotient-definition}; they require no independence
between different quotient codes.

\subsection{Orbit defects under alphabet extension}

For a permutation $\phi$ of a finite set $Z$, let $o_\phi(Z)$ be
its number of orbits. Replacing $p$ by $p^u$ replaces the permutation
$x\mapsto px$ on Fourier labels by its $u$th power. The following
elementary fact will provide bounds uniform in the extension degree.

\begin{lemma}\label{lem:orbitdefect}
Suppose $X\to Y$ is a surjection equivariant for permutations of
$X$ and $Y$, both denoted by $\phi$. For every integer $u\ge1$,
\[
 o_{\phi^u}(X)-o_{\phi^u}(Y)
       \ge o_\phi(X)-o_\phi(Y)\ge0.
\]
\end{lemma}
\begin{proof}
Fix a $\phi$-orbit of $Y$ of length $b$, and let the $\phi$-orbits
above it have lengths $a_1,\ldots,a_r$. Surjectivity gives $r\ge1$,
and equivariance gives $b\mid a_i$ for every $i$. Taking the $u$th
power splits a cycle of length $a$ into $\gcd(a,u)$ cycles. Hence
the contribution of this part of the surjection to the new defect is
\[
 \sum_{i=1}^r\gcd(a_i,u)-\gcd(b,u)
       \ge(r-1)\gcd(b,u)\ge r-1.
\]
The contribution to the original defect is $r-1\ge0$.
Summing over the original orbits of $Y$ proves both inequalities.
\end{proof}

\subsection{Affine multipliers and nondegeneracy}\label{sec:affine}

Set
\[
 \begin{aligned}
 M_q(n)&=\frac{\varphi(n)}{\ord_n(q)},\\
 a_q(n)&=\Pr\bigl(\PAut(C)\cap A(n)\ne A_q(n)\bigr),\\
 d_q(n)&=\Pr(C\text{ has least common shift period less than }n).
 \end{aligned}
\]
In particular, $d_q(1)=0$. We call the event of full shift period
$\mathcal N$. Every affine exception can be detected by a multiplier
outside $\langle q\rangle$, while selecting a primitive constituent
already guarantees $\mathcal N$.

\begin{lemma}\label{lem:affine-nondegenerate}
For every admissible length,
\begin{align}
 a_q(n)&\le
   \min\{1,(M_q(n)-1)2^{-M_q(n)/2}\},
       \label{eq:affine-probability}\\
 d_q(n)&\le2^{-M_q(n)},\qquad
 \Pr(\mathcal N)\ge\tfrac12,
       \label{eq:degeneracy-probability}\\
 f_q(n)&\le\epsilon_q(n)\le f_q(n)+a_q(n).
       \label{eq:affine-nonaffine-split}
\end{align}
Moreover, for every event $E$,
\begin{equation}\label{eq:conditioning-difference}
 \bigl|\Pr(E\mid\mathcal N)-\Pr(E)\bigr|\le d_q(n).
\end{equation}
\end{lemma}
\begin{proof}
The primitive constituents are indexed by the quotient group
\[
 H_n=(\Z/n\Z)^*/\langle q\rangle,
 \qquad |H_n|=M_q(n).
\]
This group acts regularly on their indices. Since shifts already
preserve $C$, an affine automorphism outside $A_q(n)$ gives a
multiplier represented by a nonidentity element of $H_n$.
If that element has order $e\ge2$, its orbits on the primitive
constituents all have size $e$. A selected set of these constituents
is invariant only when its selection bits are constant on each
orbit. The probability of this event is therefore
\[
 \frac{2^{M_q(n)/e}}{2^{M_q(n)}}
    \le2^{-M_q(n)/2}.
\]
Invariance of the whole code implies invariance of this selected
primitive set, because coordinate multipliers preserve additive
orders of Fourier labels. A union over the $M_q(n)-1$
nonidentity elements gives \eqref{eq:affine-probability}.

On a primitive constituent, the shift has eigenvalues of order
$n$. Thus a shift power fixing that constituent pointwise must be
a multiple of $n$. A degenerate code consequently omits all
$M_q(n)$ primitive constituents, an event of probability
$2^{-M_q(n)}$. Selecting any one specified primitive constituent
has probability $1/2$ and implies $\mathcal N$, proving
\eqref{eq:degeneracy-probability}. At $n=1$ every code already
has full shift period, so these inequalities still hold.

Because $A_q(n)\le\PAut(C)$, a code with a nonaffine
automorphism is exceptional. Conversely, an exceptional code
either has a nonaffine automorphism or has an extra affine
automorphism. This proves \eqref{eq:affine-nonaffine-split}.

Finally, put $\delta=\Pr(\mathcal N^c)=d_q(n)$. If
$0<\delta<1$, total probability gives
\[
 \Pr(E\mid\mathcal N)-\Pr(E)
 =\delta\bigl(\Pr(E\mid\mathcal N)
                    -\Pr(E\mid\mathcal N^c)\bigr).
\]
Taking absolute values proves \eqref{eq:conditioning-difference}.
If $\delta=0$, the inequality is immediate. The conditional law
is always defined because \eqref{eq:degeneracy-probability}
gives $\Pr(\mathcal N)\ge1/2$.
\end{proof}

For later use, these bounds have simple arithmetic lower parameters.
Let $r(n)$ be the number of distinct odd prime divisors of $n$,
and let $\lambda(n)$ be the exponent of $(\Z/n\Z)^*$, with
$\lambda(1)=1$. If $r(n)\ge1$, then
\begin{equation}\label{eq:affine-odd-prime-bound}
 \lambda(n)\le\frac{\varphi(n)}{2^{r(n)-1}},\qquad
 M_q(n)\ge\frac{\varphi(n)}{\lambda(n)}\ge2^{r(n)-1}.
\end{equation}
Indeed, the exponent of each odd-prime-power unit group is its
even order, and the least common multiple of $r(n)$ even integers
is at most their product divided by $2^{r(n)-1}$. The possible
power-of-two component has exponent at most its order.
Also, for $q=p^u$,
\begin{equation}\label{eq:primitive-count-extension}
 \ord_n(p^u)=\frac{\ord_n(p)}{\gcd(\ord_n(p),u)},
 \qquad M_{p^u}(n)\ge M_p(n).
\end{equation}
Thus the elementary affine and degeneracy estimates can be made
uniform over all alphabets of one characteristic.

\subsection{Dimension and rate windows}

The uniform law on cyclic codes also determines their dimension
distribution. The following elementary estimate makes explicit
which rates this ensemble samples with high probability.

\begin{proposition}\label{prop:rate-window}
Let $K(C)=\dim_{\F_q}C$ for a uniform code in $\mathcal C_q(n)$.
With the constituent degrees from \eqref{eq:constituent-law},
\begin{equation}\label{eq:dimension-moments}
 \mathbb E K(C)=\frac n2,\qquad
 \operatorname{Var}K(C)=\frac14\sum_{i=1}^{m_q(n)}d_i^2
                   \le\frac{n\ord_n(q)}4.
\end{equation}
Consequently, for every $\eta>0$,
\begin{equation}\label{eq:rate-window-bound}
 \Pr\left(\left|\frac{K(C)}n-\frac12\right|>\eta\right)
 \le\min\left\{1,\frac{\ord_n(q)}{4\eta^2n}\right\}.
\end{equation}
For fixed $q$ and $n=q^m-1$, this gives
\begin{equation}\label{eq:primitive-rate-window}
 \Pr\left(\left|\frac{K(C)}{q^m-1}-\frac12\right|>\eta\right)
 \le\min\left\{1,\frac{m}{4\eta^2(q^m-1)}\right\}.
\end{equation}
More generally, if $n\mid p^m-1$, the bound
$m/(4\eta^2n)$ in \eqref{eq:rate-window-bound} is valid
simultaneously for every $q=p^u$.
\end{proposition}
\begin{proof}
Let $B_i$ be the indicator that the $i$th simple constituent is
selected. By \eqref{eq:constituent-law}, the variables $B_i$ are
independent Bernoulli variables with parameter $1/2$, and
\[
 K(C)=\sum_i d_iB_i.
\]
Therefore
\[
 \mathbb E K(C)=\frac12\sum_i d_i=\frac n2,
 \qquad
 \operatorname{Var}K(C)=\sum_i d_i^2\operatorname{Var}(B_i)
                      =\frac14\sum_i d_i^2.
\]
Each $d_i$ is a multiplication-by-$q$ orbit length on $\Z_n$,
so $d_i\le\ord_n(q)$. Hence
\[
 \sum_i d_i^2\le\ord_n(q)\sum_i d_i=n\ord_n(q),
\]
which proves \eqref{eq:dimension-moments}. Chebyshev's inequality
with deviation $\eta n$ proves \eqref{eq:rate-window-bound}.
For $n=q^m-1$ one has $\ord_n(q)\le m$, giving
\eqref{eq:primitive-rate-window}. If $n\mid p^m-1$, then
$\ord_n(p)\mid m$ and
$\ord_n(p^u)\le\ord_n(p)\le m$, giving the last assertion.
\end{proof}

For clarity, write
\[
 \mathcal W_{q,n}(\eta)=
 \left\{C:\left|\frac{\dim C}{n}-\frac12\right|\le\eta\right\}.
\]
Whenever $\Pr(\mathcal W_{q,n}(\eta))>0$, the same
total-probability identity used in
\eqref{eq:conditioning-difference} gives, for every event $E$,
\begin{equation}\label{eq:window-conditioning}
 \bigl|\Pr(E\mid\mathcal W_{q,n}(\eta))-\Pr(E)\bigr|
       \le\Pr\bigl(\mathcal W_{q,n}(\eta)^c\bigr).
\end{equation}
Thus any probability limit is preserved upon conditioning on a
sequence of such windows whose probabilities tend to one.
For example, at $n=q^m-1$, any positive sequence $\eta_m$
satisfying $\eta_m^2(q^m-1)/m\to\infty$ has this property.
The conditioning here allows a range of dimensions; it does not
assert a result for a code sampled uniformly at one prescribed
exact dimension.

\section{Canonical coordinate obstructions}
\label{sec:canonical}\label{tk:section}\label{ab:section}
We count two kinds of coordinate actions that will be used in the
structural reduction: independent translations on prime-size blocks,
and natural alternating actions on larger blocks. For each specified translation group and each specified alternating
block size we give an exact count. We use union bounds when several
witnesses can occur for the same code.
The corresponding losses are nondecreasing under extension of the coefficient field, so a
single pair of functions depending only on the characteristic controls
all extension alphabets.

\subsection{Periodic prime-block translations}
Fix a prime power $q=p^u$ and a length $n$ with $(n,p)=1$.
We use the uniform code distribution of Section~\ref{sec:preliminaries}
and abbreviate $\sigma_n$ to $\sigma$. Throughout this section,
$\ell$ and $r$ denote primes dividing coordinate lengths; the letter
$p$ is reserved for the coefficient characteristic. It is convenient
to denote the individual terms in the constituent count by
\[
 \kappa_q(d)=\frac{\varphi(d)}{\ord_d(q)},\qquad
                      m_q(v)=\sum_{d\mid v}\kappa_q(d).
\]

For a prime $\ell\mid n$, put $b=n/\ell$. The canonical blocks of
size $\ell$ are the residue classes modulo $b$. Write $\eta_v$ for
the element of $V_\ell(n)$ with translation vector
$v=(v_j)_{j\in\Z_b}\in\F_\ell^b$; explicitly,
\[
 \eta_v(x)=x+bv_{x\bmod b}\pmod n.
\]
Composition corresponds to addition of translation vectors, so
$V_\ell(n)\cong(\F_\ell^b,+)$. These are permutations of coordinate
positions. The constant vector $\boldsymbol1$ corresponds to
$\beta=\sigma^b$. For a cyclic code $C$, set
\[
 W_\ell(C)=\{v\in\F_\ell^b:\eta_v\in\PAut(C)\}.
\]
This is an $\F_\ell$-subspace containing the constants. Conjugation
by $\sigma$ rotates the entries of $v$, so $W_\ell(C)$ is invariant
under cyclic rotation. For every divisor $t\mid b$, define the
periodic translation group
\begin{equation}\label{tk:periodic-group}
 V_{\ell,t}(n)=
 \{\eta_v:v_j=v_{j\bmod t}\text{ for every }j\in\Z_b\}.
\end{equation}
Its order is $\ell^t$, and $V_{\ell,1}(n)=\langle\beta\rangle$.

\begin{theorem}\label{tk:closure}
For every cyclic code $C$ and every prime $\ell\mid n$, there is a
unique divisor $s_\ell(C)\mid b$ such that
\begin{equation}\label{tk:closure-equality}
 \PAut(C)\cap V_\ell(n)=V_{\ell,s_\ell(C)}(n).
\end{equation}
\end{theorem}
\begin{proof}
Put $W=W_\ell(C)$ and define
\[
 j\sim j'\quad\Longleftrightarrow\quad
 v_j=v_{j'}\text{ for every }v\in W.
\]
Rotation invariance makes this equivalence relation translation-invariant
on $\Z_b$. Its classes are therefore the cosets of a subgroup, hence
the residue classes modulo a unique divisor $t\mid b$. Initially this
shows only that $W$ is contained in the full space of period-$t$ vectors.

Let $K$ be an algebraic closure of $\F_q$, and choose a primitive
$\ell$th root of unity $\zeta\in K$. Fourier transformation on each
canonical $\ell$-block simultaneously diagonalizes $V_\ell(n)$.
The weights of $W$ occurring in $K^n$ are the characters
\[
 \chi_{k,j}(v)=\zeta^{kv_j},\qquad
              k\in\F_\ell,\quad j\in\Z_b.
\]
All weights with $k=0$ coincide. Evaluation on $\boldsymbol1\in W$
shows that, for $k,k'\ne0$,
\[
 \chi_{k,j}=\chi_{k',j'}
 \quad\Longleftrightarrow\quad k=k'\text{ and }j\sim j'.
\]
No nonzero weight equals the zero weight. Consequently the decomposition
of $K^n$ into common weight spaces is the same for $W$ and for the
full period-$t$ translation group.

The image of the group algebra of a finite abelian group of order
prime to $p$ consists of arbitrary scalar operators on its distinct
weight spaces. Indeed, the projection onto the $\chi$-weight space is
\[
 e_\chi=\frac1{|W|}\sum_{v\in W}\chi(v)^{-1}\eta_v.
\]
It follows that $W$ and $V_{\ell,t}(n)$ have the same group-algebra
image on $K^n$. The space $K\otimes_{\F_q}C$ is invariant under this
image, hence under $V_{\ell,t}(n)$. These permutation matrices are
defined over $\F_q$, so they preserve $C$ itself. Thus $W$ is the
full period-$t$ space. Its dimension $t$ proves uniqueness.
\end{proof}

The constant translation is essential in this argument: it separates
the different nonzero block characters. Rotation invariance of an
arbitrary subspace of $\F_\ell^b$ alone would not imply
\eqref{tk:closure-equality}.

\begin{theorem}\label{tk:exact-count}
For $t\mid b$, the number of cyclic codes admitting $V_{\ell,t}(n)$ is
\begin{equation}\label{tk:fixed-count}
 F_{q,\ell,n}(t)
 =2^{m_q(b)+m_q(n/t)-m_q(b/t)}.
\end{equation}
Moreover,
\begin{equation}\label{tk:period-distribution}
 \#\{C:s_\ell(C)=t\}
 =\sum_{\substack{w\mid b\\t\mid w}}
                         \mu(w/t)F_{q,\ell,n}(w).
\end{equation}
\end{theorem}
\begin{proof}
Write $b=ct$, so $n=\ell ct$. The independent generators of
$V_{\ell,t}(n)$ are supported on the strips $x\equiv j\pmod t$.
On one strip, in coordinates $x=j+ty$, the corresponding generator
sends $y$ to $y+c$ in $\Z_{\ell c}$.

The averaging idempotent of a strip generator projects onto its fixed
space; the complementary idempotent projects onto the nonfixed part of
that strip and is zero outside it. These complementary projections have
disjoint supports. Their sum is the projection onto the nonfixed part
of $\beta$, so every invariant code is the direct sum of its part in
$\Fix(\beta)$ and its nonfixed parts on the individual strips.

The first part is a cyclic code of length $b$, repeated $\ell$ times,
and has $2^{m_q(b)}$ possibilities. The shift transports the nonfixed
strip parts cyclically. The first such part must be invariant under
$\sigma^t$, which is the full $\ell c$-cycle on that strip. Its allowed
simple constituents are exactly those on which the step-$c$ translation
is nontrivial. The excluded constituents are the factors of $x^c-1$.
There are therefore
\[
 2^{m_q(\ell c)-m_q(c)}
\]
choices for this part. Conversely, every such choice, transported by
$\sigma$ to the other strips, together with any choice in
$\Fix(\beta)$, gives an invariant cyclic code. This proves
\eqref{tk:fixed-count}. No coprimality among $\ell,c,t$ is needed.

Finally, $V_{\ell,t}(n)\le V_{\ell,w}(n)$ if and only if $t\mid w$.
Theorem~\ref{tk:closure} gives
\[
 F_{q,\ell,n}(t)=
 \sum_{\substack{w\mid b\\t\mid w}}\#\{C:s_\ell(C)=w\},
\]
and divisor inversion gives \eqref{tk:period-distribution}.
\end{proof}

\begin{corollary}\label{tk:prime-witness}
The group $\PAut(C)\cap V_\ell(n)$ is larger than
$\langle\sigma^{n/\ell}\rangle$ if and only if
$V_{\ell,r}(n)\le\PAut(C)$ for some prime $r\mid n/\ell$.
For this witness,
\begin{equation}\label{tk:witness-probability}
 \Pr(V_{\ell,r}(n)\le\PAut(C))
     =2^{-\Delta_q(n;\ell,r)},
\end{equation}
where
\begin{equation}\label{tk:loss-definition}
 \Delta_q(n;\ell,r)
 =m_q(n)-m_q(n/\ell)-m_q(n/r)+m_q(n/(\ell r)).
\end{equation}
\end{corollary}
\begin{proof}
If $s_\ell(C)>1$, choose a prime $r\mid s_\ell(C)$ and use
$V_{\ell,r}(n)\le V_{\ell,s_\ell(C)}(n)$. Conversely, admitting
$V_{\ell,r}(n)$ forces $s_\ell(C)>1$. Divide
\eqref{tk:fixed-count}, with $t=r$, by $2^{m_q(n)}$.
\end{proof}

The numerical loss is symmetric in $\ell,r$, although the witnessing
groups need not be equal. In the union bound below the prime pairs
are ordered.

\begin{proposition}\label{tk:distinct-loss}
If $\ell\ne r$ and $\ell r\mid n$, then
\begin{equation}\label{tk:distinct-formula}
 \Delta_q(n;\ell,r)
 =\sum_{\substack{d\mid n\\d\nmid n/\ell\\d\nmid n/r}}\kappa_q(d)
 \ge2^{\omega(n)-2}.
\end{equation}
\end{proposition}
\begin{proof}
The divisors common to $n/\ell$ and $n/r$ are precisely the divisors
of $n/(\ell r)$, so inclusion--exclusion proves the equality.
Every divisor containing the full powers of $\ell$ and $r$ in $n$
contributes. Adjoining arbitrary squarefree products of the other
prime divisors gives $2^{\omega(n)-2}$ such divisors, and each
$\kappa_q(d)$ is a positive integer.
\end{proof}

For repeated-prime witnesses, write $n=\ell^ks$, where $k\ge2$ and
$(\ell,s)=1$, and set
\[
 R_j=\ord_{\ell^j}(q),\qquad
 a_j(d)=\frac{\varphi(\ell^j d)}{\ord_{\ell^j d}(q)}
                      \quad(d\mid s).
\]
Reduction of unit groups modulo $\ell^k$ to those modulo
$\ell^{k-1}$ has kernel of order $\ell$. Hence
$R_k/R_{k-1}\in\{1,\ell\}$, also when $\ell=2$.

\begin{proposition}\label{tk:repeated-loss}
If $R_k=R_{k-1}$, then
\begin{equation}\label{tk:no-growth}
 \Delta_q(n;\ell,\ell)
      =(\ell-1)\sum_{d\mid s}a_{k-1}(d).
\end{equation}
If $R_k=\ell R_{k-1}$ and $e=v_\ell(R_k)$, then $e\ge1$ and
\begin{equation}\label{tk:growth}
 \Delta_q(n;\ell,\ell)
      =(\ell-1)
        \sum_{\substack{d\mid s\\\ell^e\mid\ord_d(q)}}a_{k-1}(d).
\end{equation}
\end{proposition}
\begin{proof}
Taking the successive differences of $m_q(\ell^js)$ gives
\[
 \Delta_q(n;\ell,\ell)
       =\sum_{d\mid s}\bigl(a_k(d)-a_{k-1}(d)\bigr).
\]
The numerator ratio in $a_k(d)/a_{k-1}(d)$ is $\ell$. Its
denominator ratio is
\[
 \frac{\operatorname{lcm}(R_k,\ord_d(q))}
      {\operatorname{lcm}(R_{k-1},\ord_d(q))}.
\]
If $R_k=R_{k-1}$ this ratio is one. Otherwise it is one when
$\ell^e\mid\ord_d(q)$ and is $\ell$ when this divisibility fails.
The two asserted formulas follow term by term.
\end{proof}

The original-code rigidity condition $Q(C)$ is equivalent to
$s_\ell(C)=1$ for every prime $\ell\mid n$.
The following explicit majorant will also permit uniform estimates
under field extension:
\begin{equation}\label{tk:envelope}
 U_q(n)=\min\left\{1,
   \sum_{\substack{\ell,r\ \text{primes}\\\ell r\mid n}}
                            2^{-\Delta_q(n;\ell,r)}\right\}.
\end{equation}
An empty sum is zero.

\subsection{Alternating actions on canonical blocks}
Write $\mathcal Q_q(v)$ for the multiplication-by-$q$ orbits on $\Z_v$.
We retain the notation and probability model of Section~\ref{sec:preliminaries}.

For $a\mid n$, put $b=n/a$. The unique $\langle\sigma_n\rangle$-invariant
partition into blocks of size $a$ is
\[
 \mathcal B_a=\{B_r:r\in\Z_b\},\qquad
 B_r=\{x\in\Z_n:x\equiv r\pmod b\}.
\]
Indeed the block containing zero in a regular cyclic action is a subgroup.
For $a\ge5$, define $\mathcal L_a(C)$ to mean that there exists
\[
 \langle\sigma_n\rangle\le G\le\PAut(C)
\]
preserving $\mathcal B_a$, whose setwise stabilizer of a block induces
a group containing the natural $\Alt(a)$. The witnessing subgroup
may depend on $C$ and $a$; the full automorphism group need not preserve
this partition. The case $a=n$ is allowed. Set
\begin{equation}\label{ab:primeparts}
 A_a=\prod_{\ell\mid a}\ell^{v_\ell(n)},\qquad
 B_a=n/A_a,\qquad T_a=A_a/a.
\end{equation}
Here $B_a$ is the largest divisor of $n$ coprime to $a$, and
$n=aB_aT_a$.

\subsubsection{Canonical diagonal factors}

We use the classical automorphism description of alternating groups:
for $a\ge5$, every automorphism of $\Alt(a)$ is induced by $\Sym(a)$
unless $a=6$. In degree six,
\[
 \operatorname{Out}(\Alt(6))\cong C_2\times C_2,
\]
and the natural $\Sym(6)$ is the inverse image of one subgroup of
order two in this quotient; see \cite[Section~1 and Lemma~2.1 of the preprint]{KOZ}.

\begin{lemma}\label{ab:untwist}
Let $D\cong\Alt(a)$, $a\ge5$, act on $ac$ points with $c$ orbits of
size $a$. Suppose its action on each orbit is faithful and has image
the full alternating group. If an $ac$-cycle $\delta$ normalizes $D$,
then $(a,c)=1$. In the cyclic coordinates obtained by following
successive powers of $\delta$, $D$ is the synchronized alternating
action on the $\Z_a$ factor of $\Z_a\times\Z_c$.
\end{lemma}
\begin{proof}
The full cycle permutes the $D$-orbits cyclically. Choose one orbit
$O_0$, let $O_j=\delta^jO_0$, and identify $D$ with $\Alt(a)$ through
its action on $O_0$. Conjugation by $\delta$ induces an automorphism
$\theta$, and the restriction $\alpha=\delta^c|_{O_0}$ is an $a$-cycle.
Consequently
\begin{equation}\label{ab:powerauto}
 \theta^c=\operatorname{conj}_{\alpha}.
\end{equation}
If $a\ne6$, let $\pi\in\Sym(a)$ induce $\theta$. If $a=6$, let
$\tau$ denote the nonidentity outer class of the natural $\Sym(6)$.
The $6$-cycle $\alpha$ is odd, so
\[
 [\theta]^c=\tau\ne1
 \quad\hbox{in }C_2\times C_2.
\]
The outer group has exponent two. Thus $c$ is odd and
$[\theta]=\tau$, so $\theta$ again is induced by a natural permutation
$\pi\in\Sym(6)$.

The centralizer of the natural $\Alt(a)$ in $\Sym(a)$ is trivial:
a commuting permutation must send a point to a point fixed by its
stabilizer, and double transitivity makes that point unique.
Equation~\eqref{ab:powerauto} therefore gives $\pi^c=\alpha$.
Choose labels $\lambda_0:O_0\to\{1,\ldots,a\}$ compatible with the
chosen identification, and on $O_j$ use
\[
 \lambda_j=\pi^j\lambda_0\delta^{-j}.
\]
Transport by $\delta^{-j}$ changes an element $g\in D$ to
$\theta^{-j}(g)$; conjugation by $\pi^j$ changes it back to $g$.
Thus these labels make $D$ literally diagonal. They are consistent
at the wrap because $\pi^c=\alpha$.

In these labels its permutation normalizer is
\[
 N_{\Sym(ac)}(D)=\operatorname{diag}\Sym(a)\times\Sym(c).
\]
A normalizing permutation permutes the $D$-orbits, and its within-orbit
bijections induce the same automorphism; triviality of the natural
centralizer makes those bijections identical. Write
$\delta=(\gamma,\beta)$ in this product action. Its transitivity forces
$\gamma$ to be an $a$-cycle and $\beta$ a $c$-cycle. Hence
\[
 ac=\ord(\delta)=\operatorname{lcm}(a,c),
\]
which proves $(a,c)=1$. Starting at any point and following $\delta$
now identifies the two coordinates with $x\bmod a$ and $x\bmod c$.
Conjugating the full alternating group by the resulting labels does
not change it. This proves the assertion in the fixed cyclic
coordinates, rather than only up to an unspecified conjugacy.
\end{proof}

For $n=act$ with $(a,c)=1$, use the fixed coordinates
\begin{equation}\label{ab:coordinates}
 x=r+ty,\quad 0\le r<t,\quad y\in\Z_{ac},\qquad
 y\longleftrightarrow(i,j)\in\Z_a\times\Z_c.
\end{equation}
Define $D_{a,c,t}$ to be the product of $t$ copies of $\Alt(a)$,
one on each strip $r$: its factor applies the same permutation to
$i$ for every $j$ and fixes other strips. Let $K_{a,c,t}$ be the
analogous product of symmetric groups. The shift cycles the strips
and on wrapping sends $(i,j)$ to $(i+1,j+1)$, so it normalizes both
groups.

\begin{lemma}\label{ab:scott}
Let $T$ be a nonabelian finite simple group. A subdirect subgroup of
$T^m$ is a product of full diagonal subgroups on disjoint collections
of coordinates.
\end{lemma}
\begin{proof}
Induct
on the number of coordinates and project away the last one. By
induction that projection is a product of diagonal simple factors.
Call that projection $P$ and the last simple factor $T$. The kernel
of the projection onto $P$, viewed inside $T$, is normal in $T$
because the subgroup projects onto $T$. If this kernel is $T$, the
subgroup is $P\times T$. If it is trivial, the subgroup is the graph
of a surjective homomorphism $P\to T$. The images of the simple
factors of $P$ are commuting normal subgroups of $T$, so precisely
one maps onto $T$, isomorphically, and all others map trivially.
Thus the new coordinate joins exactly one diagonal factor. This
proves the assertion by induction.
\end{proof}

\begin{theorem}\label{ab:kernel}
Suppose $G\le\Sym(n)$ contains $\sigma_n$, preserves $\mathcal B_a$
with $a\ge5$, and has local setwise-stabilizer action containing
$\Alt(a)$. If $H$ is its kernel on the set of blocks, then for some
$c\mid b=n/a$ with $(a,c)=1$,
\[
 [H,H]=D_{a,c,b/c}
\]
in the fixed coordinates~\eqref{ab:coordinates}.
\end{theorem}
\begin{proof}
For each block $B$, write $L_B=G_B^B\le\Sym(B)$ for the
local image of its setwise stabilizer. The subgroup $H^B$ is normal
in $L_B$, and $\Alt(B)$ is also normal in $L_B$ because it is normal
in $\Sym(B)$. Moreover $H^B$ contains the full $a$-cycle
$\sigma_n^b|_B$, so it is nontrivial. The intersection
$H^B\cap\Alt(B)$ is normal in $\Alt(B)$. If it were trivial, then
$[H^B,\Alt(B)]\le H^B\cap\Alt(B)=1$, so $H^B$ would centralize
$\Alt(B)$. The natural centralizer of $\Alt(B)$ in $\Sym(B)$ is
trivial, a contradiction. Simplicity of $\Alt(B)$ therefore gives
$\Alt(B)\le H^B\le\Sym(B)$, so $H^B$ is either $\Alt(B)$ or
$\Sym(B)$. In either case $[H^B,H^B]=\Alt(B)$, since $a\ge5$.
It follows that $[H,H]$ lies in $\prod_B\Alt(B)$ and projects
onto every factor. The subdirect-product fact gives a product of
diagonal copies $D_0,\ldots,D_{t-1}$ on disjoint collections of
blocks, allowing automorphism twists.

The group $[H,H]$ is normal in $G$. Thus $\sigma_n$ permutes its
simple factors and their supports. Coordinate transitivity makes this
permutation a single cycle. All factors involve the same number $c$
of original blocks, so $b=ct$. Their supports form a
$\sigma_n$-invariant partition with $t$ parts, hence are the residue
classes modulo $t$. On the support containing zero, $\sigma_n^t$ is
a full $ac$-cycle normalizing $D_0$.
Lemma~\ref{ab:untwist} gives $(a,c)=1$ and the asserted canonical
action on this support. Transporting by $\sigma_n^r$ gives exactly
the same action on each strip $r$.
\end{proof}

\begin{lemma}\label{ab:nesting}
If $c_1\mid c_2\mid b=n/a$ and $(a,c_2)=1$, then
\[
 D_{a,c_2,b/c_2}\le D_{a,c_1,b/c_1}.
\]
\end{lemma}
\begin{proof}
Put $t_i=b/c_i$ and $d=c_2/c_1$, so $t_1=dt_2$. A strip modulo
$t_2$ is a union of strips modulo $t_1$. On one finer strip write
\[
 r_1=r_2+t_2s,\quad0\le s<d,\qquad
 x=r_1+t_1y_1=r_2+t_2y_2.
\]
Then $y_2=s+dy_1$, so $i_2=s+di_1\pmod a$. A synchronized
permutation $\rho$ of $i_2$ induces
\[
 i_1\longmapsto d^{-1}\bigl(\rho(s+di_1)-s\bigr)\pmod a.
\]
This is conjugate to $\rho$ and hence is even whenever $\rho$ is
even. It is independent of the second coordinate. The action preserves
the finer strip because it preserves $y_2\bmod c_2$, hence
$y_2\bmod d$; it also preserves its second coordinate modulo $c_1$.
Thus it belongs to the product on the right.
\end{proof}

\begin{theorem}\label{ab:event}
For every $a\mid n$ with $a\ge5$,
\[
 \mathcal L_a(C)\quad\Longleftrightarrow\quad
 D_{a,B_a,T_a}\le\PAut(C).
\]
\end{theorem}
\begin{proof}
A witness for $\mathcal L_a(C)$ supplies $D_{a,c,b/c}$ by
Theorem~\ref{ab:kernel}. Every eligible $c$ divides $B_a$, so
Lemma~\ref{ab:nesting} supplies its subgroup $D_{a,B_a,T_a}$.
Conversely, $\langle D_{a,B_a,T_a},\sigma_n\rangle$ preserves the
blocks modulo $b=B_aT_a$. On each block its setwise stabilizer
induces a group containing $\Alt(a)$, since the corresponding factor
of $D_{a,B_a,T_a}$ already induces $\Alt(a)$. When $a$ is even,
$\sigma_n^b$ also induces an odd $a$-cycle, so in this case the local
image is $\Sym(a)$. The required witness property follows.
\end{proof}

\subsubsection{Exact counts}

\begin{lemma}\label{ab:count}
For $n=act$, $(a,c)=1$, $a\ge5$, and $(n,q)=1$,
\[
 \#\{C\text{ cyclic}:D_{a,c,t}\le\PAut(C)\}
       =2^{m_q(ct)+m_q(c)}.
\]
Exactly the same codes are invariant under $K_{a,c,t}$.
\end{lemma}
\begin{proof}
Since $a$ is invertible in $\F_q$, the natural module splits as
\[
 \F_q^a=R_a\oplus U_a,
\]
where $R_a$ is the constant space and $U_a$ the augmentation space.
The alternating group algebra contains every endomorphism of $U_a$,
extended by zero on $R_a$. To see this directly, choose $3$-cycles
$\rho,\eta$ whose supports meet in exactly one point $i$, possible
because $a\ge5$. With the usual permutation action on coordinate
vectors,
\[
 (\eta-I)(\rho-I)v
   =(v_{\rho^{-1}(i)}-v_i)(e_{\eta(i)}-e_i).
\]
Left multiplication by alternating permutations varies the image
through all coordinate differences, by double transitivity. Varying
the cycles supplies all coordinate-difference functionals. The image
vectors span $U_a$; the functionals restricted to $U_a$ span $U_a^*$,
since their common kernel in $\F_q^a$ is $R_a$ and $R_a\cap U_a=0$.
More explicitly, for any distinct $h,i$ one may take
$\rho=(h\ i\ k)$ and $\eta=(i\ l\ m)$ with $h,i,k,l,m$ distinct.
This realizes the functional $v_h-v_i$, and left multiplication
varies its image vector independently. The resulting rank-one
operators therefore span $\End(U_a)$, zero on $R_a$.
When a factor is supported on one strip, these difference operators
are zero outside that strip as well. This argument does not require
the coefficient characteristic to be coprime to $|\Alt(a)|$.

In particular, the projector onto the augmentation part of each
strip preserves $C$. These projectors, and their complementary
projector onto all constant parts, split $C$ into the corresponding
components. In strip $r$, invariance under the augmentation matrix
units makes its component
\[
 U_a\otimes E_r,\qquad E_r\le\F_q^c.
\]
Indeed, diagonal matrix units project the invariant subspace onto
the individual copies of $\F_q^c$, and off-diagonal matrix units
identify the resulting subspaces. All those copies are therefore
the same $E_r$. The shift identifies the $E_r$ for successive strips. At the wrap it
acts as the full cyclic permutation on both tensor coordinates;
the first preserves $U_a$, so the common space $E$ must be a cyclic
code of length $c$. This gives $2^{m_q(c)}$ choices.
On the sum of the constant parts, the shift is a full $ct$-cycle,
giving independently $2^{m_q(ct)}$ choices. Conversely every pair
of these choices gives a cyclic code invariant under $D_{a,c,t}$.

The image of either the alternating or symmetric group algebra on
$R_a\oplus U_a$ is $\F_q\oplus\End(U_a)$: the augmentation operators
above give the second summand, and subtracting their identity from
the total identity gives the projector onto $R_a$. Thus replacing
the alternating factors by symmetric factors does not change the
invariant codes.
\end{proof}

\begin{corollary}\label{ab:exact}
For every divisor $a\ge5$ of $n$,
\begin{align*}
 \#\{C:\mathcal L_a(C)\}
   &=2^{m_q(n/a)+m_q(B_a)},\\
 \Pr(\mathcal L_a(C))
   &=2^{-\left(m_q(n)-m_q(n/a)-m_q(B_a)\right)}.
\end{align*}
\end{corollary}
\begin{proof}
Use Theorem~\ref{ab:event} and Lemma~\ref{ab:count} with
$c=B_a$, $t=T_a$, noting that $ct=n/a$.
\end{proof}

\subsubsection{A lower bound for the alternating loss}

Let $\omega(v)$ and $\Omega(v)$ count prime factors without and with
multiplicity. Define
\[
 \mathcal P_q=\{\ell\text{ prime}:\ell\nmid q,
                              \ 2\mid\ord_\ell(q)\},\qquad
 h_q(n)=\#\{\ell\mid n:\ell\in\mathcal P_q\}.
\]

\begin{lemma}\label{ab:loss}
If $h_q(n)\ge2$, then every divisor $a\ge5$ of $n$ satisfies
\[
 m_q(n)-m_q(n/a)-m_q(B_a)\ge2^{\omega(n)-2}.
\]
\end{lemma}
\begin{proof}
Put $A=A_a$, $B=B_a$, $r=\omega(A)$, $s=\omega(B)$, and
$L=m_q(A)-m_q(A/a)$. For coprime $u,v$, counting the simultaneous
$q$-orbits in products of cyclotomic cosets gives
\[
 m_q(uv)=\sum_{X\in\mathcal Q_q(u)}
             \sum_{Y\in\mathcal Q_q(v)}\gcd(|X|,|Y|).
\]
The cosets on $A$ whose additive orders divide $A/a$ correspond
exactly to those on $A/a$. Subtract their contributions and then
$m_q(B)$ to obtain
\begin{equation}\label{ab:coarseloss}
 m_q(n)-m_q(n/a)-m_q(B)\ge(L-1)m_q(B).
\end{equation}
If $r\ge2$, fix one prime dividing $a$, use its full exponent in $A$,
and adjoin any squarefree product of the other primes of $A$.
This yields $2^{r-1}$ distinct additive orders not dividing $A/a$,
each supplying at least one coset. Thus $L\ge2^{r-1}$, and
$m_q(B)\ge\tau(B)\ge2^s$; hence~\eqref{ab:coarseloss} is at least
$2^{r+s-2}$. If $r=1$ and $L\ge2$, the same bound is stronger than
required.

It remains to consider $r=1$, $L=1$. Since $a\ge5$, this forces
$a=\ell$ to be an odd prime, $A=\ell^v$, and $q$ primitive modulo
$A$. Indeed an exponent at least two in $a$ would remove at least
two additive orders; with $a=\ell$, the unique removed order is $A$
and its coset count must be one. The sole removed coset has even
length $e=\varphi(A)$, and the exact loss is
\[
 \sum_{Y\in\mathcal Q_q(B)}\bigl(\gcd(e,|Y|)-1\bigr).
\]
Here $\ell\in\mathcal P_q$. The hypothesis $h_q(n)\ge2$ supplies
another prime $p'\in\mathcal P_q$ dividing $B$. Each divisor of $B$
divisible by $p'$ supplies at least one coset of even size. There
are at least $2^{s-1}$ such divisors, each contributing at least one
to the last sum. This is $2^{s-1}=2^{\omega(n)-2}$.
\end{proof}

\subsection{Monotonicity under field extension}
The exact counts above give two families of losses. Retain
$\Delta_q(n;\ell,r)$ and $U_q(n)$ from
\eqref{tk:loss-definition} and~\eqref{tk:envelope}, and define
\begin{equation}\label{ab:loss-envelope}
 \begin{split}
 \Gamma_q(n;a)&=m_q(n)-m_q(n/a)-m_q(B_a),\\
 V_q(n)&=\min\left\{1,\sum_{\substack{a\mid n\\a\ge5}}
                                  2^{-\Gamma_q(n;a)}\right\}.
 \end{split}
\end{equation}
As before, an empty sum is zero. The next lemma expresses both losses
as orbit counts or orbit defects, allowing us to apply
Lemma~\ref{lem:orbitdefect}.

\begin{lemma}\label{lem:lossmonotonicity}
For every $p\nmid n$, $u\ge1$, primes $\ell,r$ with $\ell r\mid n$,
and divisor $a\mid n$ with $a\ge5$,
\[
 \Delta_{p^u}(n;\ell,r)\ge\Delta_p(n;\ell,r),\qquad
 \Gamma_{p^u}(n;a)\ge\Gamma_p(n;a).
\]
Consequently $U_{p^u}(n)\le U_p(n)$ and $V_{p^u}(n)\le V_p(n)$.
\end{lemma}
\begin{proof}
Multiplication by $p$ acts on the additive character labels $\Z_n$;
its $u$th power is multiplication by $p^u$. For $\ell\ne r$,
inclusion--exclusion writes $\Delta_q$ as the number of $q$-orbits
on the labels whose additive orders divide neither $n/\ell$ nor $n/r$.
Taking a power only splits orbits, so this number cannot decrease.

For $\ell=r$, one has $\ell^2\mid n$. Reduction gives an equivariant
surjection
\[
 \{x\in\Z_n:\ell\nmid x\}\longrightarrow
 \{y\in\Z_{n/\ell}:\ell\nmid y\}.
\]
The two orbit counts are respectively
$m_q(n)-m_q(n/\ell)$ and $m_q(n/\ell)-m_q(n/\ell^2)$.
Lemma~\ref{lem:orbitdefect} proves the repeated-prime inequality.

For the alternating loss, write $A=A_a$, $B=B_a$, and let
\[
 \Omega=\{x\in\Z_A:\operatorname{ord}_{\rm add}(x)\nmid A/a\}.
\]
This is a nonempty invariant set. In CRT coordinates, the projection
$\Omega\times\Z_B\to\Z_B$ is equivariant and surjective. Its
$q$-orbit defect is exactly
$m_q(n)-m_q(n/a)-m_q(B)=\Gamma_q(n;a)$.
Apply Lemma~\ref{lem:orbitdefect} once more.
\end{proof}

\subsection{Uniform rarity of the canonical obstructions}
Recall $t_q(n)=\Pr(\neg Q(C))$ from
Section~\ref{sec:preliminaries}, and put
\[
 P_q^{\mathrm{alt}}(n)=\Pr_{C\in\mathcal C_q(n)}(\mathcal A(C)).
\]
The definitions of $\mathcal A$ and $\mathcal L_a$ give
$\mathcal A(C)=\bigcup_{a\mid n,\ a\ge5}\mathcal L_a(C)$.
The following theorem combines the exact counting formulas with one
density argument over the prime field.

\begin{theorem}\label{tk:average}\label{ab:average}
\label{prop:uniformcounts}\label{thm:canonical-uniform}
Fix a prime $p$. For every $n$ with $p\nmid n$ and every $u\ge1$,
\begin{equation}\label{tk:probability-bound}
 \begin{split}
 t_{p^u}(n)&\le U_{p^u}(n)\le U_p(n),\\
 P_{p^u}^{\mathrm{alt}}(n)&\le V_{p^u}(n)\le V_p(n).
 \end{split}
\end{equation}
If $h_p(n)\ge2$, then
\begin{equation}\label{ab:bound}
 V_p(n)\le\min\{1,\tau(n)2^{-2^{\omega(n)-2}}\}.
\end{equation}
In particular,
\begin{equation}\label{eq:uniformbad}
 \begin{split}
 \sup_{u\ge1}\Pr_{C\in\mathcal C_{p^u}(n)}
       (\neg Q(C)\text{ or }\mathcal A(C))
       &\le\min\{1,U_p(n)+V_p(n)\},\\
 \Av_{p,X}(U_p+V_p)&\longrightarrow0.
 \end{split}
\end{equation}
The same average limit for these event probabilities holds when each
code law is conditioned on full shift period $n$, uniformly over $u$.
\end{theorem}
\begin{proof}
Corollary~\ref{tk:prime-witness} and the union bound give
$t_q(n)\le U_q(n)$. Theorem~\ref{ab:event} and
Corollary~\ref{ab:exact} give
$P_q^{\mathrm{alt}}(n)\le V_q(n)$. There is no additional union over
conjugate groups: the canonical groups in those statements represent
every witness. Lemma~\ref{lem:lossmonotonicity} gives the remaining
inequalities in~\eqref{tk:probability-bound}. Lemma~\ref{ab:loss}
applied at $q=p$, followed by a union bound over at most $\tau(n)$
block sizes, gives~\eqref{ab:bound}.

It remains to prove the length-average limit for the two envelopes
$U_p$ and $V_p$. Throughout the rest of this proof, relative densities
and averages are taken among lengths coprime to $p$.

For distinct-prime pairs, Proposition~\ref{tk:distinct-loss} bounds
their capped sum by
\[
 D(n)=\min\{1,\omega(n)(\omega(n)-1)
                             2^{-2^{\omega(n)-2}}\},
\]
interpreted as zero if $\omega(n)<2$. The number $\omega(n)$ tends
to infinity in relative density among $(n,p)=1$. To see this, choose
a finite set of primes not dividing $p$. In the limiting distribution,
their divisibility indicators are independent Bernoulli variables by
the Chinese remainder theorem. Their sum has mean
$L=\sum_r1/r$ and variance at most $L$. For fixed $R<L$, the upper
relative density of $\omega(n)\le R$ is at most $L/(L-R)^2$.
The divergence of the reciprocal prime sum allows $L\to\infty$.
It follows that the bounded function $D(n)$ has vanishing average.

For the repeated-prime pairs, first fix positive cutoffs $P_0,K_0$,
and restrict to $\ell\le P_0$ and $2\le v_\ell(n)\le K_0$.
There are only finitely many pairs $(\ell,k)$ to consider. Set
$R_j=\ord_{\ell^j}(p)$. On the class $n=\ell^ks$, $(s,\ell p)=1$, a nongrowing order at level $k$
gives, by \eqref{tk:no-growth},
\[
 \Delta_p(n;\ell,\ell)
       \ge(\ell-1)\tau(s)\ge(\ell-1)2^{\omega(s)}.
\]
Thus the fixing probability has zero average on this class.

Suppose instead that $R_k=\ell R_{k-1}$, and put $e=v_\ell(R_k)$.
By Wiertelak's theorem \cite{Wiertelak}, as stated in
\cite[Theorem 1 and the paragraph following it]{Moree}, the counting
function for primes $r$ satisfying $\ell^e\mid\ord_r(p)$ is
asymptotic to $\delta\operatorname{Li}(X)$ for some $\delta>0$.
Partial summation gives a divergent sum of their reciprocals. Removing the finitely many primes dividing $\ell p$ does not
change this conclusion. The preceding finite-prime argument, now
among integers $s$ coprime to $\ell p$, shows that the set of $s$
having no such prime divisor has relative density zero.

If such a prime $r$ divides $s$, every divisor of $s$ divisible by
$r$ contributes to \eqref{tk:growth}. There are at least
$2^{\omega(s)-1}$ of these divisors. Consequently
\[
 \Delta_p(n;\ell,\ell)
           \ge(\ell-1)2^{\omega(s)-1}
\]
outside a relative density-zero subset of this fixed valuation class.
Since $\omega(s)\to\infty$ in relative density, its fixing probability
again has zero average. A finite sum handles the truncated repeated
pairs.

If a repeated pair lies outside these cutoffs, the length belongs to
\[
 \mathcal E_{P_0,K_0}=
 \bigcup_{\ell>P_0}\{n:\ell^2\mid n\}
 \ \cup\!
 \bigcup_{\ell\le P_0}\{n:\ell^{K_0+1}\mid n\},
\]
where both unions range over primes. Its upper relative density among
$(n,p)=1$ is at most
\[
 \frac1{1-1/p}\left(
       \sum_{\ell>P_0}\frac1{\ell^2}
       +\sum_{\ell\le P_0}\frac1{\ell^{K_0+1}}\right).
\]
Indeed, for every $X$ the number of $n\le X$ divisible by at least
one $\ell^2$ with $\ell>P_0$ is at most
$\sum_{P_0<\ell\le\sqrt X}\lfloor X/\ell^2\rfloor
 \le X\sum_{\ell>P_0}\ell^{-2}$. The other union is finite, and
$N_p(X)/X\to1-1/p$. In particular,
\[
 U_p(n)\le D(n)+
 \sum_{\substack{\ell\le P_0\ \mathrm{prime}\\2\le v_\ell(n)\le K_0}}
      2^{-\Delta_p(n;\ell,\ell)}
 +\boldsymbol1_{\mathcal E_{P_0,K_0}}(n).
\] Taking $X\to\infty$,
then $K_0\to\infty$ at fixed $P_0$, and finally $P_0\to\infty$
proves $\Av_{p,X}U_p\to0$.

We now estimate the alternating envelope.
The same order-density result with prescribed divisor $2$ gives
a positive relative prime density for $\mathcal P_p$.
Thus $\sum_{\ell\in\mathcal P_p}1/\ell=\infty$. For any fixed finite
subset, the divisibility indicators among lengths coprime to $p$
have limiting independent Bernoulli laws by the Chinese remainder
theorem. Their sum has mean $L=\sum1/\ell$ and variance at most $L$.
For fixed $R<L$, Chebyshev's inequality bounds the upper relative
density of $h_p(n)\le R$ by $L/(L-R)^2$. Increasing the finite subset
makes $L\to\infty$, proving $h_p(n)\to\infty$ in relative density.
In particular $\omega(n)\to\infty$ in relative density.

Furthermore,
\[
 \sum_{n\le X}(\Omega(n)-\omega(n))
 \le X\sum_{\ell\text{ prime}}\frac1{\ell(\ell-1)}=O(X).
\]
For fixed $R$, the event $\Omega(n)>2\omega(n)$ is contained in
$\{\omega(n)<R\}\cup\{\Omega(n)-\omega(n)>R\}$.
Markov's inequality therefore bounds its upper relative density by
$O_p(1/R)$; letting $R\to\infty$ makes that density zero. On the
remaining lengths,
\[
 \tau(n)\le2^{\Omega(n)}\le4^{\omega(n)}.
\]
Consequently, outside a set of relative density zero,
\[
 V_p(n)\le\min\{1,4^{\omega(n)}2^{-2^{\omega(n)-2}}\},
\]
and the right side tends to zero in relative density. Since $0\le
V_p(n)\le1$, this proves $\Av_{p,X}V_p\to0$. Together with the
translation estimate it gives the average limit in~\eqref{eq:uniformbad}.

Finally, a primitive constituent of length $n$ exists over every
$\F_{p^u}$. Selecting it forces full shift period $n$, and the uniform
cyclic-code law selects it with probability $1/2$. Thus the conditioning
event has probability at least $1/2$, independently of $n$ and $u$.
Conditional probabilities are at most twice the corresponding
unconditional probabilities. The bounds just proved therefore imply
the final assertion.
\end{proof}

Normality reflection in the next section uses $Q(C)$ on the code
being reduced. At a fixed quotient length $v$, the later structural
argument first tests $Q(D_v(C))$ separately. Failure at that stage
is one of the events counted by \eqref{tk:probability-bound} under
the uniform quotient law of Lemma~\ref{lem:quotient-law}.

\section{Cyclic cores and actual coordinate actions}
\label{st:section}

The canonical translation condition controls the kernels that arise
when a permutation group is passed to a system of coordinate blocks.
We first determine exactly which subgroups of the specified regular
cyclic group are normal by examining their images in these quotient actions. We then describe the simple factors on minimal
blocks and lift their actions to the original coordinates. Finally, we
remove a canonical divisor from an arbitrary cyclic core to reach either
a core-free action or a core whose order is coprime to its index.

Unless a statement explicitly concerns an abstract permutation group,
let $q=p^u$, let $n>1$ with $p\nmid n$, and let $C\le\F_q^n$ be cyclic and
satisfy $Q(C)$. Write
\[
 \sigma=\sigma_n,\qquad S=S_n=\langle\sigma\rangle,
 \qquad S\le G\le\PAut(C).
\]
The rigidity condition in the arguments below always concerns $C$.
No rigidity property of a quotient code is needed for the descent.

\subsection{Block kernels and reflection of normality}

For a factorization $n=ab$, define
\[
 U_{a,b}=\{x\mapsto x+bw_{x\bmod b}:w\in\Z_a^b\}.
\]
This is the group of permutations that fix each canonical block of
size $a$ and commute with $N=\langle\sigma^b\rangle$.
The constant translation vectors form its diagonal subgroup $N$.
The prime-block condition $Q(C)$ also controls this composite-block
group.

\begin{lemma}\label{st:composite}
For every factorization $n=ab$,
\[
 \PAut(C)\cap U_{a,b}=\langle\sigma^b\rangle.
\]
\end{lemma}
\begin{proof}
Suppose that an allowed translation vector $w\in\Z_a^b$ is
nonconstant. Cyclicity allows us to subtract the constant vector with
value $w_0$. The resulting allowed vector $w'$ is nonzero and has
first coordinate zero. Let $d>1$ be its additive order, choose a prime
$\ell\mid d$, and put $z=(d/\ell)w'$. Then $z$ has order $\ell$,
its first coordinate is zero, and at least one of its coordinates is
nonzero. Since $\ell\mid a$, we can write
\[
 z_r=(a/\ell)v_r,\qquad v_r\in\F_\ell.
\]
The associated permutation is
\[
 x\longmapsto x+(n/\ell)v_{x\bmod b}.
\]
The divisibility $b\mid n/\ell$ shows that its coefficient depends
only on $x\bmod(n/\ell)$, so it belongs to $V_\ell(n)$. Its
translation vector on the canonical blocks of size $\ell$ is
nonconstant, contrary to $Q(C)$. Therefore every allowed vector is
constant. Conversely, all constant vectors preserve $C$ by cyclicity.
\end{proof}

\begin{proposition}\label{st:kernel}
Let $b\mid n$, suppose that $N=\langle\sigma^b\rangle\lhd G$, and let $H$ be
the kernel of the action of $G$ on the $N$-orbits. Then
\[
 C_H(N)=N,\qquad H\le A(n).
\]
\end{proposition}
\begin{proof}
On each $N$-orbit, the centralizer of its regular cyclic action is
its cyclic translation group. Consequently,
\[
 C_H(N)=G\cap U_{n/b,b}=N
\]
by Lemma~\ref{st:composite}.
For $h\in H$, the elements $\sigma h\sigma^{-1}$ and $h$
induce the same conjugation automorphism on $N$, because $\sigma$
centralizes $N$. Since $H\lhd G$, their quotient satisfies
\[
 c=\sigma h\sigma^{-1}h^{-1}\in C_H(N)=N.
\]
Thus
\[
 h\sigma h^{-1}=c^{-1}\sigma\in S.
\]
This conjugate has order $n$ and hence generates $S$. Therefore $h$
normalizes $S$, which means $h\in A(n)$.
\end{proof}

The kernel calculation gives more than an inclusion in the affine
normalizer: it permits normality to be lifted from the quotient.

\begin{theorem}\label{st:reflection}
Suppose that $M=\langle\sigma^b\rangle\lhd G$, where $b\mid n$.
Let $\overline G$ be the induced group on the $b$ orbits of $M$,
and let $\overline\sigma$ be its induced full $b$-cycle. For every
$c\mid b$,
\[
 \langle\sigma^c\rangle\lhd G
 \quad\Longleftrightarrow\quad
 \langle\overline\sigma^c\rangle\lhd\overline G.
\]
Equivalently, for every $M\le W\le S$,
\begin{equation}\label{eq:reflection}
 W\lhd G\quad\Longleftrightarrow\quad
             \overline W\lhd\overline G.
\end{equation}
\end{theorem}
\begin{proof}
Put $W=\langle\sigma^c\rangle$. The forward implication follows
by taking the image in the quotient action. Conversely, suppose that
$\overline W\lhd\overline G$, and let $H$ be the kernel on the
$M$-orbits. The full preimage of $\overline W$ is
\[
 L=HW\lhd G.
\]
Since $M\lhd G$, its centralizer $K=C_G(M)$ is normal in $G$.
Furthermore $W\le S$ centralizes $M$, and
Proposition~\ref{st:kernel} gives $C_H(M)=M$. Therefore
\[
 L\cap K=(HW)\cap C_G(M)=C_H(M)W=MW=W.
\]
The intersection of the two normal subgroups $L$ and $K$ is normal
in $G$, proving the converse.
\end{proof}

\subsection{Elementary abelian normal subgroups}

We next show that $Q(C)$ also restricts the solvable normal structure
of every group between $S$ and $\PAut(C)$. The exceptional affine
action on four points requires a separate argument.

\begin{theorem}\label{st:elementary}
Every nontrivial elementary abelian normal subgroup $E\lhd G$ is
\[
 E=\langle\sigma^{n/\ell}\rangle
\]
for a prime $\ell\mid n$.
\end{theorem}
\begin{proof}
Let $E$ be an elementary abelian $\ell$-group. Normality and
transitivity imply that its orbits all have the same size
$a=\ell^d>1$. They form an $S$-invariant partition and hence are
the canonical residue classes modulo $b=n/a$. On any orbit $B$,
the induced group $E^B$ is transitive and abelian. A point stabilizer
in an abelian transitive group fixes every point, so the faithful
induced action is regular. Thus $E^B\cong C_\ell^d$.

The full $a$-cycle $\sigma^b|_B$ normalizes this regular elementary
abelian group. Identify $B$ with $\F_\ell^d$ via its translations.
A normalizing permutation $g$ then satisfies
$g(x+t)=g(x)+\alpha(t)$ for a linear automorphism $\alpha$ of
the translation group, and so $g(x)=g(0)+\alpha(x)$.
In particular, $\sigma^b|_B$ belongs to
$\operatorname{AGL}_d(\ell)$. Embed the latter group in
$\operatorname{GL}_{d+1}(\ell)$ using homogeneous coordinates.
An $\ell$-element is unipotent. If $J$ is nilpotent of index at
most $d+1$, then
\[
 (I+J)^{\ell^h}=I+J^{\ell^h},
\]
so its order is at most
$\ell^{\lceil\log_\ell(d+1)\rceil}$. The element of order
$\ell^d$ consequently requires
\[
 d\le\lceil\log_\ell(d+1)\rceil.
\]
For $d\ge3$, the inequality $2^{d-1}\ge d+1$ contradicts this
condition. For $d=2$, it permits only $\ell=2$. Thus either $d=1$
or $(\ell,d)=(2,2)$.

If $a=\ell$, every full $\ell$-cycle in
$\operatorname{AGL}_1(\ell)$ lies in the translation subgroup,
since the corresponding quotient has order $\ell-1$. Hence
\[
 E^B=\langle\sigma^b|_B\rangle
\]
on every block. It follows that $E\le V_\ell(n)$, and $Q(C)$
forces $E=\langle\sigma^{n/\ell}\rangle$.

It remains to exclude $a=4$. Write $n=4b$ and $z=\sigma^b$.
On each block $B$, the induced group $E^B$ is the regular Klein
four-group. Conjugation by the $4$-cycle $z|_B$ is a nontrivial
transvection on this two-dimensional vector space over $\F_2$.
Indeed, after labelling the block so that $z|_B=(0\,1\,2\,3)$,
the three nonidentity translations are
$(0\,2)(1\,3)$, $(0\,1)(2\,3)$, and $(0\,3)(1\,2)$.
Conjugation by $z$ fixes the first and interchanges the other two.
Thus its fixed line and its commutator image are both generated by
$z^2|_B$. Therefore
\[
 C_E(z)\le V_2(n),\qquad [E,z]\le V_2(n).
\]
The projection of $[E,z]$ on every block is nontrivial, so $Q(C)$
gives
\[
 [E,z]=\langle\sigma^{n/2}\rangle=\langle z^2\rangle.
\]
In particular $z^2\in E$. As $z^2$ centralizes $z$, a second
application of $Q(C)$ yields
$C_E(z)=\langle z^2\rangle$. The map $e\mapsto[e,z]$ is a
homomorphism on the abelian group $E$, with kernel $C_E(z)$.
Consequently,
\[
 |E|=|C_E(z)|\,|[E,z]|=4.
\]
Every projection $E\to E^B$ is therefore an isomorphism. Choose
$e\in E\setminus\langle z^2\rangle$. On each block,
$e|_B$ is one of the two double transpositions different from
$z^2|_B$. Its product with $z|_B$ is a transposition of one
antipodal pair and fixes the other pair. These antipodal pairs are
exactly the canonical blocks of size two. Thus $ez\in V_2(n)$,
but its translation vector is nonconstant: on each four-point block
one pair is swapped and the other is fixed. This contradicts $Q(C)$.
\end{proof}

\begin{lemma}\label{st:corefree-radical}
If $\core_G(S)=1$, then $\Rad(G)=1$.
\end{lemma}
\begin{proof}
Suppose that the solvable radical $R=\Rad(G)$ is nontrivial.
Its last nontrivial derived subgroup is a nontrivial abelian
characteristic subgroup of $R$, and hence is normal in $G$.
Choose a prime $\ell$ dividing its order and take the subgroup
of elements of order dividing $\ell$ in its $\ell$-primary
component. This gives a nontrivial elementary abelian normal subgroup
$E\lhd G$. Theorem~\ref{st:elementary} implies $E\le S$;
normality then implies $E\le\core_G(S)=1$, a contradiction.
\end{proof}

\subsection{Minimal blocks with trivial solvable radical}
\label{sec:localgroups}

The next lemma applies both to a core-free witness on the coordinates
and to the faithful quotient action obtained from a coprime cyclic core.
For a finite group $H$, write $H^{(\infty)}$ for the stable last
term of its derived series.

\begin{lemma}\label{lem:scott}
Let $L$ be a transitive permutation group of degree $b>1$ containing
a regular cyclic subgroup $\langle\rho\rangle$, and suppose that
$\Rad(L)=1$. Choose a minimal nonsingleton block system of block
size $k>1$, allowing the single block of size $b$, and let $H$ be
its kernel. Then the local action $U$ is a nonsolvable primitive
group containing a full $k$-cycle. Its socle $T$ is a nonabelian
simple group from the primitive full-cycle classification, and
\[
 D=H^{(\infty)}\cong T^t\qquad(t\ge1).
\]
The factors are full diagonal subgroups on disjoint collections of
local blocks, and $\rho$ permutes these factors transitively.
\end{lemma}
\begin{proof}
Minimality makes the local action primitive. To justify this directly,
let $B$ be an $L$-block and suppose that $B'\subset B$ is a block
for its setwise stabilizer $L_B$. If $gB'\cap B'\ne\varnothing$,
then $gB\cap B\ne\varnothing$, so $gB=B$ and $g\in L_B$.
It follows that $gB'=B'$. Thus $B'$ would also be an $L$-block,
and a proper nonsingleton choice would contradict minimality.

We record the precise canonicity property of the chosen block system,
which is what is needed when an overgroup normalizes $L$. Identify the
regular cyclic group $S=\langle\rho\rangle$ with the additive group
of $\mathbb Z_b$. If $B$ is the chosen $L$-block containing $0$, then
for every $x\in B$, translation by $x$ sends $B$ to a block meeting
$B$, hence preserves $B$. Thus $B$ is a subgroup of $S$, and its
cosets are the chosen block system. A cyclic group has a unique
subgroup of each prescribed order; in particular, this particular
subgroup $B$ is characteristic in $S$. This does not assert that the
minimal nonsingleton $L$-block system is unique. For example, let
$L=A_5\times\PSL_2(11)$ act in product action on $5\times11$
points, using the natural degree-five action of $A_5$ and the primitive
coset action of $\PSL_2(11)$ on eleven cosets of an $A_5$ subgroup.
A pair consisting of a $5$-cycle and an $11$-cycle is a regular
$55$-cycle, while the two coordinate-fiber partitions are distinct
minimal nonsingleton block systems. The group $L$ has trivial solvable
radical. What matters is that each selected block system is preserved by every
permutation normalizing $S$. Indeed, if $g$ normalizes $S$, write
$g\rho g^{-1}=\rho^u$ with $(u,b)=1$ and $g(0)=\rho^v(0)$. Then
\[
 g(B)=\{g\rho^x(0):x\in B\}
      =\{\rho^{v+ux}(0):x\in B\}
      =\rho^v B,
\]
since $uB=B$. Thus every $g\in N_{\Sym(b)}(S)$ preserves the
chosen block system $\mathcal B$. Write
$J=\prod_{B'\in\mathcal B}\Sym(B')$ for the subgroup fixing each block
setwise. Then $gJg^{-1}=J$, and the kernel of the
$L$-action on $\mathcal B$ is $H=L\cap J$. Consequently,
\[
 g\in N_{\Sym(b)}(S)\cap N_{\Sym(b)}(L)
 \quad\Longrightarrow\quad gHg^{-1}=H,\qquad
 gH^{(\infty)}g^{-1}=H^{(\infty)}.
\]
The second normalizer hypothesis matters. For example, label the seven
points of the Fano plane by $\Z_7$ and its lines by the translates of
$\{0,1,3\}$. The collineation group $L\cong\mathrm{GL}_3(2)$
contains the translation group $S$ and is primitive, so $H=L$ for
the one-block system. Its two orbits on unordered triples consist of
the seven lines and the $28$ independent triples. Negation normalizes
$S$ and sends $\{0,1,3\}$ to the nonline $\{0,4,6\}$. Since the
orbit sizes differ, negation does not normalize $L$. The same
degree-seven normalizer action is described in \cite[Example~4.6]{JZ}.

The element $\rho^{b/k}$ lies in $H$ and induces a full $k$-cycle
on every block. Hence $H\ne1$. If $U$ were solvable, then $H$,
which embeds faithfully in the product of its local projections,
would be a nontrivial solvable normal subgroup of $L$. This is
impossible when $\Rad(L)=1$.

By the primitive full-cycle classification
\cite[Theorem~1.2]{JZ}, the nonsolvable local group $U$ has one of
the following actions. Its socle is $T=\Alt(k)$ in the natural
alternating and symmetric cases. In the projective cases, for some
$d\ge2$ and prime power $s$,
\[
 \PGL_d(s)\le U\le \operatorname{P\Gamma L}_d(s),\qquad
 k=(s^d-1)/(s-1),\qquad T=\PSL_d(s),
\]
on points or on hyperplanes. In the remaining cases $U=T$ and
\[
 (T,k)\in\{(\PSL_2(11),11),\ (M_{11},11),\ (M_{23},23)\}.
\]
In every case $T$ is nonabelian simple and acts doubly transitively
and nonregularly. The quotient $U/T$ is solvable. In the projective
case its diagonal and field parts are cyclic.
For every local block $B$, the projection $H^B$ is normal in the
local setwise-stabilizer action $U$, and it is nontrivial because
it contains $\rho^{b/k}|_B$. If $H^B\cap T=1$, then
$[H^B,T]=1$. The centralizer of a nonregular doubly transitive
group is trivial, so this would force $H^B=1$. Therefore $H^B$
contains $T$.

The stable derived subgroup of $H^B$ is exactly $T$: the quotient
by $T$ is solvable, whereas $T$ is perfect. Projection commutes
with taking the stable derived subgroup of a finite group. Thus $D$
projects onto $T$ on every block and lies in the direct product of
these local simple groups. Lemma~\ref{ab:scott} expresses $D$ as
a product of full diagonal factors with disjoint block supports.
Conjugation by $\rho$ preserves this support partition. Since
$\rho$ is transitive on the local blocks, it is also transitive
on the simple factors.
\end{proof}

\subsection{Coprime cores and their coordinate covers}
\label{sec:covers}

Suppose that the cyclic core is proper and nontrivial, and that its
order and index are coprime:
\begin{equation}\label{eq:coprime}
 N=\core_G(S_n),\qquad |N|=a>1,\qquad
 n=ab,\qquad b>1,\qquad(a,b)=1.
\end{equation}
In the fixed Chinese remainder coordinates
$(i,j)\in\Z_a\times\Z_b$, the group $N$ translates the first
coordinate. Put $K=C_G(N)$. Each element of $K$ has the form
\[
 k(i,j)=(i+v_{k,j},\psi_k(j)).
\]
Because $b$ is invertible modulo $a$, define
\[
 \mu(k)=b^{-1}\sum_{j\in\Z_b}v_{k,j}\in\Z_a.
\]
Composition permutes the summands and adds the phases, so $\mu$ is
a homomorphism. Its restriction to $N$ is the identity under the
identification of $N$ with the additive group of $\Z_a$.
Consequently, with $L=\ker\mu$,
\begin{equation}\label{eq:split}
 K=N\times L.
\end{equation}
Indeed $N$ is central in $K$, $N\cap L=1$, and every $k\in K$
is the product of its phase average in $N$ and an element of $L$.

The splitting is compatible with the full group $G$. Since $N$ is
normal, every $g\in G$ has the form
\[
 g(i,j)=(u_gi+w_{g,j},\phi_g(j)),\qquad
 u_g\in\Z_a^*.
\]
At the label $\phi_g(j)$, the phase of $gkg^{-1}$ equals
\[
 u_gv_{k,j}+w_{g,\psi_k(j)}-w_{g,j}.
\]
The last two terms have equal sums because $\psi_k$ is a
permutation. Hence
\[
 \mu(gkg^{-1})=u_g\mu(k),
\]
which proves $L\lhd G$. The pure quotient cycle
$\rho:(i,j)\mapsto(i,j+1)$ belongs to $L$: it is a power of
$\sigma_n$ and has zero phase. The kernel of $K$ on the $N$-orbits
is $C_H(N)=N$, where $H$ is the corresponding kernel in $G$, by
Proposition~\ref{st:kernel}. Thus $L$ acts faithfully on the
$b$ quotient labels.

\begin{lemma}\label{lem:central}
In the setup of \eqref{eq:coprime},
\[
 \Rad(L)=1,\qquad \core_K(S_n)=N.
\]
\end{lemma}
\begin{proof}
If $\Rad(L)\ne1$, take its last nontrivial derived subgroup and
then the subgroup of elements of order dividing a prime in one of
its primary components. The resulting subgroup $E$ is nontrivial,
elementary abelian, and characteristic in $L$. Since $L\lhd G$,
it is normal in $G$. Theorem~\ref{st:elementary} gives $E\le S_n$,
and normality gives $E\le N$. This contradicts $E\le L$ and
$N\cap L=1$.

The decomposition $S_n=N\times\langle\rho\rangle$ now gives
\[
 \core_{N\times L}(N\times\langle\rho\rangle)
 =N\times\core_L(\langle\rho\rangle)=N,
\]
because $\core_L(\langle\rho\rangle)$ is an abelian normal
subgroup of the group $L$ with trivial solvable radical.
\end{proof}

Replacing $G$ by $K=N\times L$ therefore preserves both the cyclic
core and its index. The replacement is still nonaffine: any subgroup
of $A(n)$ containing $S_n$ normalizes $S_n$, whereas its core in
$K$ is the proper subgroup $N$.
Apply Lemma~\ref{lem:scott} to the faithful quotient action of $L$.
Its minimal block kernel $H$ is normal in $K$, because $H\lhd L$
and $N$ centralizes $L$. Thus
\[
 D=H^{(\infty)}\cong T^t
\]
is a subgroup of the original coordinate group, normal in $K$ and
normalized by $\sigma_n$. The following description records the
actual permutation representation of each factor.

\begin{lemma}\label{lem:covers}
Let $T_i$ be a simple factor of $D$, and let $B$ be a quotient
block of size $k$ in its support. Identify its projection on $B$
with $T$, and let $P=T_x$ be the stabilizer of a point $x\in B$.
The translations induced by $P$ on the $N$-orbit above $x$ define
a homomorphism
\[
 \chi:P\longrightarrow N,\qquad E=\chi(P),\qquad e=|E|.
\]
Every $T_i$-orbit above $B$ has size $ke$, where
\begin{equation}\label{eq:e}
 e\mid a,\qquad e\mid\exp(P^{\rm ab}),\qquad(e,k)=1.
\end{equation}
These are the $D$-orbits above $B$. Over all quotient blocks they
form a $K$-invariant coordinate-block system, with exactly one simple
factor acting nontrivially on each coordinate orbit. The action of
that factor is faithful and is the coset action on $T/\ker\chi$.
For $e=1$, it is the original degree-$k$ local action.
\end{lemma}
\begin{proof}
Since $T_i\le L\le C_G(N)$, each element stabilizing the quotient
label $x$ acts on its $N$-orbit through the centralizer of the
regular action of $N$. It therefore acts by a translation in $N$.
Restriction to this orbit is a homomorphism, which defines $\chi$.
The projection of $T_i$ onto its local action $T$ is an isomorphism.
An element fixing a coordinate above $x$ must fix $x$ in the
quotient, hence belongs to $P$ under this isomorphism.
On the regular $N$-orbit above $x$ a translation fixes a coordinate
exactly when it is the identity. The stabilizer of every coordinate
above $x$ is therefore $\ker\chi$, and its orbit has size
\[
 [T:\ker\chi]=[T:P][P:\ker\chi]=ke.
\]
Its phase image is cyclic. Hence $e\mid a$ and
$e\mid\exp(P^{\rm ab})$. Since $k\mid b$ and $(a,b)=1$,
we also have $(e,k)=1$.

Outside its quotient support, $T_i$ fixes each quotient label and
acts above that label by translations in $N$. This defines a
homomorphism from the perfect group $T_i$ to an abelian group and
is therefore trivial. Distinct factors have disjoint quotient
supports. Thus every other simple factor is trivial above $B$, and
the $T_i$-orbits there are exactly the $D$-orbits. This also proves
the assertion that exactly one simple factor acts nontrivially on
each coordinate orbit.

Normality of $D$ in $K$ makes its orbit partition $K$-invariant,
and transitivity of $K$ makes all orbit sizes equal. On each such
orbit $O$,
\[
 D^O=T_i^O\lhd K^O,
\]
where $K^O$ denotes the action of the setwise stabilizer of $O$.
The local group therefore contains and normalizes the indicated
simple group. Finally, the kernel of the nontrivial action of $T_i$
is a proper normal subgroup of a simple group, so the action is
faithful.
\end{proof}

\subsection{Stripping an arbitrary cyclic core}
\label{sec:descent}

The preceding construction requires coprimality of the core order and
its index. Exact normality reflection supplies a canonical reduction
to that situation. For a fixed divisor $c\mid n$, write $v=n/c$
and $M_c=\langle\sigma_n^v\rangle$. We use the quotient code
$D_v(C)$ and its coordinate identification from
Lemma~\ref{lem:quotient-law}. If $M_c\lhd G$, the induced group
on the $M_c$-orbits preserves this quotient code.

\begin{proposition}\label{st:core-stripping}
Suppose that $G$ is nonaffine, and let $N=\core_G(S_n)$ have
order $a$ and index $b=n/a>1$, allowing $a=1$. Put
\begin{equation}\label{eq:strip}
 c=\prod_{\ell\mid b}\ell^{v_\ell(a)},\qquad
 A=a/c,\qquad v=n/c=Ab.
\end{equation}
Then $M_c\lhd G$, $(A,b)=1$, and the induced group
$\overline G$ on the $M_c$-orbits satisfies
\[
 \core_{\overline G}(\overline S)=\overline N\cong C_A,
 \qquad |\overline S:\overline N|=b.
\]
Moreover,
\[
 c\mid\Delta(n)=n/\rad(n),\qquad v>1.
\]
Thus the induced witness on $D_v(C)$ is either core-free, when
$A=1$, or has a proper nontrivial cyclic core whose order is coprime
to its index, when $A>1$.
\end{proposition}
\begin{proof}
The subgroup $M_c$ is the unique subgroup of order $c$ in the
cyclic group $N$. It is characteristic in $N$ and hence normal in
$G$. By construction, every prime dividing $b$ has been removed
from $A$, so $(A,b)=1$.

The image $\overline N$ is a normal subgroup of $\overline G$
contained in $\overline S$, and therefore is contained in its
cyclic core. Conversely, let $\overline W$ be that core, and let
$W$ be its preimage inside $S_n$. Then
\[
 M_c\le N\le W\le S_n.
\]
Theorem~\ref{st:reflection} gives $W\lhd G$. Since
$N=\core_G(S_n)$, this forces $W\le N$. Thus $W=N$ and
$\overline W=\overline N$, proving the exact core identity.
The quotient regular cyclic group has order $v=Ab$, so its core
has order $A$ and index $b$.

If $\ell\mid c$, then $\ell\mid b$, and hence
\[
 v_\ell(c)=v_\ell(a)\le v_\ell(n)-1.
\]
This proves $c\mid n/\rad(n)$. Finally $b>1$ implies $v=Ab>1$,
and the quotient witness remains nonaffine because its cyclic core
has proper index $b$. All uses of $Q$ in this argument are through
the normality-reflection theorem for the original code $C$.
\end{proof}

The distinction between $Q(C)$ and $Q(D_v(C))$ is necessary even at
simple-root lengths. For example, over $\F_{17}$ take $n=8$,
$w_x=2^x$ for $x\in\Z_8$, and $C=\langle\boldsymbol1,w\rangle$.
The element $2$ has order eight, and $\sigma_8 w=2^{-1}w$, so $C$ is cyclic. A coordinate permutation preserving
$C$ sends $w$ to $a\boldsymbol1+bw$ with $b\ne0$. Since
$\sum_x w_x=0$, taking coordinate sums gives $8a=0$, hence $a=0$.
The eight distinct entries of $w$ are the powers of $2$, so $b$ is
also a power of $2$ and the permutation is a shift. Thus
$\PAut(C)=\langle\sigma_8\rangle$ and $Q(C)$ holds.
On the other hand, $\sigma_8^4 w=-w$, so $D_4(C)$ is the
length-four repetition code. The transposition $(0\,2)$ preserves
$D_4(C)$ and belongs to $V_2(4)\setminus\langle\sigma_4^2\rangle$.
Consequently $Q(D_4(C))$ fails.

The reduction keeps both pieces of information needed below: a
canonical quotient length and the actual coordinate action of every
simple factor constructed on that quotient. In particular, the
abstract isomorphism type of a direct power is never substituted for
its permutation representation.

\section{Projective closure and exhaustive obstructions}
\label{sec:projectiveclosure}

The cyclic-core construction in Section~\ref{sec:descent} retains
actual coordinate actions. We now determine which projective actions
can occur when translation rigidity holds and alternating local
witnesses are absent. The natural actions and their scalar covers
require separate operators. The resulting deterministic reduction has
two forms: a coarse list sufficient for the length-average theorem,
and a sharper list needed for the pointwise estimates.

\subsection{Natural projective-line actions}\label{pc:section}

The following arguments concern actual coordinate blocks.  They do not
assume that a cyclic subgroup defining those blocks is normal in the
whole automorphism group.  We first identify the natural projective-line
permutation algebra, including the case in which the coefficient
characteristic divides the group order.

\begin{lemma}\label{pc:augmentation}
Let $s\ge4$ be a prime power, put $a=s+1$, and assume $(a,q)=1$.
Let $T=\PSL_2(s)$ act naturally on
$\Omega=\mathbb P^1(\F_s)$.  In $\F_q^\Omega$, put
\[
 R=\F_q\boldsymbol1,
 \qquad U=\left\{(v_x)_{x\in\Omega}:
                         \sum_{x\in\Omega}v_x=0\right\}.
\]
Then $U$ is absolutely irreducible, and the images of the permutation
group algebras satisfy
\begin{equation}\label{pc:algebra}
 \F_q[T]=\F_q[\Sym(\Omega)]
     =\F_q\oplus\End_{\F_q}(U)
       \quad\text{on }\F_q^\Omega=R\oplus U.
\end{equation}
Here each group algebra denotes its image in
$\End_{\F_q}(\F_q^\Omega)$, and the first summand acts on $R$.
\end{lemma}
\begin{proof}
Write $e_i$ for a coordinate vector, $\boldsymbol1$ for the all-one
column, and $P_g e_i=e_{g(i)}$ for a permutation matrix.
For each $i\in\Omega$, let $U_i\le T_i$ be a conjugate of the
translation subgroup on $\F_s$. It fixes $i$ and acts regularly on
$\Omega\setminus\{i\}$. Thus the unnormalized sum
\[
 \mathcal N_i=\sum_{u\in U_i}P_u
       =s e_i e_i^{\mathsf T}
        +(\boldsymbol1-e_i)(\boldsymbol1-e_i)^{\mathsf T}
\]
belongs to the image of $\F_q[T]$ in every coefficient characteristic.
If $g(i)=j$, direct multiplication gives
\begin{equation}\label{pc:rankone}
 (P_g-I)\mathcal N_i
  =(e_j-e_i)(a e_i^{\mathsf T}-\boldsymbol1^{\mathsf T}).
\end{equation}
Since $\boldsymbol1^{\mathsf T}P_{g^{-1}}=
\boldsymbol1^{\mathsf T}$ and
$e_i^{\mathsf T}P_{g^{-1}}=e_j^{\mathsf T}$, it follows that
\begin{equation}\label{pc:transposition}
 P_{(ij)}
 =I+a^{-1}(P_g-I)\mathcal N_i(I-P_{g^{-1}}).
\end{equation}
The correction has augmentation zero as a group-algebra expression.
The transpositions involving one fixed $i$ generate $\Sym(\Omega)$,
so \eqref{pc:transposition} already proves equality of the two
permutation algebras in \eqref{pc:algebra}.

There is also a direct formula for any $\rho\in\Sym(\Omega)$.
Choose $g_i\in T$ with $g_i(i)=\rho(i)$. Summing
\eqref{pc:rankone} and using
$\sum_i(e_{\rho(i)}-e_i)=0$ gives
\begin{equation}\label{pc:permutation}
 P_\rho=I+a^{-1}\sum_{i\in\Omega}(P_{g_i}-I)\mathcal N_i.
\end{equation}
This group-algebra expression again has augmentation one.

For the stated decomposition, $a$ is invertible, so
$\F_q^\Omega=R\oplus U$. On $R$ the operators
$a^{-1}(P_g-I)\mathcal N_i$ vanish; on $U$ they act as
\[
 v\longmapsto v_i(e_{g(i)}-e_i).
\]
For fixed $i$, transitivity of $T$ makes the image vectors span $U$.
As $i$ varies, the coordinate functionals $v\mapsto v_i$ span
$U^*$. Hence these operators span every endomorphism of $U$, extended
by zero on $R$. Subtracting their identity on $U$ from $I$ supplies
the projector onto $R$. Every permutation preserves $R$ and $U$,
so the full image is exactly $\F_q\oplus\End_{\F_q}(U)$.
The same calculation holds after every field extension; it also
proves absolute irreducibility. Only $a^{-1}$ was used, with no
division by $s$ or $|T|$ and no semisimplicity assumption.
\end{proof}

\begin{lemma}\label{pc:twists}
Every automorphism of $T=\PSL_2(s)$, $s\ge4$, is induced by a
permutation of its natural projective line.  Moreover
\begin{equation}\label{pc:normalizer}
 N_{\Sym(\mathbb P^1(\F_s))}(T)
       =\operatorname{P\Gamma L}_2(s),
\end{equation}
and its quotient by $T$ is solvable.
\end{lemma}
\begin{proof}
Write $s=r^f$ with $r$ prime.  The Sylow $r$-subgroups of $T$ are
naturally and equivariantly indexed by the $s+1$ points.  Namely, the
translation group has order $s$, is a Sylow $r$-subgroup, and fixes
exactly $\infty$.  It is the unique Sylow $r$-subgroup in that point
stabilizer.  Conjugacy gives the corresponding unique subgroup at
each other point, and every Sylow subgroup is obtained in this way.
An automorphism $\theta$ of $T$ permutes these Sylow subgroups.
If $h$ is the resulting permutation of the points, equivariance gives
\[
 hgh^{-1}=\theta(g)\qquad(g\in T)
\]
in the faithful natural action.  This proves the first assertion,
including $s=9$.

Here is also a direct normalizer calculation.  Let
$h\in N_{\Sym(\mathbb P^1(\F_s))}(T)$.  Composing with an element
of $\PGL_2(s)$, which is sharply transitive on ordered triples of
distinct points and normalizes $T$, we may assume that $h$ fixes
$\infty,0,1$.  It normalizes the unique Sylow $r$-subgroup of
$T_\infty$, so its restriction to $\F_s$ is additive:
\[
 h(x+t)=h(x)+h(t)\qquad(x,t\in\F_s).
\]
It also normalizes $T_{\infty,0}$, the group of square dilations
$x\mapsto cx$, $c\in(\F_s^*)^2$.  Evaluating conjugation at $1$
shows
\[
 h(cx)=h(c)h(x)\qquad(c\in(\F_s^*)^2,\ x\in\F_s).
\]
For even $s$, every nonzero element is a square.  For odd $s$, every
$c\in\F_s$ is a difference of two squares, since
\[
 c=\left(\frac{c+1}{2}\right)^2
       -\left(\frac{c-1}{2}\right)^2.
\]
Additivity therefore extends the displayed multiplicativity to every
$c\in\F_s$.  Together with $h(1)=1$, this makes $h$ a field
automorphism.  Undoing the initial fractional linear transformation
proves \eqref{pc:normalizer}; the reverse inclusion is immediate.
Finally $\PGL_2(s)/T$ is cyclic of order $\gcd(2,s-1)$ and the
field automorphism group is cyclic of order $f$.  Hence the
normalizer quotient by $T$ is solvable.
\end{proof}

\begin{proposition}\label{pc:enlargement}
Let $C\le\F_q^n$ be a linear code invariant under a group $G$
containing a specified full $n$-cycle $\sigma$.  Suppose $G$
preserves a partition into blocks of size $a=s+1$, where $s\ge4$
is a prime power and $(a,q)=1$.  Suppose, on each block, the local
setwise-stabilizer action contains and normalizes a natural copy of
$\PSL_2(s)$.  Then there is a group $G^*\le\PAut(C)$ containing
the same $\sigma$, preserving the same block partition, and inducing
$\Sym(a)$ on every block.  In particular the same code admits an
alternating local witness on these actual coordinate blocks.
\end{proposition}
\begin{proof}
Let $H$ be the kernel of $G$ on the blocks, and write $T_B$ for the
specified natural copy of $\PSL_2(s)$ on a block $B$.  The local
group $L_B$ normalizes $T_B$; by Lemma~\ref{pc:twists},
$L_B/T_B$ is solvable.  The normal subgroup $H^B$ of $L_B$ contains
the full $a$-cycle induced by $\sigma^{n/a}$, so is nontrivial.
The intersection $H^B\cap T_B$ is either $T_B$ or trivial.  In the
latter case $[H^B,T_B]=1$.  The centralizer of $T_B$ in $\Sym(B)$
is trivial, because the natural action is doubly transitive and
nonregular.  This would imply $H^B=1$, a contradiction.  Therefore
$H^B\ge T_B$.

Let $K=H^{(\infty)}$ be the stable term of the derived series of
$H$.  Since $T_B$ is perfect and $L_B/T_B$ is solvable, the
projection of $K$ on each block is exactly $T_B$.  The action of
$H$ on the union of the blocks is faithful, so $K$ is a subdirect
subgroup of $\prod_B T_B$.  Lemma~\ref{ab:scott} decomposes it into full diagonal simple
factors on disjoint collections of blocks.

Fix one such factor $D$.  After identifying all its local groups
with the standard $T$, its elements have local actions
$(\theta_B(g))_B$, where $g\in T$ and each $\theta_B$ is an
automorphism of $T$.  Lemma~\ref{pc:twists} realizes every
$\theta_B$ by a permutation of its block.  Relabelling within these
blocks therefore makes the action of $D$ literally diagonal, with
the same permutation $g$ on each block in its collection.  Factors
have disjoint collections, so these relabellings can be made
simultaneously.  We perform the following calculation in those labels
and undo all relabellings at the end.

For any $\rho\in\Sym(a)$, formula~\eqref{pc:permutation} supplies
coefficients $\lambda_g\in\F_q$ such that, on one block,
\[
 P_\rho=\sum_{g\in T}\lambda_gP_g.
\]
Applying this identity to the nonzero constant vector gives
$\sum_g\lambda_g=1$.  The same linear combination of permutation
matrices of elements of $D$ acts by $\rho$ on every block in this
collection.  Outside the collection every element of $D$ is the
identity, so that linear combination acts as the identity there too.
It is thus exactly the matrix of the permutation that acts diagonally
by $\rho$ on the collection and fixes all other coordinates.
It preserves $C$, because each element of $D$ does and $C$ is a
linear subspace.  Consequently the full diagonal $\Sym(a)$ on
this collection preserves $C$.

Performing the construction for all diagonal factors gives a product
$K^*$ of diagonal symmetric groups preserving $C$.  Undo the blockwise
relabellings and put $G^*=\langle\sigma,K^*\rangle$ in the original
coordinate labels.  Both generators preserve the original block
partition, and $K^*$ induces the full symmetric group on every block.
The group $G^*$ has all the required properties.  No normality of
a subgroup of the given regular cyclic group was used.
\end{proof}

The construction uses explicit sums of permutation matrices over the
alphabet field. No complete reducibility of the ambient $G$-module
and no reverse implication from alternating to projective witnesses
is required.
\subsection{Scalar covers of projective-line actions}\label{sec:ranktwo}
Let $s\ge4$ be a prime power, $T=\PSL_2(s)$, and let $P$ be its
natural point stabilizer. With $U\cong(\F_s,+)$ its unipotent subgroup,
\[
 P=U\rtimes H,\qquad H\cong C_{(s-1)/\gcd(2,s-1)},\qquad
 P^{\rm ab}\cong H.
\]
Here $H$ is the diagonal torus, represented in $\PSL_2(s)$ by
$\operatorname{diag}(t,t^{-1})$ for $t\in\F_s^*$; equivalently,
it is the stabilizer of both $\infty$ and $0$.
Indeed a nonidentity torus scalar acts nontrivially on $U$, and its
commutators with $U$ fill $U$. For a cover in Lemma~\ref{lem:covers},
$k=s+1$ and therefore
\begin{equation}\label{eq:odd}
 e\mid (s-1)/\gcd(2,s-1),\qquad(e,s+1)=1.
\end{equation}
The coprimality $(e,s+1)=1$ is the general condition $(e,k)=1$
from Lemma~\ref{lem:covers}. Since $s+1$ is even when $s$ is odd,
and $s-1$ is odd when $s$ is even, these two conditions imply that
$e$ is odd. The case $e=1$ is covered by the natural
projective-line enlargement. For $e>1$, the following explicit
operator supplies the required extra translation directly over
$\F_q$, even when its characteristic divides $|T|$.

\begin{lemma}\label{lem:localizer}
Let a cyclic group $N$ act regularly on each part of a partition of
$\Omega$ indexed by $\mathbb P^1(\F_s)$. Suppose $T=\PSL_2(s)$
commutes with $N$ and induces its natural action on the parts.
For the part indexed by $\infty$, let $\chi:P\to N$ be the
translation action of its stabilizer, and suppose $E=\chi(P)$
has odd order $e>1$. Let $F$ be a field of characteristic $p$ with
$p\nmid se$. Choose $h\in H$ such that $w=\chi(h)$ generates $E$,
and choose a prime $\ell\mid e$. Put $z=w^{e/\ell}$ and
\[
 \varepsilon_U=s^{-1}\sum_{u\in U}u\in F[T].
\]
There exists a polynomial $\beta(X)\in F[X]$, computable by the
Euclidean algorithm, such that
\begin{equation}\label{eq:localizer-bezout}
 \beta(X)(X-1)(X^2-1)
       \equiv X(X^{e/\ell}-1)\pmod{X^e-1}.
\end{equation}
For any such polynomial, the group-algebra element
\begin{equation}\label{eq:localizer}
 V=1+\beta(w)(h-1)(h-w^{-1})\varepsilon_U\in F[NT]
\end{equation}
acts on $F^\Omega$ as translation by $z$ on the part indexed by
$\infty$, and as the identity on every other part.
\end{lemma}
\begin{proof}
Since $e$ is odd and $p\nmid e$, $X^e-1$ is square-free and
\[
 \gcd\bigl((X-1)(X^2-1),X^e-1\bigr)=X-1.
\]
This remains true in characteristic two: the first polynomial then
has a triple root at one, but $X^e-1$ still has a simple root there.
The right side of \eqref{eq:localizer-bezout} is divisible by $X-1$,
so the Euclidean algorithm supplies $\beta$. Evaluating at the
order-$e$ element $w$ and multiplying by $w^{-1}$ gives
\begin{equation}\label{eq:localizer-coefficient}
 \beta(w)(w-1)(w-w^{-1})=z-1.
\end{equation}

For the action identity, work over the commutative group algebra
$\Lambda=F[N]$. The permutation module on each part is free of
rank one over $\Lambda$. Choose a coordinate vector $f_\infty$
in the part at infinity. With
\[
 u_x=\begin{pmatrix}1&x\\0&1\end{pmatrix},\qquad
 \dot w=\begin{pmatrix}0&-1\\1&0\end{pmatrix}
 \quad\text{in }\PSL_2(s),
\]
put $f_x=u_x\dot w f_\infty$ for $x\in\F_s$.
These are actual coordinate vectors, one in each remaining part,
so $f_\infty,(f_x)_{x\in\F_s}$ form a $\Lambda$-basis.
The torus element $h$ normalizes $U$ and satisfies
$h\dot w=\dot w h^{-1}$. For a suitable $\kappa\in\F_s^*$,
\[
 u_y f_\infty=f_\infty,\quad u_y f_x=f_{x+y},\qquad
 h f_\infty=w f_\infty,\quad h f_x=w^{-1}f_{\kappa x}.
\]
Here $\chi(U)=1$ because $[P,U]=U$, and elements of $N$ commute
with $T$. Writing $b=\sum_{x\in\F_s}f_x$, we obtain
\[
 \varepsilon_U f_\infty=f_\infty,\qquad
 \varepsilon_U f_x=s^{-1}b,
 \qquad h b=w^{-1}b.
\]
If $P_\infty$ denotes the literal coordinate projector onto the
part at infinity, these formulas give the identity over $\Lambda$
\begin{equation}\label{eq:localizer-projector}
 (h-1)(h-w^{-1})\varepsilon_U
       =(w-1)(w-w^{-1})P_\infty.
\end{equation}
Combining \eqref{eq:localizer-coefficient} and
\eqref{eq:localizer-projector} yields
\[
 V=1+(z-1)P_\infty.
\]
This is exactly the stated permutation. The change of base coordinate
within any part only translates its $\Lambda$-basis and leaves
$P_\infty$ unchanged.
\end{proof}

The proof uses neither irreducibility of an induced module nor a
splitting field. It does not require semisimplicity of $F[N]$ or
$F[T]$. In our application $p\nmid e$ follows from $e\mid n$,
and $p\nmid s$ is precisely the cross-characteristic assumption.

\subsubsection{Synchronization and localization}\label{sec:closure}
We work in the coprime-core setup of Section~\ref{sec:covers}, with $T=\PSL_2(s)$,
$e>1$, and assume that $s$ is not a power of the coefficient prime $p$.
All group-algebra calculations in this argument are over $\F_q$.

\begin{lemma}\label{lem:cyclecentralizer}
If $\alpha$ is a full $k=s+1$ cycle normalizing the natural $T$ on
its projective line, then its induced automorphism $\widehat\alpha$
has order $k$ and
\[
 C_{\Aut(T)}(\widehat\alpha)=\langle\widehat\alpha\rangle.
\]
\end{lemma}
\begin{proof}
By Lemma~\ref{pc:twists}, every automorphism of $T$ is realized
on its natural projective line.
The permutation centralizer of the nonregular doubly transitive $T$
is trivial. Consequently an automorphism commuting with
$\widehat\alpha$, realized by $\gamma$, has $[\gamma,\alpha]=1$.
The centralizer of a full cycle in $\Sym(k)$ is its cyclic group.
The same trivial centralizer shows that $\widehat\alpha$ has order $k$.
This argument includes the small semilinear cases and does not require
that $\alpha$ be a Singer element.
\end{proof}

\begin{proposition}\label{prop:closure}
In the original-code coprime-core setup, a nontrivial projective-line
cover in cross characteristic is incompatible with $Q(C)$.
\end{proposition}
\begin{proof}
Write the pure quotient cycle as $\tau:(i,j)\mapsto(i,j+1)$.
Let a chosen Scott factor $T_1$ have support on $c$ quotient blocks,
and let the total number of factors be $t$. Then $b=kct$.
The supports form a $\tau$-invariant partition, and
$\delta=\tau^t$ normalizes $T_1$ and is a full $kc$-cycle on its
quotient support. It preserves the $N$-coordinate.
Choose one quotient block $B_0$, write $\alpha=\delta^c|_{B_0}$
for the induced $k$-cycle, and identify $T_1$ with its faithful
image on $B_0$. On the full preimage $\Z_a\times B_0$ the
restriction $\widetilde\alpha$ of $\delta^c$ is
$(i,x)\mapsto(i,\alpha x)$. Thus it is a product of $a$ disjoint
$k$-cycles and has order $k$. If $\theta$ denotes conjugation
by $\delta$ on $T_1$, faithfulness of the quotient action gives
\[
 \theta^c=\widehat\alpha.
\]
In particular $[\theta,\widehat\alpha]=1$.
Lemma~\ref{lem:cyclecentralizer} gives
$\theta=\widehat\alpha^{r}$ and $rc\equiv1\pmod k$.
Thus $(c,k)=1$.
Put $\pi=\widetilde\alpha^r$. It induces $\theta$ and satisfies
$\pi^c=\widetilde\alpha$.

For $B_j=\delta^jB_0$, identify its full preimage with that of $B_0$
by
\[
 \psi_j=\pi^j\delta^{-j},\qquad0\le j<c.
\]
These identifications conjugate all $T_1$-actions to the same action,
and are consistent at the wrap because $\pi^c=\widetilde\alpha$.
Both constituent maps preserve the $N$-coordinate. Thus this is
literal diagonalization preserving the translations, not just abstract
module equivalence.

Let $\chi:P\to N$ be the phase map on the first block, with
$E=\chi(P)$ of odd order $e>1$. Choose a torus element $h$ whose
phase $w=\chi(h)$ generates $E$, choose a prime $\ell\mid e$,
and put $z=w^{e/\ell}$. On the first lifted block, apply
Lemma~\ref{lem:localizer}, choosing the part at infinity to be
one of its $N$-orbits. Thus
\[
 V=1+\beta(w)(h-1)(h-w^{-1})\varepsilon_U
\]
translates that orbit by $z$ and fixes the other $k-1$ orbits.
The previously constructed identifications preserve the $N$-coordinate
and make the $T_1$-actions identical. Therefore the same group-algebra
element performs the same translation on the corresponding orbit in
each of the $c$ copies. Outside the support of $T_1$, its elements
act trivially by perfectness, so $h-1$ is zero there and $V$ is the
identity. Consequently $V$ is the matrix of an actual coordinate
permutation $v$. Since
\[
 V\in\F_q[NT_1],\qquad N,T_1\le\PAut(C),
\]
it preserves $C$ directly, with no scalar extension or descent.
Since $z$ has prime order $\ell$, the construction belongs to the
original canonical group $V_\ell(n)$. Indeed, its selected $N$-orbits
are unions of the canonical order-$\ell$ blocks; in cyclic coordinates
the translation is a multiple of $n/\ell$ on each such block. It is
not a constant global translation, as only one of $k>1$ labels is
selected in each copy. This contradicts $Q(C)$.
\end{proof}

The formula is an identity in the image of the group algebra over
the alphabet field. Every application of $Q$ is to the original code.

\subsection{Natural higher-rank projective actions}
\label{proj:odd-section}\label{proj:graphcentralizer-section}

For higher rank, a graph automorphism exchanges the point and
hyperplane actions. A full-cycle centralizer argument removes that
obstruction at every block multiplicity. We first establish the
permutation-algebra identity and then synchronize the actual actions.

\begin{lemma}\label{proj:pointautos}
Let $d\ge3$, let $s$ be a prime power, and let
$T=\PSL_d(s)$ act naturally on the
$k=(s^d-1)/(s-1)$ projective points. The subgroup of $\Aut(T)$
realized by permutations of these points is normal of index two
and is the image of $\operatorname{P\Gamma L}_d(s)$.
The same subgroup is realized on the hyperplanes.
\end{lemma}
\begin{proof}
The standard automorphism theorem writes every automorphism as a
product of inner, diagonal, field, and graph automorphisms; see
\cite[Section 2.1]{BurnessGuest}. The first three types are realized
by projective semilinear permutations. Their subgroup is normal of
index two; the inverse-transpose graph automorphism interchanges
point stabilizers and hyperplane stabilizers.
These two stabilizer classes are distinct for $d\ge3$.
Indeed, the stabilizer of a one-dimensional subspace $L$ contains
the full unipotent group $\operatorname{Hom}(V/L,L)$ and a group
inducing $\mathrm{SL}(V/L)$ on $V/L$. An invariant hyperplane
containing $L$ would yield a proper nonzero invariant subspace of
$V/L$, while an invariant hyperplane not containing $L$ would be
a complement fixed by the unipotent group. Both are impossible.
Thus the graph automorphism cannot be realized on the point set.
An automorphism is realized on a transitive coset action precisely
when it preserves the conjugacy class of a point stabilizer.
This proves the assertion, and the hyperplane assertion follows
in the same way. The description applies also to $\PSL_3(2)$ and
$\PSL_4(2)$; their exceptional abstract isomorphisms do not add
further automorphisms beyond the stated index-two extension.
\end{proof}

\begin{lemma}\label{proj:crossalgebra}
Let $F$ be a finite field of characteristic $p$ and assume
$p\nmid sk$. For the natural point action of $T=\PSL_d(s)$,
$d\ge3$, write $F^k=R\oplus U$, where $R$ is the constant line
and $U$ is the augmentation subspace. Then
\[
 F[T]=F[\Sym(k)]=F\oplus\End_F(U)
\]
as images in $\End_F(F^k)$. The same assertion holds for the
natural hyperplane action.
\end{lemma}
\begin{proof}
Put $X=\F_s^d$, $\Omega=\mathbb P(X)$, and
$k=|\Omega|=(s^d-1)/(s-1)$. Extend scalars temporarily to an
algebraic closure $K$ of $F$, and write $r=d-1\ge2$.
The usual coordinate pairing
\[
 \langle f,h\rangle=\sum_{x\in\Omega}f_xh_x
\]
is $T$-invariant. Since $p\nmid k$, the constant line $R_K$
is nondegenerate for this pairing and
$K^\Omega=R_K\oplus U_K$, with $U_K=R_K^\perp$.
In particular the restricted pairing on $U_K$ is nondegenerate.

Consider the subgroup of $T$ induced by
\[
 A=\left\{u_a=
 \begin{pmatrix}1&0\\a&I_r\end{pmatrix}:a\in\F_s^r\right\}.
\]
Its projective image is isomorphic to its displayed matrix group.
It acts regularly on the affine chart
$\mathcal C_0=\{[1:y]:y\in\F_s^r\}$ and fixes every point of
the complementary hyperplane. Choose a nontrivial additive
character $\psi:\F_s^r\to K^*$, which exists because $p\nmid s$.
The element
\[
 e_\psi=s^{-r}\sum_{a\in\F_s^r}\psi(a)^{-1}u_a
 \in I_K(T),\qquad
 I_K(T)=\ker(K[T]\longrightarrow K),
\]
belongs to the augmentation ideal since the sum of a nontrivial
character is zero. On $K^\Omega$ it is the rank-one operator
\[
 e_\psi=w_\psi\,\beta_\psi,\qquad
 w_\psi=\sum_{y\in\F_s^r}\psi(y)^{-1}{\bf e}_{[1:y]},
 \qquad
 \beta_\psi(f)=s^{-r}\sum_{y\in\F_s^r}\psi(y)f_{[1:y]}.
\]
It is zero on the complementary hyperplane. Both
$w_\psi\in U_K$ and $\beta_\psi(R_K)=0$, and
$\beta_\psi(w_\psi)=1$; hence its restriction to $U_K$ still
has rank one.

We next prove that the $T$-orbit of $w_\psi$ spans $U_K$.
Fix a nontrivial additive character $\psi_0:\F_s\to K^*$.
The characters of the additive group $\F_s^r$ are precisely
$a\mapsto\psi_0(c\cdot a)$, with $c\in\F_s^r$.
The subgroup $\operatorname{diag}(1,\mathrm{SL}_r(s))$ is
transitive on the nonzero labels $c$, because
$\mathrm{SL}_r(s)$ is transitive on the nonzero vectors of its
natural and dual spaces when $r\ge2$. Consequently the orbit of
$w_\psi$ contains all nontrivial Fourier vectors on
$\mathcal C_0$.

For each nonzero linear form $L\in X^*$, set
\[
 \mathcal C_L=\{[v]\in\Omega:L(v)\ne0\},
\]
and label this chart by the unique representatives satisfying
$L(v)=1$. The group $\mathrm{SL}_d(s)$ is transitive on
nonzero linear forms: a change of basis sending one such form
to another can have its determinant corrected on the kernel,
without changing that form. It therefore transports
$\mathcal C_0$ to every $\mathcal C_L$. Such a transport is
an affine isomorphism between the normalized charts, and carries
their nontrivial Fourier vectors to nonzero scalar multiples of
nontrivial Fourier vectors. Thus the span of the $T$-orbit of
$w_\psi$ contains every nontrivial Fourier vector on every
$\mathcal C_L$. These arguments include $s=2,d=3$, since
$\mathrm{SL}_2(2)$ is transitive on the three nonzero labels.

Let $W$ be that orbit span. A function $f\in K^\Omega$
orthogonal to $W$ is orthogonal to all nontrivial Fourier
vectors on each chart. As $p\nmid s$, Fourier inversion on the
additive group $\F_s^r$ implies that $f$ is constant on each
$\mathcal C_L$. Any two distinct projective points have
linearly independent representatives $v,w$; there is a linear
form $L$ with $L(v)=L(w)=1$. They therefore lie in a common
chart, so those chartwise constants make $f$ globally constant.
Conversely constant functions annihilate every such Fourier
vector. Hence $W^\perp=R_K$ and $W=U_K$.

The row orbit is full as well, independently of any irreducibility
claim. Under the coordinate pairing the covector $\beta_\psi$
corresponds to $s^{-r}w_{\psi^{-1}}$. The preceding orbit-span
argument applies equally to the nontrivial character $\psi^{-1}$.
Permutation invariance of the pairing and its nondegeneracy on
$U_K$ therefore show
\[
 \operatorname{span}_K\{g w_\psi:g\in T\}=U_K,
 \qquad
 \operatorname{span}_K\{\beta_\psi h|_{U_K}:h\in T\}=U_K^*.
\]
It follows that the rank-one operators
\[
 g e_\psi h|_{U_K}=(g w_\psi)(\beta_\psi h|_{U_K}),
 \qquad g,h\in T,
\]
span $\End_K(U_K)$. All belong to the two-sided ideal
$I_K(T)$ and annihilate $R_K$. Thus the image of that ideal
is exactly $0\oplus\End_K(U_K)$. Adding the identity gives
\[
 K[T]=K\oplus\End_K(U_K)
\]
as image algebras on $R_K\oplus U_K$.

Finally, the image of $F[T]$ and the block-diagonal space
$F\oplus\End_F(U)$ are $F$-linear subspaces of
$\End_F(F^k)$. The former is contained in the latter, and the
proved equality after extending scalars to $K$ implies equality
of their $F$-dimensions, hence equality over $F$. Every
coordinate permutation preserves both $R$ and $U$; since
$T\le\Sym(k)$, this also gives
$F[T]=F[\Sym(k)]=F\oplus\End_F(U)$.
Twisting the action by the inverse-transpose automorphism proves
the hyperplane assertion.
\end{proof}

\begin{lemma}\label{proj:nominusone}
Let $\ell$ be prime, $f\ge1$, $d\ge3$, $s=\ell^f$, and
$k=(s^d-1)/(s-1)$. Then
\[
 \ord_k(\ell)=fd,\qquad -1\notin\langle\ell\rangle
          \le(\Z/k\Z)^*.
\]
\end{lemma}
\begin{proof}
The order divides $fd$. If it were a proper divisor, it would be
at most $fd/2$, whereas
\[
 k=1+s+\cdots+s^{d-1}>s^{d/2}=\ell^{fd/2}.
\]
Thus $k$ could not divide $\ell^{\ord_k(\ell)}-1$.
This proves the first assertion. If $-1$ belonged to the cyclic
subgroup, then $fd$ would be even and
$\ell^{fd/2}\equiv-1\pmod k$. But
\[
 k\ge1+s^{d-1}>1+s^{d/2}=1+\ell^{fd/2},
\]
a contradiction.
\end{proof}

\begin{lemma}\label{proj:fullcyclecentralizer}
Let $T=\PSL_d(s)$, where $d\ge3$, and put
$k=(s^d-1)/(s-1)$. If $\alpha$ is a full $k$-cycle normalizing
the natural point or hyperplane action of $T$, let
$\widehat\alpha\in\Aut(T)$ be its induced automorphism. Then
\[
 |\widehat\alpha|=k,\qquad
 C_{\Aut(T)}(\widehat\alpha)=\langle\widehat\alpha\rangle.
\]
\end{lemma}
\begin{proof}
We use the point action; the hyperplane action follows by twisting
by a graph automorphism. Write $s=\ell^f$. By
Lemma~\ref{proj:pointautos}, the natural permutation normalizer is
$B=\operatorname{P\Gamma L}_d(s)$, viewed faithfully in $\Aut(T)$.
Jones's classification of cyclic regular subgroups \cite{JonesSinger},
in the form recorded in \cite[Example~4.5]{JZ}, says that for $d\ge3$ every
cyclic regular subgroup of $B$ is a Singer subgroup contained in
$\PGL_d(s)$. The only non-Singer cyclic exception in that
classification is $\operatorname{P\Gamma L}_2(8)$, which is not
in the present range. Hence $S=\langle\alpha\rangle$ is Singer.

Use the Singer model $V=\F_{s^d}$ over $\F_s$, in which $\alpha$
is induced by multiplication $M_\lambda:x\mapsto\lambda x$.
The symmetric trace form
\[
 b(x,y)=\operatorname{Tr}_{\F_{s^d}/\F_s}(xy)
\]
is nondegenerate and $M_\lambda$ is self-adjoint. The automorphism
$\gamma:g\mapsto(g^\dagger)^{-1}$ of $T$, where $\dagger$ is
adjoint with respect to $b$, lies in the graph coset and satisfies
\[
 \gamma\widehat\alpha\gamma^{-1}=\widehat\alpha^{-1}.
\]
Suppose a graph-coset automorphism $\theta$ centralizes
$\widehat\alpha$. Then $\theta\gamma\in B$ normalizes
$\langle\widehat\alpha\rangle$ and induces inversion on it.
The Singer normalizer description in \cite[Example 4.5]{JZ}
identifies the image of the conjugation map on $S$ as
\[
 N_B(S)/S\ \simeq\ \langle\ell\bmod k\rangle
 \ \le\ (\Z/k\Z)^*.
\]
Indeed, the field Frobenius $x\mapsto x^\ell$ conjugates
$M_\lambda$ to $M_{\lambda^\ell}$, and the displayed
multiplier subgroup has order $fd$.
Inversion is excluded by Lemma~\ref{proj:nominusone}.
Thus every automorphism centralizing $\widehat\alpha$ belongs
to $B$ and is realized by a natural permutation. Its permutation
must commute with the full cycle $\alpha$, because the natural
$T$ has trivial permutation centralizer. It therefore belongs to
$\langle\alpha\rangle$. This proves the centralizer equality.
The same trivial centralizer shows $|\widehat\alpha|=k$.
\end{proof}

\begin{corollary}\label{proj:allmultiplicities}
Let a simple group $D\cong\PSL_d(s)$, $d\ge3$, act faithfully
on $kc$ points with $c$ orbits of size $k=(s^d-1)/(s-1)$,
each inducing a natural point or hyperplane action. If a full
$kc$-cycle $\delta$ normalizes $D$, then $(k,c)=1$ and all local
actions can be made literally diagonal by permutations of their
coordinate labels. No parity condition on $c$ is needed.
\end{corollary}
\begin{proof}
Choose one orbit $B_0$. Let $\theta$ be the automorphism of $D$
induced by $\delta$, and set $\alpha=\delta^c|_{B_0}$.
Then $\theta^c=\widehat\alpha$, so $\theta$ centralizes
$\widehat\alpha$. Lemma~\ref{proj:fullcyclecentralizer} gives
\[
 \theta=\widehat\alpha^r,\qquad rc\equiv1\pmod k.
\]
In particular $(k,c)=1$. Put $\pi=\alpha^r$. Then $\pi$ induces
$\theta$ and $\pi^c=\alpha$. For $B_j=\delta^jB_0$, the
bijections $\pi^j\delta^{-j}:B_j\to B_0$, $0\le j<c$, make
all local actions identical, with consistency at the wrap ensured
by $\pi^c=\alpha$.
\end{proof}

\begin{proposition}\label{proj:oddenlargement}\label{proj:alllengthenlargement}
Let $C\le F^n$ be a linear code over a finite field of characteristic
$p$, with $p\nmid n$. Suppose that $G\le\PAut(C)$ contains a
specified full $n$-cycle $\sigma$ and preserves a partition into
blocks of size
\[
 k=\frac{s^d-1}{s-1},\qquad d\ge3,
\]
where $s$ is a prime power and $p\nmid s$. Suppose each local
setwise-stabilizer group contains and normalizes a natural point or
hyperplane action of $\PSL_d(s)$. Then there is
$G^*\le\PAut(C)$ containing the same $\sigma$, preserving the same
partition, and inducing $\Sym(k)$ on every block. In particular,
$C$ admits an alternating local witness. No parity restriction on
$n$ is required.
\end{proposition}
\begin{proof}
Let $H$ be the kernel on the blocks, and let $T_B$ denote the
specified local simple group on a block $B$. The image $H^B$ is a
nontrivial normal subgroup of the local group, because it contains
the full $k$-cycle induced by $\sigma^{n/k}$. If
$H^B\cap T_B=1$, then $[H^B,T_B]=1$. The natural action of $T_B$
is nonregular and doubly transitive, so its permutation centralizer
is trivial; this would force $H^B=1$. Thus $H^B\ge T_B$.
The local quotient by $T_B$ is solvable by
Lemma~\ref{proj:pointautos}. Therefore $H^{(\infty)}$ projects
exactly onto $T_B$ on every block. Lemma~\ref{ab:scott} decomposes
it into full diagonal simple factors on disjoint collections of
blocks.

The full cycle $\sigma$ permutes these factors transitively. Let
there be $t$ factors, each supported on $c$ blocks; thus $n=kct$.
Fix one factor $D$ and put $\delta=\sigma^t$. On the support of
$D$, the restriction of $\delta$ is a full $kc$-cycle normalizing
$D$. Corollary~\ref{proj:allmultiplicities} supplies simultaneous
coordinate labels making the actions of $D$ literally identical on
all its $c$ blocks. This step allows both natural point and
hyperplane actions and uses no assumption on the parity of $c$.

Fix $\rho\in\Sym(k)$. Since $p\nmid k$, Lemma~\ref{proj:crossalgebra}
supplies coefficients $\lambda_g\in F$, indexed by $g\in D$, such
that on one of these blocks
\[
 P_\rho=\sum_{g\in D}\lambda_gP_{g|_B}.
\]
Evaluation on the constant vector gives $\sum_g\lambda_g=1$.
In the synchronized labels, the same combination of the original
permutation matrices of $D$ acts as $\rho$ on each of its $c$
blocks. Outside its support, every element of $D$ is the identity,
so the combination is the identity there. Consequently this is the
matrix of an actual coordinate permutation preserving $C$.

Apply the construction to every diagonal factor. Their supports
are disjoint, so the blockwise relabellings can be performed
simultaneously. After undoing those relabellings, the resulting
product $K^*$ of diagonal symmetric groups lies in $\PAut(C)$.
The group $G^*=\langle\sigma,K^*\rangle$ preserves the original
partition and induces $\Sym(k)$ on every block, as required.
\end{proof}

\subsection{Higher-rank scalar covers}
\label{hr:coverlocalizer-section}

The higher-rank construction uses the augmentation ideal of the local
simple group. This ensures that the resulting correction vanishes
outside the support of that factor, even though the central cyclic
group may act on all coordinates.

\begin{lemma}\label{hr:parabolicphase}
Let $s>2$ be a prime power, $d\ge3$, and $T=\PSL_d(s)$. Let $P$ be a
point stabilizer in its natural projective action. Every homomorphism
from $P$ to an abelian group factors through the scalar by which a
lift to $\mathrm{SL}_d(s)$ acts on the stabilized line. More precisely,
it is induced by a homomorphism
\[
 \vartheta:\F_s^*\longrightarrow E
\]
that is trivial on the scalar center of $\mathrm{SL}_d(s)$.
If the homomorphism has image $E$, the corresponding transitive
cover is
\[
 \Omega=(\F_s^d\setminus\{0\})/H,\qquad H=\ker\vartheta.
\]
It has $e(s^d-1)/(s-1)$ points, where $e=|E|$, and scalar
multiplication identifies $\F_s^*/H$ with its commuting cyclic
group $E$ on each projective-point part.
\end{lemma}
\begin{proof}
Write $r=d-1$. The inverse image of $P$ in $\mathrm{SL}_d(s)$
is the semidirect product of the upper unipotent row group
$\{u_b:b\in\mathbb F_s^{1\times r}\}$ with the Levi subgroup
\[
 \left\{\operatorname{diag}\bigl((\det A)^{-1},A\bigr):
                             A\in\mathrm{GL}_{d-1}(s)\right\}.
\]
For $A\in\mathrm{SL}_r(s)$, conjugation by
$\operatorname{diag}(1,A)$ sends $u_b$ to $u_{bA^{-1}}$.
For $i\ne j$, taking $A=I+tE_{ij}$ and $b$ the $i$th
standard row produces the commutator row $-t$ in position $j$.
Thus the commutators generate the entire unipotent radical. Furthermore
$\mathrm{GL}_{d-1}(s)'=\mathrm{SL}_{d-1}(s)$ for $s>2$.
For $d-1\ge3$ this follows from the elementary commutator
relations. In dimension two, choose $a\in\F_s^*\setminus\{1\}$;
commutators of $\operatorname{diag}(a,1)$ with the upper and lower
elementary matrices generate $\mathrm{SL}_2(s)$.
Thus every abelian homomorphism depends only on $\det A$, or
equivalently on the eigenvalue on the stabilized line. Its descent
to $P$ makes it trivial on the scalar center.
Choose a nonzero vector $v$ on the base line and a coordinate
$\omega$ above that line with stabilizer $\ker\chi$, where
$\chi:P\to E$ is the given phase map. For $g\in T$ and a lift
$\widetilde g\in\mathrm{SL}_d(s)$, the map
\[
  g\omega\longmapsto (\widetilde g v)H
\]
is well defined, since $H$ contains the scalar center and the
stabilizer of $vH$ is $\ker\chi$. It is an equivariant bijection
by transitivity. It also identifies the specified phase translations:
given $t\in\F_s^*$, choose $b\in P$ with a lift sending $v$ to
$tv$. Then $\chi(b)=\vartheta(t)$ and
$\vartheta(t)g\omega=g b\omega$ maps to $(t\widetilde g v)H$.
Thus this bijection intertwines the actions of both $T$ and $E$.
\end{proof}

\begin{lemma}\label{hr:scalar-localizer}
Use the scalar-cover model of Lemma~\ref{hr:parabolicphase}. Let
$F$ be any field of characteristic $p$ with $p\nmid se$. For any
projective point $x$ and any $z\in E$, the permutation translating
the part above $x$ by $z$ and fixing every other part belongs to
the image of $F[ET]$. More precisely, its permutation matrix has
the form
\[
 I+A,\qquad A\in F[E]I_F(T),
 \qquad I_F(T)=\ker\bigl(F[T]\longrightarrow F\bigr).
\]
Here $F[T]$ and $F[E]$ denote abstract group algebras before taking
images, and the last map is the augmentation map.
\end{lemma}
\begin{proof}
Extend scalars temporarily to an algebraic closure $K$ of $F$.
Since $p\nmid e$, the commuting $E$-action gives
\[
 K^\Omega=\bigoplus_{\eta\in\widehat E} V_\eta,
 \qquad \dim V_\eta=k=\frac{s^d-1}{s-1}.
\]
We first show that, for every nontrivial $\eta$, the image of
$I_K(T)$ on $V_\eta$ is $\End_K(V_\eta)$.

Put $r=d-1$ and consider the subgroup
\[
 U=\left\{u_v=
   \begin{pmatrix}1&0\\v&I_r\end{pmatrix}:v\in\F_s^r\right\}.
\]
It fixes the hyperplane with first coordinate zero pointwise and
acts regularly on the normalized affine chart $(1,y)$.
For each projective point $x$ choose a representative
$v_x\in\F_s^d\setminus\{0\}$, taking $v_x=(1,y)$ on the affine
chart. Write $\mathbf e_{v_xH}$ for its coordinate vector and put
\[
 b_{x,\eta}=\sum_{z\in E}\eta(z)^{-1}P_z\mathbf e_{v_xH}.
\]
These vectors, one per projective point, have disjoint supports and
form a basis of $V_\eta$. The chosen lifts on the affine chart give
$u_v b_{[1:y],\eta}=b_{[1:y+v],\eta}$. Every vector indexed by a
point at infinity is fixed by $U$. Hence
\[
 V_\eta\big|_U\cong K[U]\oplus
       \mathbf 1_U^{\,(s^{d-1}-1)/(s-1)}.
\]
For a nontrivial character $\psi:U\to K^*$ put
$e_\psi=|U|^{-1}\sum_{u\in U}\psi(u)^{-1}u\in I_K(T)$;
here $p\nmid s$ and $\sum_u\psi(u)^{-1}=0$.
For a nontrivial additive character $\psi$, the idempotent
$e_\psi$ has rank one on $K[U]$ and rank zero on every trivial
summand. Thus $e_\psi$ has rank one on $V_\eta$. Its image is
the corresponding character vector on the affine chart, extended
by zero on the complementary hyperplane. The subgroup
$\operatorname{diag}(1,\mathrm{SL}_r(s))$ acts transitively on the
nontrivial additive characters of $U$. Indeed the trace pairing
identifies them with the nonzero vectors in the dual of $\F_s^r$,
and $r\ge2$. Thus the orbit span of this image contains every nontrivial Fourier
line on the given affine chart. For a nonzero linear form $L$, choose
$g\in\mathrm{SL}_d(s)$ satisfying $L\circ g=(1,0,\ldots,0)$.
It sends $(1,y)$ to a vector already normalized by $L(v)=1$.
Transporting the displayed basis by $g$ shows that the same orbit
span contains every nontrivial Fourier line on this chart as well.

For $v\ne0$, put
$b_\eta(v)=\sum_{z\in E}\eta(z)^{-1}P_z\mathbf e_{vH}$.
Then $b_\eta(tv)=\eta(\vartheta(t))b_\eta(v)$ for $t\in\F_s^*$.
Consequently a covector $\xi\in V_\eta^*$ is uniquely represented by
the function $f_\xi(v)=\xi(b_\eta(v))$, which satisfies
$f_\xi(tv)=\lambda(t)f_\xi(v)$ with
$\lambda=\eta\circ\vartheta$. This character is nontrivial since
$\vartheta$ maps onto $E$. A function annihilating all the
character vectors must be constant
on every chart $L(v)=1$, where $L$ is a nonzero linear form,
because the nontrivial Fourier characters span the augmentation
space of that chart. Write this constant as $c_L$. Homogeneity gives
\[
 f(v)=c_L\lambda(L(v))\qquad(L(v)\ne0).
\]
If $L,M$ are independent, the map $v\mapsto(L(v),M(v))$ is
surjective onto $\F_s^2$, so
\[
 c_L\lambda(a)=c_M\lambda(b)
       \qquad(a,b\in\F_s^*).
\]
Taking $a=b=1$ and then choosing $a$ with $\lambda(a)\ne1$
forces $c_L=c_M=0$. Every nonzero linear form has an independent
partner, so $f=0$. This proves the spanning assertion. The same
argument for the inverse character proves that the right translates
of the rank-one covector span $V_\eta^*$. Explicitly, the coordinate
pairing identifies the dual of $V_\eta$ with $V_{\eta^{-1}}$, and
transpose sends $e_\psi$ to $e_{\psi^{-1}}$.

Consequently the operators $g e_\psi h$, with $g,h\in T$, span
$V_\eta\otimes V_\eta^*=\End_K(V_\eta)$, and all lie in
$I_K(T)$. Let $P_x$ be the literal coordinate projector onto the
part over $x$. It commutes with $E$. On $V_\eta$, the desired
correction $(P_z-I)P_x$ is
$(\eta(z)-1)P_x$; it is zero for $\eta=1$.
For each nontrivial $\eta$, choose an element of $I_K(T)$ inducing
this endomorphism, and multiply it by the central idempotent of
$K[E]$ for $\eta$. Summing these elements gives
\[
 (P_z-I)P_x\in K[E]I_K(T).
\]
The target matrix $(P_z-I)P_x$ and the linear image map from
$F[E]I_F(T)$ to $\End_F(F^\Omega)$ are defined over $F$;
individual character-idempotent summands need not be.
Solvability of the corresponding linear system over $K$ implies
solvability over $F$. Hence
$(P_z-I)P_x\in F[E]I_F(T)$, proving the assertion without a
descent hypothesis on an individual irreducible module.
\end{proof}

The same operator works when the group $N$ in
Lemma~\ref{lem:covers} is larger than $E$: choose one $T$-orbit
over the quotient block and translate it by $N$. These copies have
identical actions because $N$ centralizes $T$, and they partition
the full preimage of the block. The operator therefore translates
the entire selected $N$-part. Moreover every element of
$F[N]I_F(T)$ acts as zero on a coordinate set where $T$ is trivial.
Thus the localized permutation is the identity outside the support
of this simple factor. Applications to several quotient blocks
require the $N$-preserving synchronization of their actual actions;
the lemma by itself does not identify differently twisted copies.

\begin{remark}\label{hr:charfree-boundary}
The localizer uses only $p\nmid se$; it does not require
$p\nmid k$, odd $e$, or characteristic two. The case $s=2$
needs no nontrivial scalar cover when $d\ge4$: writing
$R\cong\F_2^{d-1}$ for its translation subgroup, the point
stabilizer $R:\mathrm{GL}_{d-1}(2)$ is perfect, by
$[R,\mathrm{SL}_{d-1}(2)]=R$ and perfection of
$\mathrm{GL}_{d-1}(2)$ for $d-1\ge3$. The remaining case
$\PSL_3(2)$ has a point stabilizer isomorphic to $S_4$, whose
abelianization has order two. It can be retained as a fixed
simple type rather than included in the scalar-phase argument.
\end{remark}

\begin{proposition}\label{hr:odd-cover-closure}
In the original-code coprime-core setup of
Section~\ref{sec:covers}, let the alphabet have characteristic $p$.
A nontrivial scalar cover of a natural point or hyperplane action
of $\PSL_d(s)$, with $d\ge3$, $s>2$, and $p\nmid s$, is
incompatible with $Q(C)$. No parity assumption on $n$ is needed.
\end{proposition}
\begin{proof}
Write the pure quotient cycle as
$\tau:(i,j)\mapsto(i,j+1)$ in the fixed CRT coordinates
$\Z_a\times\Z_b$. Let a chosen Scott factor $T_1$ be supported
on $c$ quotient blocks, and let the number of factors be $t$.
Then $b=kct$, where $k=(s^d-1)/(s-1)$.
The element $\delta=\tau^t$ normalizes $T_1$ and induces a full
$kc$-cycle on its quotient support, while preserving the
$N$-coordinate. Choose one quotient block $B_0$, and let
$\alpha=\delta^c|_{B_0}$ be the induced full $k$-cycle.
Writing $\theta$ for the automorphism of $T_1$ induced by
$\delta$, faithfulness of its local natural action gives
\[
 \theta^c=\widehat\alpha.
\]

The automorphism $\theta$ commutes with its power
$\widehat\alpha$. Lemma~\ref{proj:fullcyclecentralizer} therefore
applies, whether the local action is on points or hyperplanes, and gives
\[
 \theta=\widehat\alpha^{\,r},\qquad rc\equiv1\pmod k
\]
for some integer $r$.

On the full preimage $\Z_a\times B_0$, let
$\widetilde\alpha$ be the restriction of $\delta^c$ and put
$\pi=\widetilde\alpha^{\,r}$. On $\Z_a\times B_0$ one has
$\widetilde\alpha(i,x)=(i,\alpha x)$, so $\widetilde\alpha$ has
order $k$. Since $rc\equiv1\pmod k$, we have
\[
 \theta^{cr}=\theta,\qquad
 \widetilde\alpha^{rc-1}=1.
\]
Thus $\pi$ preserves the $N$-coordinate, induces $\theta$ on
$T_1$, and satisfies $\pi^c=\widetilde\alpha$.
For $B_j=\delta^jB_0$, the actual bijections
\[
 \psi_j=\pi^j\delta^{-j}:
       \Z_a\times B_j\longrightarrow\Z_a\times B_0,
       \qquad 0\le j<c,
\]
conjugate all the $T_1$-actions to the same action. The identity
$\pi^c=\widetilde\alpha$ verifies consistency at the wrap.
Every $\psi_j$ preserves the $N$-coordinate, so this is a
simultaneous identification of the actual scalar covers and their
translations. For a hyperplane action, apply the scalar-cover
description to the dual vector space; the same argument applies.

Set $F=\F_q$, and let $E\le N$ be the phase image on the first
block. Write $e=|E|$; this order is prime to $p$ and exceeds one.
The full preimage $\Z_a\times B_0$ is the disjoint union of
$a/e$ transitive $T_1$-sets, each a scalar cover of degree $ke$.
Each of these sets meets the part over a fixed quotient point $x$,
and its intersection there is exactly one $E$-orbit. The regular
action of $N$ on that part permutes these $a/e$ orbits transitively.
Since $N$ centralizes $T_1$, it therefore permutes the corresponding
$T_1$-sets transitively. Its translations intertwine the actions of
both $T_1$ and $E$, so any element of $F[E]I_F(T_1)$ acts by the
same operator under these identifications.
Choose a prime $\ell\mid e$ and an element $z\in E$ of order
$\ell$. Lemma~\ref{hr:scalar-localizer} consequently supplies
an element of $1+F[N]I_F(T_1)$ which translates one chosen
$N$-part above $B_0$ by $z$ and fixes the other $k-1$ parts.
The synchronized copies have the same literal action, so the
same group-algebra element performs that translation in each
of the $c$ copies. Outside the quotient support of $T_1$, the
factor acts trivially by perfectness, as in
Lemma~\ref{lem:covers}. Its augmentation ideal is zero there,
so the constructed operator is the identity on all remaining
coordinates.

It follows that this operator is the matrix of a coordinate
permutation $v$ preserving $C$, because
$N,T_1\le\PAut(C)$. Each selected $N$-part is a union of the
canonical order-$\ell$ blocks, and translation by $z$ on it is
a power of $\sigma_n^{n/\ell}$. Hence $v\in V_\ell(n)$.
The translation is nonconstant: only one of the $k>1$ parts is
selected in each copy. This contradicts $Q(C)$.
\end{proof}

\subsection{An exhaustive obstruction theorem}
\label{prime:completion-section}

We collect the local conclusions into one deterministic statement.
For a fixed coefficient prime $p$, define $\mathcal R_p$ to consist
of the following nonabelian simple types:
\begin{enumerate}
\item $A_5$, $\PSL_2(11)$, $M_{11}$, and $M_{23}$;
\item $\PSL_d(s)$ for $d\ge3$ and a prime power $s$, retaining
      the associated degree $k(d,s)=(s^d-1)/(s-1)$;
\item $\PSL_2(p^f)$ for $p^f\ge4$, retaining the associated
      degree $p^f+1$.
\end{enumerate}
Different entries can represent isomorphic simple groups; every
parameter occurrence retains its indicated permutation degree.
For the refined reduction, put
\[
 \mathcal T_0=\{A_5,\PSL_2(11),M_{11},M_{23},\PSL_3(2)\}.
\]
The first list is useful because its unbounded degree families have
summable reciprocals. The second reduction removes every unbounded
cross-characteristic family and retains the actual scalar actions.

\begin{theorem}\label{close:exhaustion}\label{prime:refinement}
Let $q=p^u$, let $p\nmid n$, and let $C\le\F_q^n$ be cyclic
with $\PAut(C)\not\le A(n)$. There is a divisor
\[
 c\mid\Delta(n)=n/\rad(n),\qquad c<n,
\]
such that the fixed quotient code
\[
 E=D_m(C),\qquad m=n/c>1,
\]
of Section~\ref{sec:descent} has the following properties.
\begin{enumerate}
\item The code $E$ satisfies at least one of these alternatives:
$Q(E)$ fails; $\mathcal A(E)$ occurs; or $\PAut(E)$ contains a
subgroup $D\cong T^h$, $h\ge1$, normalized by $\sigma_m$, with
$T\in\mathcal R_p$. In the group alternative, each parameter
occurrence retains its associated degree, which divides $m$.
\item More precisely, outside the first two alternatives one can
choose such an actual-coordinate subgroup $D$ with either
$T\in\mathcal T_0$, or
\[
 T=\PSL_d(p^f),\qquad d\ge2,
 \qquad k_0=\frac{p^{fd}-1}{p^f-1}.
\]
In the latter case, $k_0$ divides the cyclic-core index $b$ of the
quotient witness, where $m=Ab$ and $(A,b)=1$; the core-free case
is included by $A=1$. Its transitive coordinate actions are scalar
covers of natural point or hyperplane actions, up to automorphism
twists. In both cases exactly one simple factor of $D$ acts
nontrivially on each $D$-orbit on coordinates.
\end{enumerate}
The quotient maps $D_{n/c}$ are fixed independently of the witnesses
used to establish the existence of $c$. Part~(1) uses only the
projective-line closure arguments; the higher-rank arguments are
needed for part~(2).
\end{theorem}
\begin{proof}
If $Q(C)$ fails, take $c=1$. Otherwise choose a nonaffine witness
$S_n\le G\le\PAut(C)$, and write
\[
 N=\core_G(S_n),\qquad |N|=a,
 \qquad b=n/a>1.
\]
The core is proper: if $N=S_n$, then $G$ normalizes $S_n$ and is
a subgroup of $A(n)$. Define
\[
 c=\prod_{\ell\mid b}\ell^{v_\ell(a)},\qquad
 A=a/c,\qquad m=Ab=n/c.
\]
The subgroup of $N$ of order $c$ is characteristic in $N$ and
normal in $G$. Proposition~\ref{st:core-stripping}, using the exact normality
reflection of Theorem~\ref{st:reflection}, shows that the induced
witness on $E$
has cyclic core of order exactly $A$ and index $b$. Indeed, a
larger quotient core would lift inside $S_n$ to a normal cyclic
subgroup larger than $N$. Also $(A,b)=1$, and $c\mid\Delta(n)$,
because every prime removed from $a$ occurs at least once in $b$.
Thus $m=Ab>1$ and $c<n$. These assertions use $Q$ only for the
original code. If $Q(E)$ fails or $\mathcal A(E)$ occurs, stop.
Henceforth assume $Q(E)$ and $\neg\mathcal A(E)$.

Suppose first that $A=1$, and denote the induced witness by $G_1$.
Apply Lemma~\ref{st:corefree-radical} to the code $E$ and its
witness $G_1$. Its hypotheses hold because $Q(E)$ holds in the
current branch and $G_1$ has trivial cyclic core. Hence $\Rad(G_1)=1$. Lemma~\ref{lem:scott}, applied directly to
$G_1$, therefore supplies an actual subgroup
$D\cong T^h\le\PAut(E)$ normalized by $\sigma_m$. Its factors
are diagonal on disjoint collections of minimal coordinate blocks.
Exactly one factor acts nontrivially on each $D$-orbit, and the
local actions are natural actions from the primitive full-cycle
classification.

Now suppose that $A>1$. Use the centralizer splitting of
Section~\ref{sec:covers}: the centralizer of the core is
$K=N_1\times L$, where $N_1$ has order $A$, the group $L$ is
faithful on the $b$ core orbits, and the pure quotient cycle belongs
to $L$. Lemma~\ref{lem:central} gives $\Rad(L)=1$.
Lemma~\ref{lem:scott} supplies its minimal-block stable derived
subgroup $D\cong T^h$. Viewed on the original $m$ coordinates,
this is a subgroup of $K\le\PAut(E)$. The kernel $H$ of the action
of $L$ on the chosen blocks is normal in $L$, so $D=H^{(\infty)}$
is normal in $L$. The direct factor $N_1$ centralizes $L$.
Thus $D\lhd N_1\times L=K$. Since $\sigma_m$ centralizes its own
cyclic subgroup $N_1$, it belongs to $K$ and normalizes $D$.
Lemma~\ref{lem:covers} describes each actual coordinate orbit:
if the minimal quotient-block size is $k$ and its point stabilizer
is $P$, a homomorphism $\chi:P\to N_1$ with image of order $e$
gives the faithful action on $T/\ker\chi$, of degree $ke$, with
\[
 e\mid A,\qquad e\mid\exp(P^{\rm ab}),\qquad (e,k)=1.
\]
All other simple factors are trivial on this orbit. To apply the closure propositions, take
$J=\langle D,\sigma_m\rangle$ in the core-free case and $J=K$
in the coprime-core case. In both cases $D\lhd J$ and $\sigma_m\in J$.
Consequently $J$ is transitive on the set of $D$-orbits. Each such
orbit $O$ has the same size, and its setwise stabilizer satisfies
$D^O\lhd J_O^O$. In the coprime-core case the quotient blocks
all have size $k$, so $e=|O|/k$ is independent of the chosen
block and orbit. For $e=1$ projection maps $O$ bijectively onto
its quotient block, intertwining $D^O$ with the natural local simple
action. In the core-free case the minimal coordinate blocks
themselves are the $D$-orbits and have the natural local action.
In either core case $k$ divides $b$, and hence divides $m$. In particular, since
$p\nmid n$ and $m\mid n$, we have $p\nmid k$. This coprimality
is needed when the permutation-algebra enlargement propositions are
applied below; in the cross-characteristic cases we also have
$p\nmid s$, so their full hypothesis $p\nmid sk$ is satisfied.

We first prove part~(1), using only the projective-line machinery.
The primitive full-cycle classification has natural alternating,
projective point or hyperplane, and the three exceptional types
$\PSL_2(11),M_{11},M_{23}$. For a natural alternating socle
$\Alt(k)$ with $k\ge6$, the point stabilizer $\Alt(k-1)$ is
perfect. Its phase homomorphism to the cyclic core is therefore
trivial, so each actual $D$-orbit has size exactly $k$ and carries
the natural faithful action of $\Alt(k)$. Since $D$ is normalized
by the quotient shift $\sigma_m$, its orbit partition is preserved
by $\langle D,\sigma_m\rangle$. The setwise stabilizer of each
such orbit contains the image of $D$, namely $\Alt(k)$. Thus
$\mathcal A(E)$ occurs, contrary to assumption. The remaining
alternating simple case $A_5$ is retained in the fixed list. For a projective-line action, phase $e=1$ gives
$\mathcal A(E)$ by Proposition~\ref{pc:enlargement}. If $e>1$
and the geometric field has characteristic different from $p$,
Proposition~\ref{prop:closure} contradicts $Q(E)$. Thus only the
defining-characteristic projective-line family remains. All
higher-rank projective types may be retained without using their
closure results. This is exactly $\mathcal R_p$, and the asserted
degree divisibility follows from $k\mid b$.

For part~(2), consider a remaining higher-rank occurrence
$T=\PSL_d(s)$ with $p\nmid s$. If its phase is trivial,
Proposition~\ref{proj:alllengthenlargement} produces
$\mathcal A(E)$. If its phase is nontrivial and $s>2$,
Lemma~\ref{hr:parabolicphase} identifies its actual action as a
scalar cover, and Proposition~\ref{hr:odd-cover-closure}
contradicts $Q(E)$. When $s=2$ and $d\ge4$, the point stabilizer
is perfect by Remark~\ref{hr:charfree-boundary}; the phase is
trivial and the natural-action argument applies. The remaining
$\PSL_3(2)$ occurrence is retained in $\mathcal T_0$.

Every remaining unbounded projective family consequently has
geometric field $p^f$. In rank at least three and field size greater
than two, Lemma~\ref{hr:parabolicphase} gives precisely the scalar
covers. At field size two and rank at least four the phase is
trivial. In rank two, the point stabilizer has cyclic abelianization
$(\F_{p^f}^*)/Z(\mathrm{SL}_2(p^f))$, as established at the start
of Section~\ref{sec:ranktwo}, so its phase kernel gives the same
scalar-cover description. The exceptional rank-three field-two
case is already in $\mathcal T_0$. Duality treats hyperplane
actions. The one-active-factor assertion is supplied in both core
cases by the disjoint-support Scott decomposition and
Lemma~\ref{lem:covers}. This proves all claims.
\end{proof}

The theorem is entirely deterministic. The probability argument may
therefore apply the unconditional law of every fixed quotient map
and take a union over $c\mid\Delta(n)$; it need not condition on a
witness-dependent quotient choice. For later composition estimates,
we record the defining-characteristic coordinate model explicitly.

\begin{lemma}\label{consolidated:scalar-actions}
Consider a projective point or hyperplane occurrence of geometric
type $(d,p^a)$ in the defining-characteristic alternative of
Theorem~\ref{close:exhaustion}, part~(2), with
$T=\PSL_d(p^a)\not\cong\PSL_3(2)$.
Its transitive coordinate actions are scalar covers of natural point
or hyperplane actions, up to automorphism twists. The scalar subgroup
contains the center of $\mathrm{SL}_d(p^a)$.
\end{lemma}
\begin{proof}
For rank at least three and geometric field size greater than two,
the point parabolic has scalar abelian quotient by
Lemma~\ref{hr:parabolicphase}. For geometric field size two and
rank at least four its point stabilizer is perfect by
Remark~\ref{hr:charfree-boundary}, so the phase is trivial.
The omitted rank-three case is in $\mathcal T_0$.
In rank two the geometric field size $s=p^a$ exceeds three because
$T$ is simple. Its point parabolic has unipotent radical and a
cyclic torus of order $(s-1)/\gcd(2,s-1)$. Commutators with a
torus element acting nontrivially generate the unipotent radical,
so its abelian quotient is that torus. A cyclic phase of order $e$
has the unique scalar kernel corresponding to a subgroup
$H\le\mathbb F_s^*$ containing the scalar center of
$\mathrm{SL}_d(s)$; its actual coset action is
$(\mathbb F_s^d\setminus\{0\})/H$.
The hyperplane assertion follows by duality. All identifications
refer to the actual cover construction of Lemma~\ref{lem:covers}.
\end{proof}

\section{The uniform length-average theorem}\label{sec:average}\label{sec:uniform}
We now combine the coarse structural reduction with common
probability bounds. The argument is uniform in the coefficient
extension from the outset. It does not use the higher-rank
refinement needed for the pointwise theorem.

\subsection{Amplification for a fixed simple type}\label{sec:fixedT}
For coprime $A,B$ with $(AB,q)=1$,
write
\[
 R_{q,A}(B)=\#\{X\text{ a $q$-cyclotomic coset modulo }B:
                                       \ord_A(q)\mid|X|\}.
\]

\begin{lemma}\label{lem:amplification}
In the fixed Chinese remainder coordinates
$\Z_{AB}=\Z_A\times\Z_B$, if $\pi\in\Sym(A)$ is not a translation,
then
\begin{equation}\label{eq:amp}
 \Pr(\pi\times1\in\PAut(C))\le(3/4)^{R_{q,A}(B)}.
\end{equation}
For fixed $q,A$, the quantity $R_{q,A}(B)$ tends to infinity in relative
natural density as $B$ varies through $(B,qA)=1$.
\end{lemma}
\begin{proof}
Decompose the group algebra of the second cyclic factor:
\[
 \F_q[C_B]=\prod_X E_X,\qquad E_X=\F_{q^{|X|}}.
\]
A uniform length-$AB$ cyclic code thereby becomes independent uniform
length-$A$ cyclic codes over the $E_X$. If $\ord_A(q)\mid|X|$, the
field $E_X$ splits $x^A-1$. In a Fourier basis its cyclic subspaces are
independent selections of the $A$ eigenlines. Were the permutation
matrix of $\pi$ diagonal in this basis, it would commute with the regular
$A$-cycle, hence would be a translation. Thus it has a nonzero
$i\to j$ entry, $i\ne j$. Selecting line $i$ and omitting line $j$
is incompatible with invariance, excluding one quarter of selections.
Multiplying these independent losses proves \eqref{eq:amp}.

Put $h=\ord_A(q)$. If $h=1$, then
$R_{q,A}(B)=m_q(B)\ge2^{\omega(B)}$, which diverges in density.
If $h>1$, Wiertelak's density theorem and the positivity of its
constant, as recorded in \cite[Theorem 1 and the paragraph following it]{Moree},
give a positive relative density among the primes for
\[
 \mathcal P_{q,h}=\{r\text{ prime}:r\nmid qA,
                                         h\mid\ord_r(q)\}.
\]
Partial summation gives $\sum_{r\in\mathcal P_{q,h}}1/r=\infty$.
For any finite $S\subset\mathcal P_{q,h}$, the Chinese remainder theorem
gives relative density $\prod_{r\in S}(1-1/r)$ for the admissible $B$
coprime to all $r\in S$, since $(r,qA)=1$. These products tend to zero.
Thus almost every admissible $B$ has a prime divisor $r\in\mathcal P_{q,h}$. For every divisor $d\mid B$ divisible by $r$,
$h\mid\ord_d(q)$, and the characters of additive order $d$ supply at
least one counted coset. Consequently
$R_{q,A}(B)\ge2^{\omega(B)-1}$ on these lengths.
Finally $\omega(B)\to\infty$ in relative density, by the usual finite
prime-divisibility indicators and their divergent reciprocal sum.
Excluding a fixed finite set of primes does not affect either argument.
\end{proof}

\begin{lemma}\label{lem:uniformR}
For coprime $A,B$ with $p\nmid AB$ and every $u\ge1$,
\[
 R_{p^u,A}(B)\ge R_{p,A}(B).
\]
\end{lemma}
\begin{proof}
Put $h=\ord_A(p)$. A $p$-coset of length $L$ splits into
$\gcd(L,u)$ cosets of length $L/\gcd(L,u)$ under $p^u$.
If $h\mid L$, then
\[
 \ord_A(p^u)=\frac{h}{\gcd(h,u)}
       \ \Bigm|\ \frac{L}{\gcd(L,u)}.
\]
Prime by prime, this follows from
$\max\{v_r(h)-v_r(u),0\}\le\max\{v_r(L)-v_r(u),0\}$.
Every previously counted coset therefore supplies at least one
counted coset after extending the coefficient field.
\end{proof}

\begin{theorem}\label{thm:fixedT}
Fix a prime $p$ and a nonabelian finite simple group $T$. For $p\nmid n$, put
\[
 A=\prod_{\ell\mid|\Aut(T)|}\ell^{v_\ell(n)},\qquad B=n/A,
\]
and define
\begin{equation}\label{eq:uniformfixedT}
 F_{p,T}(n)=\min\{1,A!(3/4)^{R_{p,A}(B)}\}.
\end{equation}
For every $u\ge1$, the event
\begin{equation}\label{eq:fixedTevent}
 \exists D\le\PAut(C):\quad D\cong T^r\ (r\ge1),\qquad
                      \sigma_nD\sigma_n^{-1}=D
\end{equation}
has probability at most $F_{p,T}(n)$ for a uniform cyclic code
over $\F_{p^u}$. Moreover,
\[
 \Av_{p,X}F_{p,T}\longrightarrow0.
\]
There is no restriction on the multiplicity, supports, permutation
representation, or cyclic core in this assertion.
\end{theorem}
\begin{proof}
Write $q=p^u$. Let $\mathcal P$ be the finite set of primes dividing $|\Aut(T)|$,
and put
\[
 A=\prod_{\ell\in\mathcal P}\ell^{v_\ell(n)},\qquad B=n/A,
 \qquad \beta=\sigma_n^A.
\]
Then $(A,B)=1$ and $(B,|\Aut(T)|)=1$. Suppose \eqref{eq:fixedTevent}
holds. Conjugation by $\beta$, of order dividing $B$, permutes the
simple factors of $D$. On a cycle of $d$ factors, $d\mid B$, its
return automorphism has order dividing both $B/d$ and $|\Aut(T)|$.
It is therefore the identity. The fixed subgroup on this factor cycle
is a diagonal copy of $T$, possibly with identifications by automorphisms.
It follows that
\[
 K=C_D(\beta)\cong T^s\quad\text{for some }s\ge1.
\]
Since $\sigma_n$ normalizes $D$ and commutes with $\beta$, it normalizes
$K$. Thus
\[
 J=\langle K,\sigma_n^B\rangle=K\langle\sigma_n^B\rangle
\]
has order dividing $|K|A$, so $(|J|,B)=1$, and $J$ centralizes $\beta$.

Use the fixed coordinates $\Z_n=\Z_A\times\Z_B$. The $\beta$-orbits
are the sets $F_i=\{i\}\times\Z_B$. Their $J$-action on the
set of fibers is transitive, because $\sigma_n^B$ induces a full
$A$-cycle. The setwise stabilizer $J_{F_i}$ acts on $F_i$ through
the centralizer of the regular $B$-cycle $\beta|_{F_i}$, which is
cyclic of order $B$. Its image has order dividing both $|J|$ and
$B$, and is therefore trivial. Thus every element of $J_{F_i}$ fixes
$F_i$ pointwise. It follows that a $J$-orbit meets each fiber in at
most one point; transitivity on the $A$ fibers then shows that every
$J$-orbit has exactly $A$ points. The $\sigma_n^B$-orbit of any
point is a row $R_j=\Z_A\times\{j\}$ of size $A$, so it equals
that $J$-orbit. Hence every element of $J$ preserves each row.
Writing its action as $(i,j)\mapsto(\pi_j(i),j)$, commutation
with $\beta$ gives $\pi_{j+A}=\pi_j$. Since $(A,B)=1$, these
permutations are all equal, and therefore
\[
 J\le\Sym(A)\times1.
\]
The nontrivial perfect subgroup $K$ cannot act only by translations
of $\Z_A$, an abelian group. Some element of $K$ is therefore
$\pi\times1$ with $\pi$ a nontranslation. In particular the event
is impossible if $A=1$.

Lemma~\ref{lem:amplification} and a union over all $\pi$ give the
pointwise bound
\begin{equation}\label{eq:fixedTbound}
 \Pr(\eqref{eq:fixedTevent})
 \le\min\{1,A!(3/4)^{R_{q,A}(B)}\}.
\end{equation}
This single bound includes every embedding and every multiplicity.
Lemma~\ref{lem:uniformR} bounds its right side by $F_{p,T}(n)$.
It remains to average this one common majorant, which involves only
the base alphabet $p$.
Fix $K_0$ and restrict to lengths with $v_\ell(n)\le K_0$ for all
$\ell\in\mathcal P$. Only finitely many values of $A$ remain. For
each such value, $B$ is additionally coprime to the fixed product of
all primes in $\mathcal P$. This restriction has positive relative
density, so the zero-density exceptional set from
Lemma~\ref{lem:amplification}, used with $q=p$, remains zero-density
after imposing it.
For clarity, fix such an $A$ and an integer $L\ge1$. Among the
corresponding lengths $n=AB\le X$, the proportion with
$R_{p,A}(B)<L$ tends to zero as $X\to\infty$. On all the others,
$F_{p,T}(n)\le A!(3/4)^L$. Letting $L\to\infty$ proves vanishing
of the conditional average for this $A$. There are only finitely many
retained $A$ for the fixed $K_0$. The discarded lengths have upper
relative density at most
\[
 O_p\left(\sum_{\ell\in\mathcal P}\ell^{-K_0-1}\right).
\]
First let $X\to\infty$, then $K_0\to\infty$.
\end{proof}

\subsection{The removed core and summable degree tails}
\begin{lemma}\label{lem:coremoment}
For every fixed prime $p$ and $0<\alpha<1$, the convergent product
\[
 C_{p,\alpha}=\prod_{\substack{\ell\ \mathrm{prime}\\\ell\ne p}}
 \left(1+\frac{\ell^\alpha-1}
 {\ell^2(1-\ell^{\alpha-1})}\right)
\]
satisfies
\begin{equation}\label{eq:coremoment}
 \Av_{p,X}\Delta(n)^\alpha\longrightarrow C_{p,\alpha}.
\end{equation}
In particular, for the Riemann zeta function $\zeta$,
\[
 C_{p,1/2}=\frac{\zeta(3/2)}{(1+p^{-3/2})\zeta(3)}.
\]
Consequently, for every $K>0$,
\begin{equation}\label{eq:coretail}
 \limsup_{X\to\infty}\Av_{p,X}\mathbf1_{\Delta(n)>K}
       \le C_{p,\alpha}K^{-\alpha}.
\end{equation}
\end{lemma}
\begin{proof}
The function $f(n)=\Delta(n)^\alpha$ is multiplicative. If
$g=f*\mu_{\rm Mob}$, where $\mu_{\rm Mob}$ is the M\"obius function,
then
\[
 g(1)=1,\qquad g(\ell)=0,\qquad
 g(\ell^j)=(\ell^\alpha-1)\ell^{\alpha(j-2)}\quad(j\ge2).
\]
Thus $g\ge0$ and
\[
 \sum_{\substack{d\ge1\\p\nmid d}}\frac{g(d)}d=C_{p,\alpha}<\infty,
\]
because the excess of each Euler factor over one is
$O_\alpha(\ell^{\alpha-2})$. The convolution identity gives
\[
 \Av_{p,X}f
 =\sum_{\substack{d\le X\\p\nmid d}}
       g(d)\frac{N_p(X/d)}{N_p(X)}.
\]
For each fixed eligible $d$, the ratio tends to $1/d$.
For all sufficiently large $X$ it is at most
$2/((1-1/p)d)$, and terms with $d>X$ can be set to zero.
Dominated convergence proves \eqref{eq:coremoment}.
At $\alpha=1/2$, each Euler factor is $1+\ell^{-3/2}$, so
\[
 \prod_{\ell\ne p}(1+\ell^{-3/2})
 =\frac{1}{1+p^{-3/2}}\prod_{\ell}
   \frac{1-\ell^{-3}}{1-\ell^{-3/2}}
 =\frac{\zeta(3/2)}{(1+p^{-3/2})\zeta(3)}.
\]
The pointwise inequality
$\mathbf1_{\Delta(n)>K}\le\Delta(n)^\alpha/K^\alpha$
then proves \eqref{eq:coretail}.
\end{proof}

For $r\ge3$ and a prime power $s$, write
\[
 k(r,s)=\frac{s^r-1}{s-1}.
\]
The relevant reciprocal sums are
\begin{align}
 \sum_{s\ \text{prime power}}\sum_{r\ge3}\frac1{k(r,s)}
 &\le\sum_{s=2}^{\infty}\sum_{r=3}^{\infty}s^{-(r-1)}
 =\sum_{s=2}^{\infty}\frac1{s(s-1)}=1,\label{eq:highertail}\\
 \sum_{f\ge1}\frac1{p^f+1}&\le\frac1{p-1}.\label{eq:linetail}
\end{align}
In particular, each series has tails tending to zero. There are only
finitely many parameter choices below any fixed degree cutoff.

\subsection{Affine exceptions and degeneracy in length average}
The elementary bounds in Section~\ref{sec:affine} already hold at
each length and for every coefficient field. We now average common
upper bounds. Recall from \eqref{eq:affine-odd-prime-bound} that,
if $r(n)$ counts the distinct odd prime divisors of $n$ and $r(n)\ge1$,
\[
 M_p(n)=\frac{\varphi(n)}{\ord_n(p)}\ge2^{r(n)-1}.
\]

The function $r(n)$ tends to infinity in relative density among
$p\nmid n$. Indeed, for a finite set of odd primes different from
$p$, the corresponding divisibility indicators have independent
limiting Bernoulli laws by the Chinese remainder theorem. Their
sum has mean $L=\sum_\ell1/\ell$ and variance at most $L$.
For fixed $R<L$, the upper relative density of $r(n)\le R$ is
at most $L/(L-R)^2$. The divergent reciprocal prime sum permits
$L\to\infty$. Thus $M_p(n)\to\infty$ in relative density, and
bounded functions of $M_p(n)$ tending to zero have vanishing
length average. This applies both to $2^{-M_p(n)}$ and to the
affine envelope used below.

\subsection{A common finite-cutoff bound}
For an integer $H\ge2$, let $\mathcal T_{p,H}$ be the finite set
of simple isomorphism types comprising the four fixed types in the coarse part of
Theorem~\ref{close:exhaustion} and every parameter choice in $\mathcal R_p$ of degree at most $H$.
Let $E_{p,H}(v)$ indicate that $v$ is divisible by a degree greater
than $H$ in either parameter family, and put
\begin{align*}
 \Theta_p(H)
 &=\sum_{\substack{s\ \mathrm{prime\ power},\ r\ge3\\k(r,s)>H}}
       \frac{1}{k(r,s)}
   +\sum_{\substack{f\ge1,\ p^f\ge4\\p^f+1>H}}
       \frac{1}{p^f+1},\\
 W_{p,H}(v)
 &=\min\left\{1,U_p(v)+V_p(v)
       +\sum_{T\in\mathcal T_{p,H}}F_{p,T}(v)+E_{p,H}(v)\right\}.
\end{align*}
Parameter repetitions in $\Theta_p(H)$ merely overcount the tail.
Equation~\eqref{eq:uniformfixedT} defines the finite simple-type bounds
$F_{p,T}$. The canonical union bounds $U_p,V_p$ are given in \eqref{tk:envelope} and \eqref{ab:loss-envelope},
and their uniform limits were proved in
Theorem~\ref{thm:canonical-uniform}.
For each fixed $H$, their uniformity and zero averages give
\begin{equation}\label{close:Wmean}
 \limsup_{X\to\infty}\Av_{p,X}W_{p,H}
     \le\frac{\Theta_p(H)}{1-1/p},\qquad
 \Theta_p(H)\longrightarrow0\quad(H\to\infty).
\end{equation}
Indeed, a degree $d$ divides at most $X/d$ positive integers up to $X$;
apply the union bound and \eqref{eq:highertail}--\eqref{eq:linetail}.

Define the affine tail envelope
\[
 \mathfrak a_p(n)=
  \sup_{\substack{M\in\N\\M\ge M_p(n)}}
       \min\{1,(M-1)2^{-M/2}\},\qquad
 M_p(n)=\varphi(n)/\ord_n(p).
\]
Since $M_{p^u}(n)\ge M_p(n)$, Section~\ref{sec:affine} gives
$\sup_u a_{p^u}(n)\le\mathfrak a_p(n)$ and
$\Av_{p,X}\mathfrak a_p\to0$.

\begin{proposition}\label{close:majorant}
For all integers $K,H\ge2$ and all $p\nmid n$,
\begin{equation}\label{close:pointwise}
 \sup_{u\ge1}\epsilon_{p^u}(n)
 \le\min\left\{1,\mathfrak a_p(n)
       +\mathbf1_{\Delta(n)>K}
       +\sum_{\substack{c\mid n\\1\le c\le K,\ c<n}}
                     W_{p,H}(n/c)\right\}.
\end{equation}
In particular, for every fixed $0<\alpha<1$, with
$C_{p,\alpha}$ from Lemma~\ref{lem:coremoment} and
$h_K=\sum_{c=1}^K1/c$, one has
\begin{equation}\label{close:limit}
 \limsup_{X\to\infty}\Av_{p,X}
      \bigl(\sup_{u\ge1}\epsilon_{p^u}\bigr)
 \le C_{p,\alpha}K^{-\alpha}
       +\frac{h_K}{1-1/p}\Theta_p(H).
\end{equation}
The same limiting upper bound holds for the nondegenerate laws.
\end{proposition}
\begin{proof}
Fix $u$. If a code has only affine automorphisms, any exception to
$A_{p^u}(n)$ is bounded by $a_{p^u}(n)$. Otherwise apply the coarse part of
Theorem~\ref{close:exhaustion}. When $\Delta(n)\le K$, its divisor
$c$ occurs in the displayed finite union. For a fixed $c$, Lemma~\ref{lem:quotient-law} makes the unconditional
quotient $D_{n/c}(C)$ uniform among cyclic codes of length $n/c$. Its canonical obstruction probabilities are bounded
by $U_p+V_p$. Each of the four fixed types, and each parameter-family
occurrence of degree at most $H$, is covered by its $F_{p,T}$ bound.
Any remaining parameter-family occurrence forces $E_{p,H}=1$.
Thus the probability of the entire obstruction union at that fixed
quotient is at most $W_{p,H}(n/c)$.

All these bounds use the same right side for every $u$, so the union
bound proves \eqref{close:pointwise} before and after taking the
supremum. The choice of a witnessing divisor is never conditioned
on when using the uniform quotient law. In particular, the term
$c=1$ already includes failure of the original $Q(C)$.

For fixed $c$ with $p\nmid c$, a change of variable gives
\[
 \frac{1}{N_p(X)}\sum_{\substack{n\le X\\p\nmid n,\ c\mid n}}
       W_{p,H}(n/c)
 =\frac{N_p(X/c)}{N_p(X)}\Av_{p,X/c}W_{p,H},\qquad
 \frac{N_p(X/c)}{N_p(X)}\longrightarrow\frac1c.
\]
Allowing $c=n$ in this calculation gives an upper bound for the
sum with $c<n$, since $W_{p,H}$ is nonnegative.
There is no term if $p\mid c$. Lemma~\ref{lem:coremoment} yields
\[
 \limsup_{X\to\infty}\Av_{p,X}\mathbf1_{\Delta(n)>K}
          \le C_{p,\alpha}K^{-\alpha}.
\]
Now combine \eqref{close:Wmean}, the vanishing affine average,
and $\sum_{c\le K,p\nmid c}1/c\le h_K$.
This proves \eqref{close:limit}.

Selecting any primitive constituent implies nondegeneracy. Therefore
\[
 \sup_{u\ge1}\Pr_{p^u}(C\text{ degenerate})\le2^{-M_p(n)},
\]
whose average tends to zero. Conditioning changes the probability of any
event by at most this degeneracy probability, by the total-probability
identity. This proves the last assertion.
\end{proof}

\begin{proof}[Proof of Theorem~\ref{thm:main}]
In \eqref{close:limit}, first fix $K$ and let $H\to\infty$.
The summable degree tails give $\Theta_p(H)\to0$. Next let
$K\to\infty$. For any fixed $0<\alpha<1$,
$C_{p,\alpha}K^{-\alpha}\to0$. The left side is nonnegative and independent of
both cutoffs; it must be zero. The conditional assertion follows
from the same inequality.
\end{proof}

The order of operations is $X\to\infty$, then $H\to\infty$ with
$K$ fixed, then $K\to\infty$. No bound uniform in a growing simple
type, multiplicity, or removed divisor is assumed. The pointwise
inequality is a common majorant for all alphabets, not a claim that
the full exceptional probability is monotone under alphabet extension.

\subsection{The universal intersection}
The theorem also treats the universal cyclotomic-scheme baseline
identified in \cite[Theorem A]{WX}.
\begin{corollary}\label{close:universal}
For each fixed prime $p$, put
\[
 \eta_q(n)=\Pr_{C\in\mathcal C_q(n)}
                 \bigl(\PAut(C)\ne B_q(n)\bigr),\qquad
 e_p(n)=\mathbf1_{\exists u\ge1:\ B_{p^u}(n)\ne A_{p^u}(n)}.
\]
Then
\[
 \Av_{p,X}\bigl(\sup_{u\ge1}\eta_{p^u}\bigr)\to0,
 \qquad \Av_{p,X}e_p\to0.
\]
The first conclusion also holds after conditioning on nondegeneracy.
\end{corollary}
\begin{proof}
For every cyclic code,
$A_q(n)\le B_q(n)\le\PAut(C)$, so
$\eta_q(n)\le\epsilon_q(n)$ under either distribution.
If $B_q(n)\ne A_q(n)$, then $\epsilon_q(n)=1$.
Consequently $e_p(n)\le\sup_u\epsilon_{p^u}(n)$.
Apply Theorem~\ref{thm:main}.
\end{proof}

The fixed-alphabet density-zero assertion for universal exceptional
lengths was proved by Wu and Xiang \cite[Theorem C]{WX} and is not
claimed as new. The corollary records its relation to the stronger
code-proportion statement and the uniform characteristic quantifier.

\section{Quantitative concentration from a retained prime}
\label{gl:section}

The structural reduction in Theorem~\ref{close:exhaustion} leaves
canonical translation and alternating events, a fixed finite family
of simple groups, and defining-characteristic projective groups on
actual coordinate covers. We now bound these alternatives at a
specified length. The arithmetic input is a prime divisor whose
multiplicative order equals the order of the entire length. Because
canonical stripping preserves every prime divisor, one such prime
controls all quotient lengths simultaneously.

\subsection{Full-order primes and canonical quotient losses}
\label{sec:retained-primes}\label{pp:quotients-section}

Stripping never removes an entire prime divisor. A sufficiently large
prime of full multiplicative order gives many character labels whose
orbits under multiplication by the coefficient prime $p$ have full length. The loss estimates below apply to general
lengths; the first lemma supplies the required prime on divisors of
primitive lengths.

\begin{lemma}\label{pp:retainedprime}
Fix a prime $p$. For every sufficiently large integer $m$, there is
a prime $z\mid p^m-1$ such that
\[
 \ord_z(p)=m,\qquad
 z>m\exp\left(\frac{\log m}{104\log\log m}\right)>4m.
\]
Consequently $z\mid d$ whenever
$\rad(p^m-1)\mid d\mid p^m-1$.
\end{lemma}
\begin{proof}
Apply Stewart's bound for the homogeneous cyclotomic value
$\Phi_m(\alpha,\beta)$ with $(\alpha,\beta)=(p,1)$
\cite[Theorem~1.1, Eq.~(1.7)]{Stewart}. Since
$\Phi_m(p,1)=\Phi_m(p)$, it gives the displayed lower bound for
$z=P(\Phi_m(p))$. The second inequality holds for sufficiently
large $m$ because $\log m/\log\log m\to\infty$.
In particular $z>m$, so $z\nmid m$. The polynomial $X^m-1$
is squarefree over $\F_z$, and its factorization
$\prod_{j\mid m}\Phi_j(X)$ therefore has pairwise coprime
factors modulo $z$. Since $\Phi_m(p)=0\pmod z$, the order of
$p$ modulo $z$ cannot be a proper divisor of $m$.
The final assertion follows from $z\mid\rad(p^m-1)$.
\end{proof}

\begin{lemma}\label{pp:TK}
Let $d\mid p^m-1$ and suppose a prime $z\mid d$ satisfies
$\ord_z(p)=m$ and $z>4m$.
For primes $\ell,t$ with $\ell t\mid d$, including $\ell=t$,
\[
 \Delta_p(d;\ell,t)\ge\frac{d}{8m}.
\]
For distinct primes, and for a repeated prime with no order growth,
the stronger bound $d/(4m)$ holds.
\end{lemma}
\begin{proof}
All multiplicative orders modulo divisors of $d$ divide $m$.
For distinct primes, the positive divisor sum in
Proposition~\ref{tk:distinct-loss} therefore gives
\[
 \Delta_p(d;\ell,t)
 \ge\frac1m
 \sum_{\substack{a\mid d\\a\nmid d/\ell\\a\nmid d/t}}
       \varphi(a)
 =\frac dm(1-1/\ell)(1-1/t)\ge\frac d{4m}.
\]

For repeated primes write $d=\ell^k s$, $k\ge2$, $(\ell,s)=1$,
and $R_j=\ord_{\ell^j}(p)$.
If $R_k=R_{k-1}$, Proposition~\ref{tk:repeated-loss} gives
\[
 \Delta_p(d;\ell,\ell)
 \ge\frac{(\ell-1)\varphi(\ell^{k-1})}{m}
                    \sum_{a\mid s}\varphi(a)
 =\frac dm(1-1/\ell)^2\ge\frac d{4m}.
\]
Otherwise $R_k=\ell R_{k-1}$. Put $e=v_\ell(R_k)$.
Since $R_k\mid m$, one has $\ell^e\mid m$ and $\ell\le m$.
The assumed prime $z$ satisfies $z>m$, hence $z\ne\ell$
and $z\mid s$.
For $z\mid a\mid s$ we have
$\ord_z(p)\mid\ord_a(p)\mid m$, hence $\ord_a(p)=m$.
As $\ell^e\mid m$, each such $a$ meets the order condition in
\eqref{tk:growth}. Retaining these terms gives
\begin{align*}
 \Delta_p(d;\ell,\ell)
 &\ge\frac{(\ell-1)\varphi(\ell^{k-1})}{m}
             \sum_{\substack{a\mid s\\z\mid a}}\varphi(a)\\
 &=\frac dm(1-1/\ell)^2(1-z^{-v_z(s)})
 \ge\frac d{8m}.
\end{align*}
\end{proof}

\begin{lemma}\label{pp:AB}
Under the hypotheses of Lemma~\ref{pp:TK}, for every divisor
$a\mid d$ with $a\ge5$,
\[
 \Gamma_p(d;a)
 =m_p(d)-m_p(d/a)-m_p(B_a(d))\ge\frac d{4m}.
\]
\end{lemma}
\begin{proof}
Put $A=\prod_{\ell\mid a}\ell^{v_\ell(d)}$ and $B=d/A$, so
$(A,B)=1$, $a\mid A$, and $A\ge a\ge5$. Define the invariant set
\[
 \Omega=\{x\in\Z_A:\ord_{\rm add}(x)\nmid A/a\},
 \qquad |\Omega|=A-A/a.
\]
As in Lemma~\ref{lem:lossmonotonicity}, $\Gamma_p(d;a)$ is the
orbit defect of the equivariant surjection
$\Omega\times\Z_B\longrightarrow\Z_B$ under multiplication by
$p$. Every invariant subset of the target contributes a nonnegative
orbit defect.

Let $z$ be the prime in the hypotheses.
If $z\mid B$, take the target subset $Y\subseteq\Z_B$ of labels
whose additive order is divisible by $z$. Every orbit on $Y$ has
length exactly $m$, as does every orbit in its inverse image.
Moreover $|Y|=B(1-z^{-v_z(B)})\ge B/2$. Its contribution to the
defect is therefore
\[
 \frac{(|\Omega|-1)|Y|}{m}
 \ge\frac{3A}{5}\frac{B}{2m}
 =\frac{3d}{10m}\ge\frac d{4m},
\]
where $|\Omega|-1=A-A/a-1\ge3A/5$ follows from $A\ge a\ge5$.

If $z\mid A$, then $z\mid a$. All labels of $\Z_A$ whose
additive orders contain the full $z$-part of $A$ lie in $\Omega$.
There are $A(1-1/z)$ such labels, and their products with arbitrary
labels in $\Z_B$ all have orbit length $m$. Hence the domain has
at least $d(1-1/z)/m$ orbits. The target has at most $B\le d/z$
orbits. Since $z>4m$,
\[
 \Gamma_p(d;a)
 \ge\frac d m(1-1/z)-\frac d z
 =\frac d m\left(1-\frac{m+1}{z}\right)
 \ge\frac d{2m}.
\]
This proves the claim in both cases.
\end{proof}

\begin{theorem}\label{pp:QA-all-exponents}
Let $p\nmid n$, put $m=\ord_n(p)$ and $R=\rad(n)$, and suppose
that a prime $z\mid n$ satisfies $\ord_z(p)=m$ and $z>4m$.
There is an absolute constant $K$ such that
\[
 \sup_{u\ge1}\Pr_{C\in\mathcal C_{p^u}(n)}
 \left(\exists c\mid\Delta(n):
       \neg Q(D_{n/c}(C))\quad\text{or}\quad
       \mathcal A(D_{n/c}(C))\right)
 \le K m^3\exp\left(-\frac{(\log2)R}{16m}\right).
\]
Conditioning the original code on nondegeneracy costs at most a
factor of two.
\end{theorem}
\begin{proof}
Every stripped quotient length $d=n/c$ retains $z$ and divides
$p^m-1$. The preceding lemmas and the exact canonical counts
therefore bound the two events by
\[
 \omega(d)^2 2^{-d/(8m)},\qquad
 \tau(d)2^{-d/(4m)}.
\]
These bounds are uniform over coefficient extensions by
Lemma~\ref{lem:lossmonotonicity}. Each fixed quotient map takes
the unconditional uniform code law to the uniform law at length
$d$. Since $c\mid\Delta(n)$ implies $R\mid d$, the union bound
and the elementary exponential-series estimate give
\[
 2\sum_{d\ge R}d^2e^{-(\log2)d/(8m)}
 \le K m^3e^{-(\log2)R/(16m)}.
\]
Here one splits off $e^{-(\log2)R/(16m)}$ and uses
$\sum_{d\ge1}d^2x^d=x(1+x)/(1-x)^3$ at
$x=e^{-(\log2)/(16m)}$.
The original code has nondegeneracy probability at least $1/2$;
conditioning the entire union event proves the final assertion.
\end{proof}

\subsection{Invariant cyclic subspaces and composition length}
\label{sec:composition-counts}

We use the multiplicity-free argument for a regular cyclic subgroup
that Bardoe and Sin apply to nonzero vectors
\cite[Lemma~2.1 and the paragraph following it]{BardoeSin}.
Here the simple factors are counted after adjoining the prescribed
full cycle to an arbitrary group $D$.

\begin{lemma}\label{odd:cycle-composition}
Let $F$ be a finite field of characteristic $p$, let $p\nmid v$,
and let $D\le\Sym(v)$. Suppose that the permutation module over
an algebraic closure $K$ of $F$ has composition length at most
$L$ as a $KD$-module. Then there are at most $2^L$ subspaces
of $F^v$ invariant under both $D$ and a specified full $v$-cycle
$\sigma_v$.
\end{lemma}
\begin{proof}
Put $H=\langle D,\sigma_v\rangle$ and $S=\langle\sigma_v\rangle$.
The restriction of $K^v$ to $S$ is the regular module of a cyclic
group whose order is invertible in $K$. Hence every irreducible
$KS$-module occurs exactly once.
If a simple $KH$-module occurred twice among the composition
factors of $K^v$, its restriction to $S$ would contribute some
irreducible $KS$-module at least twice. Restriction is exact and
$KS$ is semisimple, so this is impossible. Thus each simple composition factor of $K^v$ as a $KH$-module
has multiplicity one.

Every $KH$-submodule is determined by its set of composition
factors. To see this, write $[M:S_0]$ for the multiplicity of a
simple $KH$-module $S_0$ in $M$. If $U,W$ have the same factors,
then for every $S_0$ occurring in either one,
\[
 [U\cap W:S_0]=[U:S_0]+[W:S_0]-[U+W:S_0]\ge 2-1=1 .
\]
Here the equality follows from the exact sequence for intersection
and sum, and the bound follows from $U+W\le K^v$. Thus $U\cap W$
has the same composition length as $U$ and $W$, giving $U=W$.
The $KH$-composition length is at most the $KD$-composition
length: restricting a composition series gives a $KD$-filtration
with nonzero successive quotients. Hence there are at most $2^L$
submodules over $K$. Extension of scalars injects subspaces of
$F^v$ into subspaces of $K^v$, proving the assertion.
No semisimplicity assumption is imposed on $KD$ or $KH$.
\end{proof}

\begin{lemma}\label{odd:normalized-conjugates}
For any fixed $D\le\Sym(v)$, at most $v$ conjugates of $D$
are normalized by a specified full $v$-cycle $\sigma_v$.
\end{lemma}
\begin{proof}
Put $N=N_{\Sym(v)}(D)$ and let $c_v(N)$ count the $v$-cycles
in $N$. The number of $g\in\Sym(v)$ with
$\sigma_v\in N_{\Sym(v)}(D^g)=N^g$ is $v c_v(N)$.
Indeed, all $v$-cycles are conjugate in $\Sym(v)$, and each
prescribed image of $\sigma_v$ under conjugation has exactly
$|C_{\Sym(v)}(\sigma_v)|=v$ conjugators.
Each conjugate of $D$ has $|N|$ representatives, so the desired
number is
\[
 \frac{v c_v(N)}{|N|}\le v.
\]
\end{proof}

\label{pf:defining-section}

We next apply the composition count to scalar covers of a natural
projective action. The required estimate concerns composition length,
so it remains useful even when the permutation module is not
semisimple.

\begin{lemma}\label{pf:bs-length}
Let $s=p^a$, $d\ge2$, $T=\PSL_d(s)$, and
$z=\gcd(d,s-1)$. Let $H\le\F_s^*$ contain the scalar center of
$\mathrm{SL}_d(s)$, put $e=[\F_s^*:H]$, and let
\[
 \Omega_H=(\F_s^d\setminus\{0\})/H,
 \qquad k=\frac{s^d-1}{s-1}.
\]
Over an algebraically closed field $K$ of characteristic $p$,
\[
 \ell_{KT}(K^{\Omega_H})
 \le 1+(d-1)^a+(e-1)d^a\le e d^a.
\]
For $p=d=2$, one also has
\[
 \ell_{KT}(K^{\Omega_H})\le 2e\phi_{\mathrm{gr}}^a,
 \qquad \phi_{\mathrm{gr}}=(1+\sqrt5)/2.
\]
These bounds are unchanged by an automorphism twist of the action.
\end{lemma}
\begin{proof}
Use first $G=\mathrm{GL}_d(s)$ and its scalar center
$Z\cong\F_s^*$. Since $|Z|$ is prime to $p$, the permutation
module on nonzero vectors splits into central-character summands.
In the notation of Bardoe--Sin, their projective dimension is
$n=d-1$ and their extension degree is $t=a$.
The trivial-character summand is the projective permutation module.
Bardoe--Sin \cite[Theorem A]{BardoeSin} gives its composition
length as $1+|\mathcal H|\le1+(d-1)^a$.
For each nontrivial central character, their Theorem C gives at
most $d^a$ composition factors. Exactly $e$ central characters
are trivial on $H$, and taking $H$-fixed functions produces
$K^{\Omega_H}$.
Here
\[
 K^{\Omega_H}\cong
       \bigl(K^{\F_s^d\setminus\{0\}}\bigr)^H.
\]
The origin, if included in the ambient vector permutation module,
contributes only to the trivial central character; it has no effect
on the nontrivial-character application of Theorem C. Moreover,
taking $H$-fixed vectors is exact because $p\nmid|H|$, equivalently
because $|H|^{-1}\sum_{h\in H}h$ is an idempotent. Thus precisely
the indicated central-character summands are retained.

In the authors' notation the projective factors are indexed by the
constant factor and a subset of $\{1,\ldots,d-1\}^a$. For a
nontrivial central weight the indexing set is a subset of
$\{0,\ldots,d-1\}^a$. This gives the two bounds even when $d=2$.
The constituents in Theorems~A and~C remain simple on restriction
to the finite group $\mathrm{SL}_d(s)$. This is proved in
\cite[\S2.3, equations (14)--(15)]{BardoeSin}: the homogeneous
components of the truncated polynomial ring are restricted simple
modules, and Steinberg's tensor product theorem shows that the
corresponding Frobenius-twisted tensor products remain simple for
$\mathrm{SL}_d(p^a)$. Thus a composition series for any selected
central-character summand, viewed as a $G$-module, remains a
composition series after restriction to $\mathrm{SL}_d(s)$.
In particular, its composition length is preserved exactly, not
merely bounded above. Distinct $G$-factors may become isomorphic
after restriction, but they must still be counted with their
original multiplicities.

Because $H$ contains the scalar center of $\mathrm{SL}_d(s)$,
that center acts trivially on the $H$-fixed module. Hence every
restricted composition factor is trivial on the center and factors
through $T=\mathrm{PSL}_d(s)$. Passing from $\mathrm{SL}_d(s)$
to this quotient does not change simplicity or composition length.
The elementary inequality $1+(d-1)^a\le d^a$ proves the first
bound. No multiplicity-free claim for the $KT$-module is needed.

For $p=d=2$, the projective summand has length exactly two.
In the convention of \cite[Eq.~(5)]{BardoeSin}, write the
nontrivial central character as $t\mapsto t^{-w}$, with
$1\le w\le 2^a-2$ and
$w=\sum_{i=0}^{a-1}w_i2^i$, $w_i\in\{0,1\}$.
Theorem C indexes its composition factors by cyclic tuples
$(b_0,\ldots,b_{a-1})\in\{0,1\}^a$ satisfying
\[
 0\le w_i+2b_{i+1}-b_i\le2
 \quad (i\bmod a).
\]
For $w_i=0$ only the transition $1\longrightarrow0$ is
forbidden, whereas for $w_i=1$ only
$0\longrightarrow1$ is forbidden. The number of tuples is
therefore
\[
 \operatorname{tr}(A_{w_0}\cdots A_{w_{a-1}}),\qquad
 A_0=\begin{pmatrix}1&1\\0&1\end{pmatrix},\quad
 A_1=\begin{pmatrix}1&0\\1&1\end{pmatrix}.
\]
For either matrix $A_i$, the Gram matrix $A_i^{\mathsf T}A_i$
has trace $3$ and determinant $1$. Its largest eigenvalue is
$(3+\sqrt5)/2$, so $\|A_i\|_2=(1+\sqrt5)/2=\phi_{\mathrm{gr}}$.
The inequality $|\operatorname{tr} B|\le2\|B\|_2$ bounds
this count by $2\phi_{\mathrm{gr}}^a$. The same upper bound includes
the trivial-character summand; summing the $e$ selected
characters proves the assertion. Twisting a module by a group
automorphism preserves its composition length.
\end{proof}

Set
\[
 L(p,a,d)=
 \begin{cases}
  2\phi_{\mathrm{gr}}^a,&(p,d)=(2,2),\\
  d^a,&\text{otherwise}.
 \end{cases}
\]
We count all scalar-cover actions and their automorphism twists
at once. In particular, graph twists between different coordinate
orbits need not be removed.

\subsection{Counting actual groups normalized by a full cycle}
\label{pf:actualtypes-section}

The full cycle correlates the representation twists on the different
orbits of one simple factor. Recording a single outer automorphism
therefore gives a polynomial count. This argument does not require
synchronizing the local representations or choosing roots in their
permutation normalizers.

\begin{lemma}
\label{audit:generic-cycle-types}
Let $T$ be a nonabelian finite simple group. Let $\mathcal R$ be
representatives for a collection of transitive faithful $T$-sets,
where two sets are identified if one is permutation-equivalent to an
automorphism twist of the other. Write $h_\rho$ for the degree of
$\rho\in\mathcal R$.
Consider subgroups $D\cong T^t\le\Sym(v)$ normalized by a specified
full $v$-cycle $\sigma$, with each $D$-orbit acted on nontrivially
by exactly one factor, through an automorphism twist of some member
of $\mathcal R$. Then the number of permutation-conjugacy types of
such $D$ is at most
\[
 |\operatorname{Out}(T)|\sum_{\substack{\rho\in\mathcal R\\h_\rho\mid v}}
                 \tau(v/h_\rho).
\]
The number of actual such subgroups normalized by the specified
cycle is at most $v$ times this bound.
\end{lemma}
\begin{proof}
The cycle $\sigma$ permutes the $D$-orbits because it normalizes
$D$, and its transitivity on the coordinates makes this permutation
transitive. Every factor acts somewhere because the action of $D$ is
faithful. Sending each orbit to its unique active factor is therefore
an equivariant surjection,
so $\sigma$ is also transitive on the factors. Write their number as
$t$, their common local orbit degree as $h$, and the number of local
orbits per factor as $j$. Thus $v=htj$.

Choose one factor $T_0$ and identify it with $T$. The permutation
$\delta=\sigma^t$ cycles its $j$ local orbits and normalizes $T_0$.
Let $\theta\in\Aut(T)$ be the induced automorphism and choose the
identification of $T_0$ so that its action on the first orbit is
represented by some $\rho\in\mathcal R$. Transporting the other
orbits to the first one by powers of $\delta$ gives the sequence
of representations
\[
 \rho,\quad \rho\circ\theta^{-1},\quad\ldots,\quad
 \rho\circ\theta^{-(j-1)}.
\]
Only the class of $\theta$ in $\operatorname{Out}(T)$ is required to determine
the permutation-equivalence class of each term. Indeed, replacing
$\theta$ by another representative of that outer class changes every
power by an inner automorphism, and an inner twist of a permutation
representation is conjugate to the original representation by a
permutation in its image. These conjugacies may be performed
independently on disjoint coordinate orbits.
For example, if $\theta'=\operatorname{Inn}(x)\theta$, then for
$r\ge1$,
\[
 (\theta')^r
 =\operatorname{Inn}\bigl(x\theta(x)\cdots\theta^{r-1}(x)\bigr)
      \theta^r.
\]
The analogous statement for negative powers follows by inversion.
Thus the outer class really suffices for every member of the
displayed representation sequence.

The other factors are transported from $T_0$ by powers of $\sigma$;
reparametrizing each factor separately gives the same multiset of
local representations on its support. Therefore the subgroup $D$,
as an unmarked permutation group, is determined up to conjugacy by
$\rho$, the outer class of $\theta$, and $t\mid v/h_\rho$.
The number $j=v/(h_\rho t)$ is then determined. We are counting
the unmarked subgroup $D$ up to permutation conjugacy; distinct
return data can change the marked pair $(D,\sigma)$ without
creating a new conjugacy type of $D$.
The return condition on $\theta^j$ and the existence of the full
cycle can only remove choices from this description. This proves
the first bound.

Lemma~\ref{odd:normalized-conjugates} gives at most $v$ actual
normalized subgroups of each permutation-conjugacy type, proving
the last assertion.
\end{proof}

\begin{corollary}\label{audit:scalar-cycle-types}
Let $T=\PSL_d(s)$ be simple, with natural projective degree $k$.
Consider full-cycle-normalized direct powers whose coordinate orbits
are scalar covers of the natural point or hyperplane action, with
exactly one active factor on each orbit. Every orbit degree is a
multiple of $k$, so existence requires $k\mid v$. For $k\mid v$
the number of actual normalized subgroups on $v$ coordinates is at most
\[
 v|\operatorname{Out}(T)|
 \sum_{\substack{e\mid v/k\\e\mid s-1}}
                 \tau\bigl(v/(ke)\bigr)
 \le v|\operatorname{Out}(T)|\tau(v/k)^2.
\]
This conclusion does not require the actual coprime-core structure,
any parity condition, Singer centralizers, or the elimination of graph
twists between coordinate orbits.
\end{corollary}
\begin{proof}
For each permissible scalar phase $e$, its subgroup in $\F_s^*$
is unique. The cover exists as a $T$-set only when this scalar
subgroup contains the center of $\mathrm{SL}_d(s)$; retaining other
indices merely overcounts. Point and hyperplane covers are related
by the inverse-transpose automorphism, so at most one representative
in $\mathcal R$ is required for each $e$, of degree $ke$.
Apply Lemma~\ref{audit:generic-cycle-types} and
$\sum_{e\mid M}\tau(M/e)\le\tau(M)^2$.
\end{proof}

\begin{corollary}\label{fixedbase:lengthinterface}
Let $T=\PSL_d(p^a)$ be nonabelian simple and put
$k=(p^{ad}-1)/(p^a-1)$. For each permissible phase $e_0$, let
$\Omega_{e_0}=(\F_{p^a}^d\setminus\{0\})/H_{e_0}$, where
$H_{e_0}$ is the unique scalar subgroup of index $e_0$ containing
the scalar center of $\mathrm{SL}_d(p^a)$.
Suppose that the composition length over an algebraic closure $K$
of characteristic $p$ of each such cover satisfies
\[
 \operatorname{length}(K^{\Omega_{e_0}})\le e_0 L(a,d).
\]
Then, at any fixed quotient length $k\mid v\mid p^m-1$, the
probability that a cyclic code admits a full-cycle-normalized Scott
group of this geometric type with the scalar-cover actions and
condition on the active factors in Corollary~\ref{audit:scalar-cycle-types}
is at most
\[
 v|\operatorname{Out}(T)|\tau(v/k)^2\,
 2^{\,vL(a,d)/k-m_p(v)}
 \le v|\operatorname{Out}(T)|\tau(v/k)^2\,
 2^{\,vL(a,d)/k-v/m},
\]
uniformly over all coefficient alphabets of characteristic $p$.
\end{corollary}
\begin{proof}
On an orbit $O$ of phase order $e_0$ the single acting factor gives
composition length at most $e_0L(a,d)=|O|L(a,d)/k$. The other
factors act trivially. Summing over $O$ gives total length at most
$vL(a,d)/k$. The total number of codes
over $\F_{p^u}$ is $2^{m_{p^u}(v)}$, and
$m_{p^u}(v)\ge m_p(v)$ because taking a power of Frobenius only
splits its orbits. Apply
Lemma~\ref{odd:cycle-composition} and
Corollary~\ref{audit:scalar-cycle-types}. Since $v\mid p^m-1$,
every multiplication-by-$p$ orbit on $\Z_v$ has size at most $m$,
so $m_p(v)\ge v/m$.
\end{proof}

\subsection{Small defining-characteristic parameters}
\label{sec:allm}\label{allm:small-defining-section}

Fix a prime $p$, let $p\nmid n$, and write
\[
 m=\ord_n(p),\qquad R=\rad(n).
\]
Estimates involving $\log m$ are used only for sufficiently large
$m$, in particular $m\ge2$. For fixed $p$, the case $m=1$ allows
only divisors of $p-1$ and cannot affect any limit with $n\to\infty$.
When a full-order prime is needed, it is specified as a prime
$z\mid n$ with $\ord_z(p)=m$. Every canonical stripped quotient
$v=n/c$, $c\mid\Delta(n)$, satisfies $R\mid v\mid n\mid p^m-1$.

\begin{lemma}\label{am:supportedpart}
Fix a finite set $S$ of primes not containing $p$. There is a
constant $C_{p,S}$ such that for every $m\ge1$,
\[
 \prod_{\ell\in S}\ell^{v_\ell(p^m-1)}
 \le C_{p,S}\prod_{\ell\in S}\ell^{v_\ell(m)}
 \le C_{p,S}m.
\]
\end{lemma}
\begin{proof}
For odd $\ell\in S$, put $h_\ell=\ord_\ell(p)$ and
$c_\ell=v_\ell(p^{h_\ell}-1)$. If $h_\ell\nmid m$, the
valuation of $p^m-1$ is zero; otherwise lifting the exponent gives
\[
 v_\ell(p^m-1)=c_\ell+v_\ell(m),
\]
since $\ell\nmid h_\ell$. If $2\in S$, then $p$ is odd and
one may take $c_2=v_2(p^2-1)-1$. For odd $m$ the valuation is
$v_2(p-1)\le c_2$; for even $m$ it is $c_2+v_2(m)$.
Multiplication gives the assertion with
$C_{p,S}=\prod_{\ell\in S}\ell^{c_\ell}$.
\end{proof}

Write $\gamma=\log(4/3)$. The scalar-cover count records all
normalized groups without requiring synchronization of the local
representations. In the eventual structural application the finite
set $\mathcal T_0$ is removed first; Lemma~\ref{consolidated:scalar-actions}
then identifies every remaining actual defining-characteristic cover.

\begin{lemma}\label{allm:fixed-embedding-amplification}
Assume $p\nmid v$, and let $D\le\Sym(v)$ be a specified subgroup isomorphic to a
nontrivial direct power of a nonabelian simple group $T$,
normalized by the specified full $v$-cycle. Let $A$ be the part
of $v$ supported on the prime divisors of $|\Aut(T)|$, and put
$B=v/A$. For every $u\ge1$,
\[
 \Pr_{p^u}(D\le\PAut(C))
 \le(3/4)^{R_{p^u,A}(B)}
 \le(3/4)^{R_{p,A}(B)}.
\]
Here $C$ is uniform in $\mathcal C_{p^u}(v)$.
If $A=1$, the event is impossible.
\end{lemma}
\begin{proof}
The deterministic construction in the proof of
Theorem~\ref{thm:fixedT}, applied to this specified embedding,
puts $K=C_D(\sigma_v^A)$ inside $\Sym(A)\times1$ in the
fixed CRT coordinates $\Z_v=\Z_A\times\Z_B$.
It also shows that $K$ is a nontrivial direct power of $T$.
Consequently choose once and for all an element
$\pi_D\times1\in K$ for which $\pi_D$ is not a translation.
The choice depends on $D$, not on the random code.
Invariance under $D$ implies invariance under this permutation.
Lemma~\ref{lem:amplification} gives the first inequality and
Lemma~\ref{lem:uniformR} gives the second. There is no union
over permutations of $A$ for a fixed embedding $D$, and hence
no factorial prefactor. The same deterministic construction
rules out $A=1$.
\end{proof}

\begin{lemma}\label{allm:small-A}
Let $T=\PSL_d(p^a)$ be simple, let $b=ad$, and let $A$ be
the part of a divisor $v\mid p^m-1$ supported on the primes
dividing $|\Aut(T)|$. Then
\[
 A\le m\prod_{j=1}^{b}(p^j-1)
 <m p^{b(b+1)/2}.
\]
If $b<m$ and $v$ contains a prime $z$ with $\ord_z(p)=m$,
then $z$ divides $B=v/A$. In particular,
\[
 R_{p,A}(B)\ge\frac{B}{2m}.
\]
\end{lemma}
\begin{proof}
Every prime $\ell\ne p$ dividing $|\Aut(T)|$ has
$\ord_\ell(p)\le b$. Indeed, the primes in the order of the
linear group divide some $p^{ai}-1$ with $1\le i\le d$.
The additional primes in the field automorphism group divide
$a$, and therefore have order at most $\ell-1<a\le b$.
For $\ell=2\ne p$, the order is $\ord_2(p)=1$.
Indeed, writing $s=p^a$, the diagonal, field, and graph factors give
\[
 |\operatorname{Out}(T)|=\gcd(d,s-1)\,a\,g_d,\qquad
 g_d=\begin{cases}1,&d=2,\\ 2,&d\ge3.\end{cases}
\]
In particular $|\operatorname{Out}(T)|\le2ad$ and the diagonal
factor contributes only primes dividing $s-1$.
See \cite[\S2.1, pp.~40--41]{BurnessGuest}.

For each odd such prime put
\[
 d_\ell=\ord_\ell(p),\qquad
 c_\ell=v_\ell(p^{d_\ell}-1),
\]
and, when $p$ is odd and two occurs, put
$c_2=v_2(p^2-1)-1$. The order-lifting calculation in
Lemma~\ref{am:supportedpart} gives
\[
 A\le m C_{p,T},\qquad
 C_{p,T}=\prod_{\ell\mid|\Aut(T)|,\ \ell\ne p}
                  \ell^{c_\ell}.
\]
Each odd-prime factor in $C_{p,T}$ divides the factor
$p^{d_\ell}-1$ in $\prod_{j=1}^{b}(p^j-1)$, because
$d_\ell\le b$. The two-part divides its $j=2$ factor;
here $b\ge2$. Since these are distinct prime-power factors,
$C_{p,T}$ divides that product. This proves the bound on $A$.

For the last assertion, $z\mid v$ and $\ord_z(p)=m>b$.
The preceding order bound shows that $z$ is not among the
primes supporting $A$, and hence $z\mid B$.
The $B$-labels whose additive orders are divisible by $z$
number $B(1-z^{-v_z(B)})\ge B/2$; each has multiplication-by-$p$
orbit length exactly $m$. Since $\ord_A(p)\mid m$, all these
orbits are counted by $R_{p,A}(B)$.
\end{proof}

\begin{theorem}\label{allm:small-defining}
Fix a real constant $C_1>0$, put $H=\lceil C_1\log m\rceil$ and
$U=m p^{H(H+1)/2}$, and suppose that a prime $z\mid n$ has
$\ord_z(p)=m$. For all sufficiently large $m$, uniformly over
$u\ge1$, the probability that a stripped quotient $D_v(C)$
admits a full-cycle-normalized group of simple type
$\PSL_d(p^a)$ with $2\le ad\le C_1\log m$, acting through scalar
covers with exactly one active factor on each coordinate orbit,
is at most
\[
 K H^2m^5U^4\exp\left(-\frac{\gamma z}{4m}\right).
\]
Here $C$ is a uniform cyclic code of length $n$, and $K$ is an
absolute constant. All scalar phases, multiplicities, and actual
normalized embeddings are included.
\end{theorem}
\begin{proof}
Put $H=\lceil C_1\log m\rceil$ and
\[
 A_* = m p^{H(H+1)/2}.
\]
For large $m$ one has $H<m$. Fix one eligible geometric pair
$(a,d)$, and put $k=(p^{ad}-1)/(p^a-1)$.
Put $T=\PSL_d(p^a)$. Consider a quotient length $v$ with $R\mid v\mid n$ and $k\mid v$.
Since $z\mid R$, this quotient retains $z$ and
$\ord_z(p)\mid\ord_v(p)\mid\ord_n(p)$, so $\ord_v(p)=m$. We have
$|\operatorname{Out}(T)|\le2ad\le2H<2m$.
Corollary~\ref{audit:scalar-cycle-types} gives at most
$v|\operatorname{Out}(T)|\tau(v/k)^2\le2mv^3$ groups
normalized by the specified cycle. For each specified group,
Lemmas~\ref{allm:fixed-embedding-amplification} and
\ref{allm:small-A} give the probability bound
$\exp(-\gamma B/(2m))$, where $v=AB$, $A\le A_*$, and
$B\ge z$. Each fixed quotient has the uniform code law before
conditioning. Thus the union over quotient lengths is bounded by
\[
 2m\sum_{\substack{1\le A\le A_*\\ B\ge z}}
        A^3B^3e^{-\gamma B/(2m)}
 \le2m A_*^4\sum_{B\ge z}B^3e^{-\gamma B/(2m)}.
\]
For fixed $T$, a quotient length $v$ determines its pair $(A,B)$,
and the pair determines $v=AB$ and $c=n/v$. Thus summing over all
the displayed positive integers counts every admissible quotient at
least once. The inequality uses $\sum_{1\le A\le A_*}A^3\le A_*^4$.

The elementary exponential-series estimate gives an absolute
constant $K$ (depending only on $\gamma$) such that
\[
 \sum_{B\ge z}B^3e^{-\gamma B/(2m)}
 \le e^{-\gamma z/(4m)}
          \sum_{B\ge1}B^3e^{-\gamma B/(4m)}
 \le K m^4 e^{-\gamma z/(4m)}.
\]
There are at most $H^2$ geometric pairs. Consequently the
entire event has probability at most
\[
 K H^2 m^5 A_*^4\exp\left(-\frac{\gamma z}{4m}\right).
\]
Since $A_*=U$, this is the asserted bound. All estimates are
uniform in $u$. Conditioning the original code on nondegeneracy
costs at most a factor of two.
\end{proof}

\subsection{Large defining-characteristic parameters}
\label{allm:large-section}

The complementary range is controlled by a uniform composition-density
estimate. The actual group count is polynomial in the quotient length,
so the estimates can be summed without a factor exponential in $m$.

\begin{lemma}
\label{allm:cover-density}
Put
\[
 \eta=\frac14\log\frac43,
 \qquad \phi_{\mathrm{gr}}=\frac{1+\sqrt5}{2}.
\]
For a prime $p$, integers $a\ge1$ and $d\ge2$, set
\[
 k=\frac{p^{ad}-1}{p^a-1},\qquad
 L(p,a,d)=
 \begin{cases}
  2\phi_{\mathrm{gr}}^a,&(p,d)=(2,2),\\
  d^a,&\text{otherwise}.
 \end{cases}
\]
Then, uniformly in all these parameters,
\begin{equation}\label{allm:density-bound}
 \frac{L(p,a,d)}{k}\le 2\exp(-\eta ad).
\end{equation}
In particular, if $T=\mathrm{PSL}_d(p^a)$ is simple and $D\cong T^t$
acts on $v$ coordinates so that each coordinate orbit is a scalar
cover from Lemma~\ref{pf:bs-length}, acted on by exactly one factor,
then over an algebraically closed field $K$ of characteristic $p$,
\[
 \ell_{KD}(K^v)\le \frac{v}{k}L(p,a,d)
                 \le 2v\exp(-\eta ad).
\]
The statement permits arbitrary multiplicities and automorphism twists
of these scalar-cover actions.
\end{lemma}
\begin{proof}
First $k\ge p^{a(d-1)}$. For $d\ge3$ we have
\[
 \frac{d}{2^{d-1}}\le\left(\frac34\right)^{d/4}.
\]
Indeed the inequality holds at $d=3$, and on increasing $d$ the ratio
of successive terms on the left is $(d+1)/(2d)\le2/3$, whereas the
corresponding ratio on the right is $(3/4)^{1/4}>2/3$. Consequently
\[
 \frac{d^a}{k}\le\left(\frac{d}{p^{d-1}}\right)^a
             \le\left(\frac34\right)^{ad/4}
             =\exp(-\eta ad).
\]
If $d=2$ and $p\ge3$, then
\[
 \frac{2^a}{k}\le(2/3)^a
       \le(3/4)^{a/2}=\exp(-\eta ad).
\]
Finally, if $p=d=2$, the inequality $\phi_{\mathrm{gr}}<\sqrt3$ gives
\[
 \frac{2\phi_{\mathrm{gr}}^a}{k}\le2(\phi_{\mathrm{gr}}/2)^a
       \le2(3/4)^{a/2}=2\exp(-\eta ad).
\]
This proves the uniform estimate.

A coordinate orbit of phase order $e$ has degree $ek$ and
composition length at most $eL(p,a,d)$ by
Lemma~\ref{pf:bs-length}. Inflating its module from its single
acting simple factor to $D$ does not change its submodule lattice.
Composition length is additive over the direct sum of the orbit
modules, so summing their degrees proves the assertion. Twisting
by an automorphism preserves composition length.
\end{proof}

\begin{proposition}
\label{allm:large-defining}
Use the general length $n$, $m=\ord_n(p)$, and $R=\rad(n)$
fixed at the start of this section, and put $C_0=4/\eta$.
Choose a cyclic code $C$ of length $n$ uniformly over
$\mathbb F_{p^u}$, where $u\ge1$ may vary with $m$.
For each $c\mid\Delta(n)$ write $v=n/c$, and consider the actual
Scott subgroups in the core-free or coprime-core constructions for
the stripped quotient $D_{n/c}(C)$.
Let $\mathcal E_m$ be the event that one such subgroup has simple
type $T=\mathrm{PSL}_d(p^a)$, with natural degree
$k=(p^{ad}-1)/(p^a-1)$ dividing $v$, actual scalar covers as in
Corollary~\ref{audit:scalar-cycle-types}, and
\[
 ad>C_0\log m.
\]
Then there is an absolute constant $K_0$ such that, for all
sufficiently large $m$,
\begin{equation}\label{allm:large-union}
 \sup_{u\ge1}\Pr(\mathcal E_m)
 \le2m^3\sum_{v\ge R}v^3\,2^{-v/(2m)}
 \le K_0m^7\exp\left(-\frac{(\log2)R}{4m}\right).
\end{equation}
After conditioning $C$ to be nondegenerate, at most twice the
displayed upper bound applies.
\end{proposition}
\begin{proof}
We first count possible parameters. Put $h=\ord_k(p)$. Since $k\mid p^{ad}-1$,
we have $h\mid ad$. If $h<ad$, then $h\le ad/2$. But $d\ge2$
gives
\[
 k\le p^h-1\le p^{ad/2}-1
 <1+p^{a(d-1)}\le k.
\]
Thus $h=ad$.
Since $k\mid p^m-1$, this proves $ad\mid m$. Hence there are
at most $m^2$ possible ordered pairs $(a,d)$, including all the
simple types relevant here.

Fix one such pair above the stated threshold and a length $v$.
Lemma~\ref{allm:cover-density} gives, for each actual subgroup $D$,
\[
 \frac{\ell_{KD}(K^v)}{v}
 \le2\exp(-\eta ad)<2m^{-4}\le\frac1{2m}
 \qquad(m\ge2).
\]
Lemma~\ref{odd:cycle-composition} therefore gives at most
$2^{v/(2m)}$ cyclic codes invariant under this particular $D$.
Every irreducible factor of $X^v-1$ over $\mathbb F_{p^u}$ has
degree at most $m$, because $v\mid p^m-1$ implies that the
$p^u$-power Frobenius on its roots has order dividing $m$.
As $p\nmid v$, there are at least $2^{v/m}$ cyclic codes of
length $v$. Thus the probability for a specified $D$ is at most
$2^{-v/(2m)}$, independently of $u$.

Corollary~\ref{audit:scalar-cycle-types} bounds the number of actual
subgroups of the fixed type normalized by the specified full
$v$-cycle by
\[
 v|\operatorname{Out}(T)|\tau(v/k)^2\le2mv^3,
 \qquad |\operatorname{Out}(T)|\le2ad\le2m.
\]
This is a count of actual groups, already including the union
over phase orders, multiplicities, Scott-factor counts and
permutation embeddings. Thus the probability for this pair at
length $v$ is at most $2mv^3 2^{-v/(2m)}$.

Each fixed stripped quotient is a uniform cyclic code of its
length. Furthermore $c\mid\Delta(n)$ implies $R\mid n/c$,
so its length is at least $R$. The union bound over all stripped
quotients and at most $m^2$ parameter pairs is therefore bounded
by the first sum in~\eqref{allm:large-union}. Enlarging the sum
from the relevant divisors of $n$ to all integers $v\ge R$
introduces no assumption of independence between quotients.

For completeness, put $\alpha=(\log2)/(2m)$ and
$\beta=\alpha/2$. Since $0<\beta\le1$,
\begin{align*}
 \sum_{v\ge R}v^3e^{-\alpha v}
 &\le e^{-\alpha R/2}\sum_{v\ge1}v^3e^{-\beta v}\\
 &\le e^{-\alpha R/2}\int_0^\infty(x+1)^3e^{-\beta x}\,dx\\
 &=e^{-\alpha R/2}
   (6\beta^{-4}+6\beta^{-3}+3\beta^{-2}+\beta^{-1})\\
 &\le16\beta^{-4}e^{-\alpha R/2}.
\end{align*}
For the integral comparison, integrate over $[v-1,v]$, on which
$(x+1)^3\ge v^3$ and $e^{-\beta x}\ge e^{-\beta v}$.
Consequently one may take $K_0=8192/(\log2)^4$.

All estimates are uniform in the coefficient extension degree $u$.

Finally the probability that $C$ is nondegenerate is at least
$1/2$: including any fixed irreducible constituent containing a
primitive $n$th root already ensures nondegeneracy. Therefore
conditioning multiplies any unconditional event bound by at most
two.
\end{proof}

\subsection{A common quantitative bound and primitive divisors}

The preceding estimates isolate a sufficient arithmetic hypothesis:
one sufficiently large prime divisor with full multiplicative order.
A prime of full order $m$ means a prime $z$ satisfying
$\ord_z(p)=m$, equivalently a primitive prime divisor of $p^m-1$:
it divides none of $p^j-1$ for $1\le j<m$. No primality assumption
on the exponent $m$ is intended. The remaining finite simple types
are controlled by counting their actual transitive coordinate actions.

Throughout this section, $p$ is fixed, $p\nmid n$, and
\[
 m=\ord_n(p),\qquad R=\rad(n),\qquad
 z\mid n\text{ is prime},\qquad \ord_z(p)=m.
\]
Every stripped quotient length $v=n/c$, $c\mid\Delta(n)$, satisfies
$R\mid v\mid n\mid p^m-1$, and in particular retains $z$.

\begin{lemma}\label{gl:fixed-actual-types}
Fix a finite set $\mathcal T$ of nonabelian simple groups.
For all sufficiently large $m$, the probability that some stripped
quotient has a subgroup $D\cong T^t$ of its permutation automorphism
group normalized by the specified quotient cycle, with $t\ge1$,
$T\in\mathcal T$, and exactly one active simple
factor on each coordinate orbit, is at most
\[
 K_{p,\mathcal T}m^6
       \exp\left(-\frac{\beta z}{4m}\right),
 \qquad \beta=\log(4/3),
\]
uniformly over all alphabets of characteristic $p$.
\end{lemma}
\begin{proof}
Fix $T$. There are finitely many faithful transitive $T$-sets up to
permutation equivalence and automorphism twist: they are coset
actions on proper subgroups of the finite group $T$. Taking all of
them in Lemma~\ref{audit:generic-cycle-types}, the number of actual
groups normalized by a specified $v$-cycle is at most
\[
 C_Tv\tau(v)\le C_Tv^2
\]
for a constant $C_T$ depending only on $T$. This includes every
permitted multiplicity, local degree, and twist in the statement.

Let $A$ be the part of $v$ supported on primes dividing $|\Aut(T)|$
and put $B=v/A$. Since $p\nmid v$, apply
Lemma~\ref{am:supportedpart} to the primes dividing $|\Aut(T)|$
other than $p$. The divisibility $v\mid p^m-1$ gives
$A\le C_{p,T}m$.
For large $m$, the prime $z>m$ exceeds every prime divisor of
$|\Aut(T)|$, so $z\mid B$. At least $B/2$ character labels of
$B$ have additive order divisible by $z$, and all have $p$-orbit
length $m$. Since $\ord_A(p)\mid m$, this gives
\[
 R_{p,A}(B)\ge\frac{B}{2m}.
\]
Lemma~\ref{allm:fixed-embedding-amplification} bounds the probability
for each specified group by $e^{-\beta B/(2m)}$.

Every fixed quotient code is uniform before conditioning. We may
therefore sum over all positive integers $A\le C_{p,T}m$ and
$B\ge z$, overcounting the admissible quotient lengths:
\begin{align*}
 \Pr(\text{one such $T$-occurrence})
 &\le C_T\sum_{A\le C_{p,T}m}\sum_{B\ge z}
                     A^2B^2e^{-\beta B/(2m)}\\
 &\le K_{p,T}m^3e^{-\beta z/(4m)}
                     \sum_{B\ge1}B^2e^{-\beta B/(4m)}\\
 &\le K'_{p,T}m^6e^{-\beta z/(4m)}.
\end{align*}
The finite sum over $T\in\mathcal T$ proves the claim. No
independence of the different quotient maps is asserted or needed.
\end{proof}

We include the nondegenerate law in the same quantitative estimate.
Write
\[
 \epsilon_q^{\mathrm{nd}}(n)=
 \Pr\bigl(\PAut(C)\ne A_q(n)\mid C\text{ is nondegenerate}\bigr),
 \qquad C\in\mathcal C_q(n),
\]
and define
\begin{equation}\label{gl:common-error}
 \mathfrak E_p(n)=\sup_{u\ge1}
       \max\{\epsilon_{p^u}(n),\epsilon_{p^u}^{\mathrm{nd}}(n)\}.
\end{equation}
The conditioning event has probability at least $1/2$, so this
quantity is defined at every eligible length.

\begin{theorem}\label{gl:quantitative}
Fix a prime $p$ and put $c_*=(\log2)/32$.
There are constants $C_p>0$ and $m_0(p)\ge2$ such that the following
holds. Let $p\nmid n$, set $m=\ord_n(p)\ge m_0(p)$, and suppose
that a prime $z\mid n$ satisfies
\[
 \ord_z(p)=m,\qquad z>4m.
\]
Then
\begin{equation}\label{gl:quantitative-bound}
 \mathfrak E_p(n)
 \le \min\left\{1,
       \exp\left(C_p(\log m)^2-c_*\frac zm\right)\right\}.
\end{equation}
In particular, the estimate holds simultaneously for every coefficient
field of characteristic $p$ and for both the unconditional and
nondegenerate-conditioned code laws. No further growth assumption
on $z/m$ is required for this inequality.
\end{theorem}
\begin{proof}
Let $\mathcal T_0$ be the finite family in
Theorem~\ref{close:exhaustion}, in its defining-characteristic
refinement. The reduction retains the actual coordinate actions:
each coordinate orbit is acted on nontrivially by exactly one simple
factor. Thus Lemma~\ref{gl:fixed-actual-types} applies to the
occurrences from $\mathcal T_0$.

Outside the canonical translation and alternating events and these
finite-type occurrences, every nonaffine code supplies an actual
defining-characteristic group $T=\PSL_d(p^a)$ on a stripped quotient
of length $v$, with
\[
 k=\frac{p^{ad}-1}{p^a-1}\mid v\mid n.
\]
Lemma~\ref{consolidated:scalar-actions} identifies its actual
coordinate actions as scalar covers. As proved in
Proposition~\ref{allm:large-defining}, $\ord_k(p)=ad$, so $ad\mid m$.
Put
\[
 \beta=\gamma=\log(4/3),\qquad \eta=\beta/4,\qquad C_0=4/\eta,
 \qquad H=\lceil C_0\log m\rceil,\qquad
 U=m p^{H(H+1)/2}.
\]
For sufficiently large $m$, the estimates of
Theorems~\ref{pp:QA-all-exponents} and~\ref{allm:small-defining},
Lemma~\ref{gl:fixed-actual-types}, and
Proposition~\ref{allm:large-defining} apply simultaneously. Their
small and large parameter ranges cover all defining-characteristic
occurrences. Every probability in those union bounds is taken under
the unconditional uniform law of a fixed quotient code.

To handle affine multipliers, define
\[
 \Psi(t)=\sup_{\substack{M\in\mathbb N\\M\ge t}}
          \min\{1,(M-1)2^{-M/2}\}\qquad(t>0).
\]
Write $r=v_z(n)$. Since $(z,n/z^r)=1$,
\[
 \varphi(n)=\varphi(z^r)\varphi(n/z^r)\ge z-1,
 \qquad M_{p^u}(n)\ge M_p(n)=\frac{\varphi(n)}m
                              \ge\frac{z-1}{m}.
\]
The estimates in Section~\ref{sec:affine} therefore bound the affine
exceptional event by $\Psi((z-1)/m)$ and degeneracy by
$2^{-(z-1)/m}$. Conditioning any event on nondegeneracy changes its
probability by at most the degeneracy probability. Consequently,
with $R=\rad(n)\ge z$, the following is a common bound for the two
laws in~\eqref{gl:common-error}:
\begin{align}
 \mathfrak E_p(n)
 \le{}&K m^3e^{-(\log2)R/(16m)}
       +K_pm^6e^{-\beta z/(4m)}\notag\\
 &+K H^2m^5U^4e^{-\beta z/(4m)}
       +K_0m^7e^{-(\log2)R/(4m)}\notag\\
 &+\Psi((z-1)/m)+2^{-(z-1)/m}.
 \label{gl:bound}
\end{align}
The last term accounts for conditioning the original code. No
conditional uniformity of its quotient codes is used.

It remains to simplify the prefactors and the affine envelope.
For every $M\ge1$,
\[
 (M-1)2^{-M/2}
 \le M e^{-(\log2)M/2}
 \le \frac{4}{e\log2}\,e^{-(\log2)M/4},
\]
because the maximum of $x e^{-(\log2)x/4}$ on $x\ge0$ is
$4/(e\log2)$. Hence
\[
 \Psi(t)\le\frac{4}{e\log2}\,e^{-(\log2)t/4}.
\]
For $m\ge2$, this gives
\begin{align*}
 \Psi((z-1)/m)
   &\le\frac{8}{e\log2}\,e^{-(\log2)z/(4m)},\\
 2^{-(z-1)/m}&\le2e^{-(\log2)z/m}.
\end{align*}
Set $\lambda=(\log2)/16=2c_*$. All four exponential rates
in~\eqref{gl:bound}, and the two rates just obtained, are at least
$\lambda$; in particular $\beta/4>\lambda$. Replacing $R$ by $z$
therefore yields
\[
 \mathfrak E_p(n)\le B_p(m)e^{-\lambda z/m},
\]
where one may take
\[
 B_p(m)=K m^3+K_pm^6+K H^2m^5U^4+K_0m^7
                       +\frac{8}{e\log2}+2.
\]
Now $H=O(\log m)$ and
\[
 \log U=\log m+\frac{H(H+1)}2\log p=O_p((\log m)^2).
\]
It follows that, after increasing $m_0(p)$ and $C_p$ if necessary,
$B_p(m)\le\exp(C_p(\log m)^2)$. Thus
\[
 \mathfrak E_p(n)
 \le\exp\left(C_p(\log m)^2-2c_*\frac zm\right)
 \le\exp\left(C_p(\log m)^2-c_*\frac zm\right).
\]
Combining this inequality with $\mathfrak E_p(n)\le1$ proves
\eqref{gl:quantitative-bound}.
\end{proof}

The arithmetic criterion stated in the introduction is an immediate
consequence of this quantitative estimate.

\begin{proof}[Proof of Theorem~\ref{gl:criterion}]
For fixed $p$, the relation $n\mid p^m-1$ implies $m\to\infty$
whenever $n\to\infty$. The assumed condition
$z/[m(\log m)^2]\to\infty$ implies $z>4m$ eventually and
\[
 C_p(\log m)^2-c_*z/m\longrightarrow-\infty.
\]
Theorem~\ref{gl:quantitative} therefore gives
$\mathfrak E_p(n)\to0$, which proves both asserted limits.
\end{proof}

For the primitive lengths, Stewart's estimate gives a rate stronger
than every fixed inverse power of the exponent. This also holds
uniformly over all divisor lengths retaining the selected prime.

\begin{corollary}\label{gl:superpolynomial}
Fix a prime $p$. For every sufficiently large integer $m$, let
$z_m=P(\Phi_m(p))$. For every fixed real $A>0$,
\begin{equation}\label{gl:superpolynomial-bound}
 \sup_{\substack{n\mid p^m-1\\z_m\mid n}}
       \mathfrak E_p(n)=o(m^{-A})\qquad(m\longrightarrow\infty).
\end{equation}
The rate is in the exponent $m$. The supremum covers the individual
code distributions at all indicated lengths and all coefficient
fields of characteristic $p$; it is not the probability of a union
of experiments at different lengths.
\end{corollary}
\begin{proof}
Lemma~\ref{pp:retainedprime} gives $\ord_{z_m}(p)=m$ and
\[
 \frac{z_m}{m}>
 E(m):=\exp\left(\frac{\log m}{104\log\log m}\right)>4
\]
for sufficiently large $m$. Whenever $z_m\mid n\mid p^m-1$,
\[
 m=\ord_{z_m}(p)\mid\ord_n(p)\mid m,
\]
so $\ord_n(p)=m$. Theorem~\ref{gl:quantitative} then gives a bound
independent of the selected divisor length:
\[
 \sup_{\substack{n\mid p^m-1\\z_m\mid n}}\mathfrak E_p(n)
 \le\exp\bigl(C_p(\log m)^2-c_*E(m)\bigr).
\]
Finally,
\[
 \log\frac{E(m)}{(\log m)^2}
 =\frac{\log m}{104\log\log m}-2\log\log m
 \longrightarrow\infty.
\]
For any fixed $A>0$, it follows that
\[
 C_p(\log m)^2+A\log m-c_*E(m)\longrightarrow-\infty.
\]
Multiplying the preceding common bound by $m^A$ proves
\eqref{gl:superpolynomial-bound}.
\end{proof}

\begin{proof}[Proof of Corollary~\ref{gl:primitive-divisors}]
Corollary~\ref{gl:superpolynomial} implies the claimed uniform
convergence over all divisors retaining $z_m$, for both code laws.
Every length satisfying
$\rad(p^m-1)\mid n\mid p^m-1$ retains $z_m$ and is therefore
included. For a fixed prime power $q=p^f$, apply that corollary
with exponent $fm$ and choose the coefficient field $p^f$.
Then $p^{fm}-1=q^m-1$, so the entire primitive-length sequence over
$\F_q$ is covered. In fact the exceptional probability on that
sequence is $o(m^{-A})$ for every fixed $A>0$, since $f$ is fixed.
\end{proof}

\subsection{Conditioning on a rate window}

The uniform constituent model concentrates near rate $1/2$ on the
lengths considered above. Combining this concentration with the
pointwise error estimate gives an automorphism theorem for codes
restricted to a rate window. We state the bound for both choices of
whether full shift period is also required.

\begin{proposition}\label{prop:rate-conditioning}
Fix a prime $p$ and a real number $0<\eta<1/2$. For $p\nmid n$,
put $m=\ord_n(p)$ and
\[
 b_{p,\eta}(n)=\frac{m}{4\eta^2n}+2^{-M_p(n)},
 \qquad M_p(n)=\frac{\varphi(n)}m.
\]
Let $\mathcal W_{q,n}(\eta)$ be the rate window defined in
Section~\ref{sec:preliminaries}, and let $\mathcal N$ denote full
shift period. Whenever $b_{p,\eta}(n)<1$, both conditional laws
below are defined, and
\begin{align}
 &\sup_{u\ge1}\max\Bigl\{
 \Pr\bigl(\PAut(C)\ne A_{p^u}(n)\mid
                   \mathcal W_{p^u,n}(\eta)\bigr),\notag\\[-1mm]
 &\hspace{38mm}
 \Pr\bigl(\PAut(C)\ne A_{p^u}(n)\mid
                   \mathcal W_{p^u,n}(\eta)\cap\mathcal N\bigr)
                         \Bigr\}\notag\\
 &\hspace{15mm}\le
 \min\left\{1,\frac{\mathfrak E_p(n)}{1-b_{p,\eta}(n)}\right\},
 \qquad C\in\mathcal C_{p^u}(n).
 \label{eq:rate-conditioned-error}
\end{align}
Under the hypotheses of Theorem~\ref{gl:criterion}, the left side
tends to zero. For $z_m=P(\Phi_m(p))$ and every fixed $A>0$, its
supremum over $z_m\mid n\mid p^m-1$ is $o(m^{-A})$.
\end{proposition}
\begin{proof}
By Proposition~\ref{prop:rate-window} and
Lemma~\ref{lem:affine-nondegenerate}, for $q=p^u$,
\[
 \Pr(\mathcal W_{q,n}(\eta)^c)\le\frac{m}{4\eta^2n},
 \qquad
 \Pr(\mathcal N^c)\le2^{-M_q(n)}\le2^{-M_p(n)}.
\]
The union bound therefore gives
\[
 \Pr(\mathcal W_{q,n}(\eta))\ge1-b_{p,\eta}(n),
 \qquad
 \Pr(\mathcal W_{q,n}(\eta)\cap\mathcal N)
                              \ge1-b_{p,\eta}(n).
\]
For an event $E$ and either of these conditioning events $F$,
\[
 \Pr(E\mid F)=\frac{\Pr(E\cap F)}{\Pr(F)}
       \le\frac{\Pr(E)}{1-b_{p,\eta}(n)}.
\]
Apply this inequality to $E=\{\PAut(C)\ne A_q(n)\}$ and use
$\Pr(E)=\epsilon_q(n)\le\mathfrak E_p(n)$ to obtain
\eqref{eq:rate-conditioned-error}.

If a prime $z\mid n$ has order $m$ as in
Theorem~\ref{gl:criterion}, then $n\ge z$ and
$\varphi(n)\ge z-1$. Consequently
\[
 b_{p,\eta}(n)
 \le\frac{m}{4\eta^2z}+2^{-(z-1)/m}\longrightarrow0.
\]
The theorem gives $\mathfrak E_p(n)\to0$, proving convergence
under both conditional laws. For the divisor family in
Corollary~\ref{gl:superpolynomial}, the same bound with $z=z_m$
tends to zero uniformly in $n$ and $u$. Multiplying that
corollary's $o(m^{-A})$ bound by a factor tending uniformly to
one proves the asserted rate.
\end{proof}

The width $\eta$ is fixed in this statement. More generally, both
conditioned exceptional probabilities tend to zero uniformly in
$u$ for a shrinking width $\eta_n$ whenever
$m/(\eta_n^2n)\to0$ and
$\sup_u\epsilon_{p^u}(n)\to0$.
Indeed, the window has probability tending to one, and its
intersection with $\mathcal N$ has probability at least
$1/2-o(1)$ by \eqref{eq:degeneracy-probability}.
An unconditional $o(m^{-A})$ error is likewise preserved when
these window bounds hold uniformly over the lengths in question.
These statements concern a range of dimensions around $n/2$;
they do not imply concentration for every exact dimension.

\section{Exact nonconcentration and the scope of uniformity}
\label{tree:section}
\subsection{Periodic codes and their coordinate partitions}

The standard automorphism formula for a periodic cyclic code is
\begin{equation}\label{tree:periodic-law}
 \PAut(C)=\Sym(n/d)\wr\PAut(D),
\end{equation}
where $d$ is the least common shift period and $D$ is the resulting
length-$d$ code. It is stated for general nonzero cyclic codes in
\cite[Lemma~2.4]{FHLXClassification}. Its proof attributes the periodic
decomposition and group formula to Hollmann
\cite[Lemma~5.22 and Corollary~5.24]{Hollmann}. The symmetric base group permutes
the coordinates in each residue class modulo $d$. These classes are
canonical: the coordinate functionals at $x$ and $y$ agree on a cyclic
code exactly when $x-y$ is a common period. Thus
\eqref{tree:periodic-law} is an equality in the actual coordinate
action, not only an abstract group isomorphism.
For completeness, let $\lambda_x:C\to\F_q$ be the coordinate
functional $c\mapsto c_x$. Cyclicity gives
\[
 \lambda_x=\lambda_y
 \quad\Longleftrightarrow\quad
 c_r=c_{r+y-x}\ \text{for every }c\in C\text{ and }r\in\Z_n
 \quad\Longleftrightarrow\quad d\mid(y-x).
\]
Indeed, equality at $x,y$ may be applied to every shift of each
codeword, and the common shift periods form the subgroup
$d\Z_n$. Every automorphism therefore permutes these canonical
classes and induces an element of $\PAut(D)$. Its kernel consists
of all independent permutations inside the classes. Conversely,
every such internal permutation, and every lift of a permutation
preserving $D$, preserves $C$. This proves
\eqref{tree:periodic-law} directly. For the zero code, $d=1$ and
both sides are $\Sym(n)$, so that boundary case is covered as well.

The calculation below is a corollary of this formula, duality, and
the elementary cyclotomic decomposition. In the binary case the
same wreath groups also follow from the explicit constructions of
Bienert--Klopsch \cite[Theorem 5.1]{BienertKlopsch}. We use the exact
distribution to delimit the probability statements of this paper;
no separate novelty claim is made for the group classification.

Fix an odd prime $\ell$ and a prime power $q$ such that
\begin{equation}\label{tree:primitive-root}
 \ord_{\ell^2}(q)=\ell(\ell-1).
\end{equation}
In particular, the characteristic $p$ of $\F_q$ differs from $\ell$.
For $j\ge1$ put $n=\ell^j$, label coordinates by $\Z_n$, and let
$\mathcal P_i$ be the partition into residue classes modulo $\ell^i$,
where $0\le i\le j$. Thus $\mathcal P_0$ has one part and
$\mathcal P_j$ consists of singletons. If
$I\subseteq\{1,\ldots,j-1\}$, write
\[
 G_j(I)=\{\pi\in\Sym(n):\pi\text{ preserves }\mathcal P_i
                                      \text{ for every }i\in I\}.
\]
Here $G_j(\varnothing)=\Sym(n)$.

\begin{lemma}\label{tree:projectors}
For $0\le i\le j$, define a matrix over $\F_q$ by
\[
 (Q_i)_{x,y}=\ell^{i-j}\mathbf1_{x\equiv y\pmod{\ell^i}},
 \qquad x,y\in\Z_n,
\]
where powers of $\ell$ are interpreted in $\F_q$. Put
$E_0=Q_0$ and $E_i=Q_i-Q_{i-1}$ for $i\ge1$.
The spaces $\operatorname{im}(E_i)$, $0\le i\le j$, are precisely
the simple constituents of the cyclic permutation module $\F_q^n$.
Every cyclic code has a unique form
\begin{equation}\label{tree:code-bits}
 C_b=\bigoplus_{i=0}^j b_i\operatorname{im}(E_i),
 \qquad b_i\in\{0,1\}.
\end{equation}
Its dimension is
\[
 \dim C_b=b_0+\sum_{i=1}^j b_i(\ell^i-\ell^{i-1}),
\]
and duality complements the constituent bits:
\[
 C_b^\perp=C_{1-b},\qquad \PAut(C_b)=\PAut(C_{1-b}).
\]
\end{lemma}
\begin{proof}
Condition~\eqref{tree:primitive-root} gives
$\ord_\ell(q)=\ell-1$ and $v_\ell(q^{\ell-1}-1)=1$.
Lifting the exponent therefore gives
\[
 \ord_{\ell^i}(q)=\ell^{i-1}(\ell-1)\qquad(i\ge1).
\]
There is exactly one $q$-cyclotomic coset of additive character
labels of each order $\ell^i$, including order one. Since
$p\nmid n$, these $j+1$ cosets give all simple cyclic constituents.

The matrix $Q_i$ averages over each orbit of
$\langle\sigma_n^{\ell^i}\rangle$; it is the projector onto the
subspace of vectors with period dividing $\ell^i$. Equivalently,
its Fourier support is the set of characters of order dividing
$\ell^i$. Thus $E_i$ selects precisely the characters of order
$\ell^i$, proving the simple-constituent assertion and the dimension
formula. Directly,
\[
 Q_i^T=Q_i,\qquad Q_iQ_h=Q_{\min(i,h)},\qquad Q_j=I_n.
\]
The $E_i$ are therefore symmetric pairwise orthogonal idempotents
whose sum is the identity. Their images are mutually orthogonal nondegenerate subspaces, so
taking the orthogonal complement replaces each $b_i$ by $1-b_i$.
A coordinate permutation preserves a linear code if and only if
it preserves its orthogonal complement, proving the last equality.
\end{proof}

\begin{corollary}\label{tree:exact-groups}
For the code $C_b$ in~\eqref{tree:code-bits}, put
\[
 I(b)=\{i:1\le i<j,\ b_i\ne b_{i+1}\}.
\]
Then, as a permutation group on the fixed coordinates,
\begin{equation}\label{tree:group-classification}
 \PAut(C_b)=G_j(I(b)).
\end{equation}
If $I(b)=\{t_1<\cdots<t_r\}$, set $t_0=0$,
$t_{r+1}=j$, and $a_h=\ell^{t_h-t_{h-1}}$ for
$1\le h\le r+1$. The group is the automorphism group of the
rooted regular tree with successive branching numbers
$a_1,\ldots,a_{r+1}$, and
\begin{equation}\label{tree:order}
 |\PAut(C_b)|=\prod_{h=1}^{r+1}(a_h!)^{\ell^{t_{h-1}}}.
\end{equation}
Equivalently it is the iterated imprimitive wreath product
$\Sym(a_{r+1})\wr\cdots\wr\Sym(a_1)$, with the factors
ordered from the leaves toward the root.
The groups $G_j(I)$ for distinct subsets $I$ are distinct as
subgroups of $\Sym(n)$ in the fixed coordinate labels.
\end{corollary}
\begin{proof}
We use induction on $j$, with the four elementary codes giving the
case $j=1$. If $I(b)=\varnothing$, the bits $b_1,\ldots,b_j$ are
constant. The resulting codes are zero, full, repetition, and the
dual of repetition, all with group $\Sym(\ell^j)$.

Suppose $I(b)\ne\varnothing$ and put $t=\max I(b)$. Duality
preserves both the group and the switch set. We may therefore
complement all bits so that
\[
 b_t=1,\qquad b_{t+1}=\cdots=b_j=0.
\]
The code is now the periodic repetition of a length-$\ell^t$
cyclic code $D$ with bits $b_0,\ldots,b_t$. Its least common
shift period is $\ell^t$, because its order-$\ell^t$ constituent
is selected. By \eqref{tree:periodic-law} and induction,
\[
 \PAut(C_b)
 =\Sym(\ell^{j-t})\wr\PAut(D)
 =\Sym(\ell^{j-t})\wr
      G_t\bigl(I(b)\cap\{1,\ldots,t-1\}\bigr).
\]
In the canonical residue coordinates this is precisely $G_j(I(b))$:
the base action is inside each residue class modulo $\ell^t$,
and the top action preserves exactly the selected shorter levels.
This proves \eqref{tree:group-classification}.

The stabilizer of this nested partition chain allows arbitrary
permutations of the children of each tree node, independently
at different nodes. At level $h-1$ there are $\ell^{t_{h-1}}$
nodes and each has $a_h$ children. This proves the stated wreath
product and its order.

Finally, the only canonical partitions $\mathcal P_i$ preserved
by $G_j(I)$ are those with $i\in I\cup\{0,j\}$.
Indeed, if $i$ is omitted, let $a<b$ be consecutive indices in
$\{0\}\cup I\cup\{j\}$ with $a<i<b$.
Within one $\mathcal P_a$ part, the group permits arbitrary
permutations of its $\ell^{b-a}$ children in $\mathcal P_b$.
The partition $\mathcal P_i$ groups these children into
$\ell^{i-a}$ families, each of size $\ell^{b-i}$.
Transposing one child from each of two families, while fixing
all other children, does not preserve that grouping. The
transposition lifts to an element of $G_j(I)$ by transporting
the identical descendant trees of the two exchanged children.
Thus $G_j(I)$ does not preserve $\mathcal P_i$. This property
distinguishes the groups for different subsets $I$.
\end{proof}

\begin{corollary}\label{tree:universal-probability}
Let
\[
 B_q(n)=\bigcap_{C\in\mathcal C_q(n)}\PAut(C).
\]
For $n=\ell^j$ under~\eqref{tree:primitive-root}, one has
\[
 B_q(\ell^j)=G_j(\{1,\ldots,j-1\}),\qquad
 |B_q(\ell^j)|=(\ell!)^{(\ell^j-1)/(\ell-1)}.
\]
Exactly four of the $2^{j+1}$ labelled cyclic codes satisfy
$\PAut(C)=B_q(\ell^j)$. Therefore
\begin{equation}\label{tree:probability}
 \Pr_{C\in\mathcal C_q(\ell^j)}
       \bigl(\PAut(C)=B_q(\ell^j)\bigr)
       =2^{1-j}\longrightarrow0.
\end{equation}
Conditioning the code on having least common shift period
$\ell^j$ gives exactly the same probability. In fact each subset
$I\subseteq\{1,\ldots,j-1\}$ occurs for exactly four codes,
or for exactly two codes under this conditioning.
Consequently the automorphism group is uniformly distributed
over the $2^{j-1}$ distinct canonical groups $G_j(I)$, and
\begin{equation}\label{tree:no-concentration}
 \sup_{H\le\Sym(\ell^j)}
 \Pr_{C\in\mathcal C_q(\ell^j)}\bigl(\PAut(C)=H\bigr)
 =2^{1-j}\longrightarrow0.
\end{equation}
The same equality holds under the full shift-period conditioning.
\end{corollary}
\begin{proof}
Every group in~\eqref{tree:group-classification} contains the
stabilizer of the complete chain. Taking $b_1,\ldots,b_j$
alternating attains that stabilizer, proving the intersection
formula. If a level is omitted, one branching number $a_h$ is
at least $\ell^2$. The full symmetric action on its $a_h$
children does not preserve an omitted intermediate partition
into $\ell$-power-sized parts. It supplies an element of
$G_j(I)$ outside the complete-chain stabilizer. Thus equality
with $B_q(\ell^j)$ holds precisely when every interior level
is selected.

For any prescribed $I$, choose $b_0$ and $b_1$ arbitrarily;
the conditions $b_i\ne b_{i+1}$ exactly at the indices in
$I$ determine all remaining bits uniquely. This gives four
codes. Their least common shift period is $\ell^j$ exactly
when $b_j=1$, because $E_j$ is the unique primitive constituent.
For fixed $I$, changing $b_1$ complements every bit from $b_1$
through $b_j$, while $b_0$ is independent. Hence $b_j=1$ selects
two of the four codes for each $I$, out of $2^j$ conditioned codes
in total. The displayed probabilities
follow, using the pairwise distinctness of the groups proved in
Corollary~\ref{tree:exact-groups}.
\end{proof}

For example, $(q,\ell)=(2,3)$ satisfies \eqref{tree:primitive-root}.
At length nine there are eight labelled binary cyclic codes. Four have
group $\Sym(9)$ of order $362880$, and four have group
$\Sym(3)\wr\Sym(3)$ of order $(3!)^4=1296$. Of the four codes with
full shift period, two have each group. Thus even replacing the affine
group by the universal permutation group does not make a
pointwise ``almost every code'' assertion valid along all
eligible lengths. This obstruction persists after conditioning
on the full shift period and after the four elementary codes
are removed. For $j\ge2$, removing those four codes deletes
exactly the group $G_j(\varnothing)=\Sym(\ell^j)$ from the
uniform distribution; each remaining group has probability
$4/(2^{j+1}-4)$, which still tends to zero. Thus no sequence
of deterministic subgroups $H_j\le\Sym(\ell^j)$ can be the
typical exact automorphism group in any of these labelled-code
models.

\subsection{Why the characteristic must remain fixed}

The alphabet supremum in Theorem~\ref{thm:main} cannot be extended
to all characteristics. The following consequence of the arithmetic
obstruction in \cite[Theorem B]{WX} makes this distinction precise.
We include a direct coding-theoretic proof; no novelty is claimed
for the underlying obstruction.

\begin{proposition}\label{scope:varyingchar}
For $n\ge1$, put
\[
 e_*(n)=\sup_{\substack{q\ \mathrm{prime\ power}\\(q,n)=1}}
                    \epsilon_q(n).
\]
If $n$ has a prime divisor $\ell\ge5$ with $v_\ell(n)=1$, then
there is a prime $p\nmid n$ such that every cyclic code of length
$n$ over $\F_p$ has an automorphism outside $A_p(n)$.
Consequently $e_*(n)=1$ on a set of natural density one and
\[
 \frac1X\sum_{n\le X}e_*(n)\longrightarrow1.
\]
The same statements hold when each code law is conditioned on
nondegeneracy.
\end{proposition}
\begin{proof}
Write $n=\ell B$, with $(\ell,B)=1$, and choose a primitive root
$g$ modulo $\ell$. The Chinese remainder theorem specifies a
reduced residue class by
\[
 p\equiv g\pmod\ell,\qquad p\equiv1\pmod B.
\]
Dirichlet's theorem supplies a prime in this class
\cite{Davenport}. Multiplication by $p$ on $\Z_\ell$ has the two
orbits $\{0\}$ and $\Z_\ell\setminus\{0\}$. Therefore
\[
 \F_p[C_\ell]=R\oplus U,
\]
where $R$ is the constant subspace and $U$ the augmentation
subspace, and both are irreducible cyclic constituents.
Since $p\equiv1\pmod B$, the algebra $\F_p[C_B]$ splits into
$B$ one-dimensional summands. In the fixed CRT coordinates,
every cyclic code of length $n$ independently chooses $0,R,U$,
or $R\oplus U$ in each summand of the second factor.
It follows that the diagonal group
$\Sym(\ell)\times1$ preserves every such code.

A transposition in this diagonal group is outside $A(n)$.
Indeed an affine permutation of $\Z_n$ induces an affine
permutation of $\Z_\ell$, while a nonidentity affine permutation
of a prime field fixes at most one point. A transposition fixes
$\ell-2\ge3$ points. Hence every code is exceptional, including
every nondegenerate code.

For a finite set $S$ of primes at least five, the integers having
$v_\ell(n)\ne1$ for every $\ell\in S$ have natural density
\[
 \prod_{\ell\in S}\left(1-\frac1\ell+\frac1{\ell^2}\right),
\]
by the Chinese remainder theorem modulo $\prod_{\ell\in S}\ell^2$.
This product tends to zero as $S$ increases, since
$\sum_{\ell\ge5}1/\ell$ diverges and $\sum_\ell1/\ell^2$ converges.
Thus the lengths covered by the first assertion have density one.
Since $0\le e_*(n)\le1$, the averaged conclusion follows.
\end{proof}

\section{Composite order carriers and the limits of the pointwise hypothesis}
\label{sec:composite-carriers}

The full-order-prime condition in Theorem~\ref{gl:criterion} is a
convenient sufficient hypothesis, but its exact role is to provide
many Fourier labels on which the relevant Frobenius orbit length is
large. A single prime is not essential for that mechanism. On the
other hand, the exact prime-power obstruction in
Corollary~\ref{tree:universal-probability} shows that no replacement
can yield pointwise convergence along every simple-root length.
We make both statements precise.

\subsection{A composite-carrier lower bound}

For coprime positive integers $A,B$ with $p\nmid AB$, put
$h=\ord_A(p)$ and retain the notation $R_{p,A}(B)$ from
Section~\ref{sec:fixedT}. For each prime $r\mid B$, write
$r^{e_r}\Vert B$.

\begin{lemma}[Composite order-carrier bound]
\label{lem:composite-carrier}
Let $w\mid B$ satisfy
\[
 h\mid\ord_w(p),
 \qquad w=\prod_{r\mid w}r^{f_r},\quad 1\le f_r\le e_r.
\]
Then
\begin{equation}\label{eq:composite-carrier-bound}
 R_{p,A}(B)\ge
 \frac{B}{m_{AB}}
 \prod_{r\mid w}
 \left(1-r^{-(e_r-f_r+1)}\right),
 \qquad m_{AB}=\ord_{AB}(p).
\end{equation}
In particular, if $w$ is squarefree, the product in
\eqref{eq:composite-carrier-bound} is
$\prod_{r\mid w}(1-r^{-e_r})$.
\end{lemma}
\begin{proof}
The additive-order-$b$ labels in $\mathbb Z_B$ form a set of size
$\varphi(b)$, and every $p$-orbit on this set has length
$\ord_b(p)$. Consequently
\begin{equation}\label{eq:R-exact-coset-sum}
 R_{p,A}(B)=
 \sum_{\substack{b\mid B\\h\mid\ord_b(p)}}
       \frac{\varphi(b)}{\ord_b(p)}.
\end{equation}
If $w\mid b$, then $h\mid\ord_w(p)\mid\ord_b(p)$, so every
such divisor $b$ occurs in the sum. Also
$\ord_b(p)\mid\ord_{AB}(p)=m_{AB}$, whence
\[
 R_{p,A}(B)\ge\frac1{m_{AB}}
       \sum_{\substack{b\mid B\\w\mid b}}\varphi(b).
\]
The divisor sum is multiplicative. For $r^{e_r}\Vert B$, its local
factor is $r^{e_r}$ if $r\nmid w$, and, if $r^{f_r}\Vert w$,
\[
 \sum_{j=f_r}^{e_r}\varphi(r^j)
 =r^{e_r}-r^{f_r-1}
 =r^{e_r}\left(1-r^{-(e_r-f_r+1)}\right).
\]
Multiplying these local factors proves
\eqref{eq:composite-carrier-bound}.
\end{proof}

The prime witness used in Theorem~\ref{gl:criterion} is the special
case $w=z$, $f_z=1$, and $\ord_z(p)=m_{AB}$. The lemma also applies
when the full order is assembled from several prime-power factors.
For example, if $w$ is squarefree and
$\operatorname{lcm}_{r\mid w}\ord_r(p)=m_{AB}$, then
\[
 R_{p,A}(B)\ge
 \frac{B}{m_{AB}}\prod_{r\mid w}(1-r^{-e_r}).
\]
Thus the amplification argument does not intrinsically require one
prime of full order. What it requires is a carrier outside the
$A$-part whose order covers $h=\ord_A(p)$, together with enough
weighted labels to dominate the logarithm of the number of actual
embeddings being counted.

For later use, note the particularly simple case $A=1$. With the
standard convention $\ord_1(p)=1$, taking $w=B$ in
Lemma~\ref{lem:composite-carrier} gives
\begin{equation}\label{eq:composite-carrier-A1}
 R_{p,1}(B)=m_p(B)\ge\frac{\varphi(B)}{\ord_B(p)}.
\end{equation}
Thus the relevant orbit mass can be large even when the full order is
assembled from several proper prime orders.

\subsection{An infinite pointwise family with no full-order prime}

We next construct lengths on which several prime orders are needed.
The lower bound on the number of primitive constituents controls the
affine event. The possible projective degrees in the structural
reduction can be excluded by divisibility.

\begin{theorem}[Composite-carrier pointwise family]
\label{thm:composite-carrier-family}
Fix a prime $p$. There exists a sequence of squarefree integers
$n_k$ such that, with $m_k=\ord_{n_k}(p)$,
\begin{enumerate}
\item $m_k\to\infty$ and $n_k\mid p^{m_k}-1$;
\item no prime divisor $r$ of $n_k$ satisfies
      $\ord_r(p)=m_k$;
\item nevertheless,
\[
 \epsilon_p(n_k)\longrightarrow0,
 \qquad
 \epsilon_p^{\rm nd}(n_k)\longrightarrow0.
\]
\end{enumerate}
Here $\epsilon_p^{\rm nd}$ denotes the exceptional probability
conditioned on nondegeneracy, equivalently on full shift period.
\end{theorem}
\begin{proof}
Put $c_p=\operatorname{lcm}(2,p-1)$. For each integer $k\ge2$,
choose $k$ distinct primes $s_1<\cdots<s_k$ exceeding
$\max\{6,k,p,c_p\}$. The infinitude of primes suffices for this choice.
Put $d_i=c_ps_i$. Since every $d_i>6$, Zsigmondy's theorem
\cite{Zsigmondy} gives a prime $r_i$ dividing $p^{d_i}-1$ which does not divide $p^j-1$
for any $1\le j<d_i$. Thus
\[
 \ord_{r_i}(p)=d_i,
 \qquad d_i\mid r_i-1.
\]
The primes $r_i$ are distinct because their multiplicative orders
are distinct. Define
\[
 m_k=c_p\prod_{i=1}^k s_i,
 \qquad n_k=\prod_{i=1}^k r_i.
\]
Since $(c_p,s_i)=1$ and the $s_i$ are pairwise distinct,
$\operatorname{lcm}(d_1,\ldots,d_k)=m_k$. Hence
\[
 \ord_{n_k}(p)=m_k,
 \qquad n_k\mid p^{m_k}-1.
\]
For $k\ge2$, every $d_i<m_k$, so no prime divisor of $n_k$
has full order $m_k$. Moreover, since $n_k$ is squarefree,
\[
 \frac{\varphi(n_k)}{m_k}
 =\frac{\prod_i(r_i-1)}{c_p\prod_i s_i}
 \ge\frac{\prod_i(c_ps_i)}{c_p\prod_i s_i}
 =c_p^{k-1}\ge2^{k-1}.
\]
Also $\min_i d_i\to\infty$ with $k$.

We now apply the exhaustive structural reduction of
Theorem~\ref{close:exhaustion}. First consider the canonical
translation obstructions. Since $n_k$ is squarefree, there are no
repeated-prime witnesses. For any distinct primes $\ell,r\mid n_k$,
the divisor $n_k$ itself occurs in the positive sum defining
$\Delta_p(n_k;\ell,r)$. Hence
\[
 \Delta_p(n_k;\ell,r)
 \ge\frac{\varphi(n_k)}{\ord_{n_k}(p)}
 =\frac{\varphi(n_k)}{m_k}
 \ge2^{k-1}.
\]
There are at most $k(k-1)$ ordered pairs, and therefore
\[
 U_p(n_k)\le k(k-1)2^{-2^{k-1}}\longrightarrow0.
\]
Furthermore $\operatorname{ord}_{r_i}(p)=d_i$ is even for every
$i$, because $c_p$ is even. Thus $h_p(n_k)=k$. Lemma~\ref{ab:loss} gives
\[
 \Gamma_p(n_k;a)\ge2^{k-2}
 \quad(a\mid n_k,\ a\ge5).
\]
As $n_k$ is squarefree, it has $2^k$ divisors. Thus
\[
 V_p(n_k)\le2^k2^{-2^{k-2}}\longrightarrow0.
\]
The affine and degeneracy bounds in
Section~\ref{sec:affine} also tend to zero, since
$M_p(n_k)=\varphi(n_k)/m_k\ge2^{k-1}$.

Since $n_k$ is squarefree, $\Delta(n_k)=1$, so the structural
reduction is applied at length $n_k$ itself. For large $k$, every
$r_i\ge d_i+1$ exceeds every prime dividing the order of a group in
the fixed finite family $\mathcal T_0$. Thus $(n_k,|T|)=1$ for each
$T\in\mathcal T_0$. If a subgroup $D\cong T^t$ of $\Sym(n_k)$ were
normalized by $\sigma_{n_k}$, then the cycle would permute its
orbits transitively. Their common size would divide both $n_k$
and $|D|=|T|^t$, hence would be one. This contradicts $D\ne1$.

We exclude every defining-characteristic projective degree directly.
Suppose for a contradiction that
\[
 v=\frac{p^{ad}-1}{p^a-1}\ \Bigm|\ n_k,\qquad a\ge1,\quad d\ge2.
\]
Since $v>1$ and $v\mid n_k$, some $r_i$ divides $v$.
For each such prime, $r_i\mid p^{ad}-1$ gives $d_i\mid ad$.
Thus $ad\ge\min_i d_i>6$ for large $k$. Zsigmondy's theorem gives a prime of
order $ad$ dividing $p^{ad}-1$. It does not divide $p^a-1$, so
it divides $v$ and equals some $r_j$. Hence $ad=c_ps_j$.
If $r_i\mid v$, the divisibility $c_ps_i\mid c_ps_j$ forces $i=j$.
Since $v>1$ and $n_k$ is squarefree, $v=r_j$ is prime.

We claim that primality of $v$ forces $ad$ to be a prime power.
Write $S_d(X)=1+X+\cdots+X^{d-1}$. For $d=uv'$ with $u,v'>1$,
$S_d(X)=S_u(X)S_{v'}(X^u)$ has two factors greater than one at
$X=p^a$. Thus $d=t$ is prime. Write $a=t^b h$ with $t\nmid h$
and put $x=p^{t^b}$. Modulo $S_t(x)$ one has $x^t\equiv1$, and
multiplication by $h$ permutes the residues modulo $t$. Therefore
\[
 S_t(x^h)=\sum_{j=0}^{t-1}x^{hj}
 \equiv \sum_{j=0}^{t-1}x^j=0\pmod{S_t(x)}.
\]
If $h>1$, then $1<S_t(x)<S_t(x^h)=v$, contradicting primality.
Thus $h=1$ and $ad=t^{b+1}$. But $ad=c_ps_j$, where $c_p\ge2$
and $s_j$ is a prime coprime to $c_p$. This contradiction excludes
$v$ for every $(a,d)$.

Every projective occurrence in the defining-characteristic alternative
of Theorem~\ref{close:exhaustion} would have one of the excluded
degrees as a divisor of $n_k$. Consequently
\[
 \epsilon_p(n_k)\le U_p(n_k)+V_p(n_k)+a_p(n_k)\longrightarrow0.
\] Since the probability
of full shift period is at least $1/2$, conditioning gives the
second asserted limit.
\end{proof}

\begin{remark}
For each prime field $\F_p$, Theorem~\ref{thm:composite-carrier-family}
gives pointwise concentration on a sequence with no full-order prime.
On this sequence a composite squarefree carrier can
assemble the full order from many smaller prime orders. The prime-power sequence $3^j$ in
Corollary~\ref{tree:universal-probability} shows at the same time
that no arithmetic hypothesis can be deleted altogether while
retaining the pointwise conclusion for every simple-root length.
The correct scope is therefore a family of alternative order-carrier
and orbit-mass conditions, not an unconditional all-length theorem.
\end{remark}

\section{Conclusion}\label{sec:conclusion}

For uniform labelled simple-root cyclic codes, the shift and Frobenius
multiplier generate the typical permutation automorphism group in a
characteristic-uniform length-average sense. A separate criterion based
on one large prime divisor of full multiplicative order gives pointwise
convergence. Stewart's theorem makes this criterion applicable to every
exponent in the primitive-length sequence and uniformly to the divisors
retaining the selected prime. The quantitative bound is smaller than
every fixed inverse power of the exponent. These conclusions persist
under full shift-period conditioning, and the pointwise conclusion also
holds in every fixed positive-width rate window about one half.

The argument treats selected sums of cyclic constituents directly.
Its structural steps preserve the actual coordinate actions through
cyclic-core descent and local projective covers; its probability steps
use the unconditional law of each fixed quotient map. This distinction
allows the group counts and composition estimates to be combined without
assuming that a permutation preserving a code preserves its individual
cyclic constituents.

The exact distribution at primitive-root prime-power lengths shows why
restrictions on the length sequence are necessary. At binary length
$3^j$, every individual automorphism group has probability at most
$2^{1-j}$, even after conditioning on full shift period. Thus no choice
of deterministic groups repairs a pointwise concentration statement at
all simple-root lengths.

Several questions remain for other sampling laws. Conditioning on one
exact dimension may substantially change the constituent distribution,
and the rate-window argument does not resolve that problem. Biased
constituent selection, sampling by permutation-equivalence classes, and
length distributions weighted by the number of codes require separate
estimates. Theorem~\ref{thm:composite-carrier-family} treats an explicit family
in which several prime divisors supply the full order. A general
arithmetic criterion covering other such length sequences remains open. The present theorems establish the stated probability
formulations; they do not classify all exceptional cyclic codes.


\end{document}